\documentclass[preprint,amsmath,amssymb,aps,pra,longbibliography]{revtex4-1}

\usepackage[margin=0.75in, top=1in, bottom=0.85in]{geometry}

\usepackage[final,letterspace=-0]{microtype}
\usepackage[colorlinks=true, linkcolor=blue, urlcolor=blue, citecolor=blue, anchorcolor=blue]{hyperref}

\usepackage{mathpazo} 
\usepackage{upgreek}

\usepackage{graphicx}
\usepackage{dcolumn}
\usepackage{bm}
\usepackage{helvet}

\begin{document}

\title{Space-time wave packets localized in all dimensions}

\author{Murat Yessenov$^{1,\dagger}$, Justin Free$^{2}$, Zhaozhong Chen$^{3}$, Eric G. Johnson$^{2}$, Martin P.~J. Lavery$^{3}$, Miguel A. Alonso$^{4,5}$, and Ayman F. Abouraddy$^{1,}$ }
\email{Corresponding authors: raddy@creol.ucf.edu}
\email{$^{\dagger}$yessenov@knights.ucf.edu}
\affiliation{$^{1}$ CREOL, The College of Optics \& Photonics, University of Central Florida, Orlando, Florida 32816, USA \\
$^{2}$ Micro-Photonics Laboratory, the Holcombe Department of Electrical and Computer Engineering, Clemson University, Clemson, South Carolina 29634, USA \\
$^{3}$ James Watt School of Engineering, University of Glasgow, UK\\
$^{4}$ CNRS, Centrale Marseille, Institut Fresnel, Aix Marseille Univ., Marseille, France\\
$^{5}$ The Institute of Optics, University of Rochester, Rochester, NY, USA }

\begin{abstract}
Optical wave packets that are localized in space and time, but nevertheless overcome diffraction and travel rigidly in free space, are a long sought-after field structure with applications ranging from microscopy and remote sensing, to nonlinear and quantum optics. However, synthesizing such wave packets requires introducing non-differentiable angular dispersion with high spectral precision in two transverse dimensions, a capability that has eluded optics to date. Here, we describe an experimental strategy capable of sculpting the spatio-temporal spectrum of a generic pulsed beam by introducing arbitrary radial chirp via two-dimensional conformal coordinate transformations of the spectrally resolved field. This procedure yields propagation-invariant `space-time' wave packets localized in all dimensions, with tunable group velocity in the range from $0.7c$ to $1.8c$ in free space, and endowed with prescribed orbital angular momentum. By providing unprecedented flexibility in sculpting the three-dimensional structure of pulsed optical fields, our experimental strategy promises to be a versatile platform for the emerging enterprise of space-time optics.
\end{abstract}

\maketitle

\noindent

\section*{Introduction}

Creating spatio-temporally localized optical wave packets that overcome diffraction and propagate rigidly in free space has been a long-standing yet elusive goal in optics. Such wave packets can have applications ranging from remote optical sensing and biological imaging, to nonlinear and quantum optics. To date, this challenge has been addressed via nonlinear optical effects that sustain solitons \cite{Malomed05JOB}, waveguiding structures \cite{SalehBook07}, or by exploiting particularly shaped waveforms such as Bessel-Airy wave packets in linear dispersive media \cite{Chong10NP}. Propagation invariance in a linear non-dispersive medium necessitates inculcating a precise spatio-temporal spectral structure into the field by introducing angular dispersion (AD) \cite{Fulop10Applications,Torres10AOP}; i.e., associating each wavelength with a single propagation direction \cite{Donnelly93ProcRSLA,Turunen10PO}. Examples of such wave packets date back to Brittingham’s focus-wave mode \cite{Brittingham83JAP}, X-waves \cite{Saari97PRL,Grunwald2003PRA}, and more recently the general class of `space-time' (ST) wave packets \cite{Kondakci16OE,Parker16OE,Wong17ACSP2,Porras17OL,Efremidis17OL,Kondakci17NP,Yessenov19PRA,Yessenov19OPN,Wong21OE}. The challenge of producing the AD necessary for propagation-invariant wave packets localized in all dimensions (referred to hereon as 3D ST wave packets) is twofold. First, the AD must be inculcated in two transverse dimensions rather than in one as typically realized via gratings or prisms \cite{Fulop10Applications,Torres10AOP}. Second, non-differentiable AD is required \cite{Hall21OE1}; i.e., it is necessary that the derivative of the wavelength-dependent propagation angle not be defined at some wavelength \cite{Yessenov21ACSPhot,Hall21OL} -- a field configuration that cannot be directly produced with conventional optical components. Consequently, with the exception of X-waves that are AD-free, no propagation-invariant optical wave packets that are localized in all dimensions have been observed in free space \cite{Turunen10PO}.

The challenge of introducing arbitrary AD into a generic pulsed beam along one transverse dimension has been recently addressed by constructing a universal AD synthesizer \cite{Hall21OE2}. This experimental strategy has enabled the realization of ST wave packets in the form of light sheets \cite{Kondakci17NP} (referred to hereon as 2D ST wave packets), which exhibit a broad host of sought-after effects, such as long-distance propagation invariance \cite{Bhaduri19OL}, tunable group velocities \cite{Salo01JOA,Valtna07OC,ZamboniRached2008PRA,Wong17ACSP2,Kondakci19NC,Yessenov20NC}, anomalous refraction at planar interfaces \cite{Bhaduri20NatPhot}, and the space-time Talbot effect \cite{Hall21APL}. Although this arrangement produces non-differentiable AD with high spectral resolution, these features cannot be extended to both transverse dimensions. Crucially, the centerpiece of this configuration is a spatial light modulator that modifies the temporal spectrum along one dimension, leaving only one dimension to manipulate the field spatially -- a limitation that is shared by other recently investigated spatio-temporal field structures \cite{Vaughan03OL,Jhajj2016PRX,Hancock2019Optica,Chong20NP,Cao2021PR,Wan2021arxiv,Hancock2021Optica,Hancock2021PRL}. Therefore, the fundamental challenge of producing non-differentiable AD encompassing both transverse dimensions remains outstanding.

Here, we demonstrate a spatio-temporal modulation strategy that efficiently produces arbitrary yet precise AD in two transverse dimensions, and thus yields ST wave packets localized in all dimensions -- while preserving all the key attributes of its reduced-dimension counterpart. This modulation scheme is implemented in three stages. In the first stage, the spectrum of a generic plane-wave pulse is spatially resolved along one dimension after a double-pass through a volume chirped Bragg grating. In the second stage, a spectral transformation `reshuffles' the wavelengths into a prescribed sequence. In the third stage, a log-polar-to-Cartesian conformal coordinate transformation converts the spatial locus of each wavelength from a line into a circle \cite{Bryngdahl74JOSA,Hossack87JOMO}. A lens finally converts the spectrally resolved wave front into a 3D ST wave packet localized in all three dimensions. Utilizing this approach, we produce 3D ST wave packets with $\approx\!30$~$\mu$m transverse beam width and $\approx\!6$~ps pulse width that propagate for over $50$~mm. Moreover, by modulating the spatio-temporal spectral structure, we realize group velocities extending from the subluminal to the superluminal regimes over the range from $0.7c$ to $1.8c$ ($c$ is the speed of light in vacuum). Furthermore, by providing access to both transverse dimensions in a ST wave packet, new degrees of freedom of the optical field can be accessed, such as orbital angular momentum (OAM) \cite{Allen92PRA,Ornigotti15PRL,Porras2022PRA}. Specifically, by encoding a helical phase structure in the spatio-temporal spectrum, we demonstrate propagation-invariant pulsed OAM wave packets with controllable group velocity in free space, which we refer to as ST-OAM wave packets. In addition to the propagation-invariance and arbitrary group velocities of ST-OAM wave packets, their underlying spatio-temporal structure may lead to variations of some of the recently uncovered behaviors of conventional OAM pulses, such as the trade-off between the topological charge and pulse duration  \cite{Ornigotti15PRL,Porras19PRL,Porras20PRA}. Such 3D ST wave packets that are fully localized in all dimensions have potential uses in areas such as free-space optical communications, imaging, and nonlinear optics.


\section*{Results}

\subsection*{Theory of 3D space-time wave packets}

\begin{figure*}[t!]
\centering
\includegraphics[width=8cm]{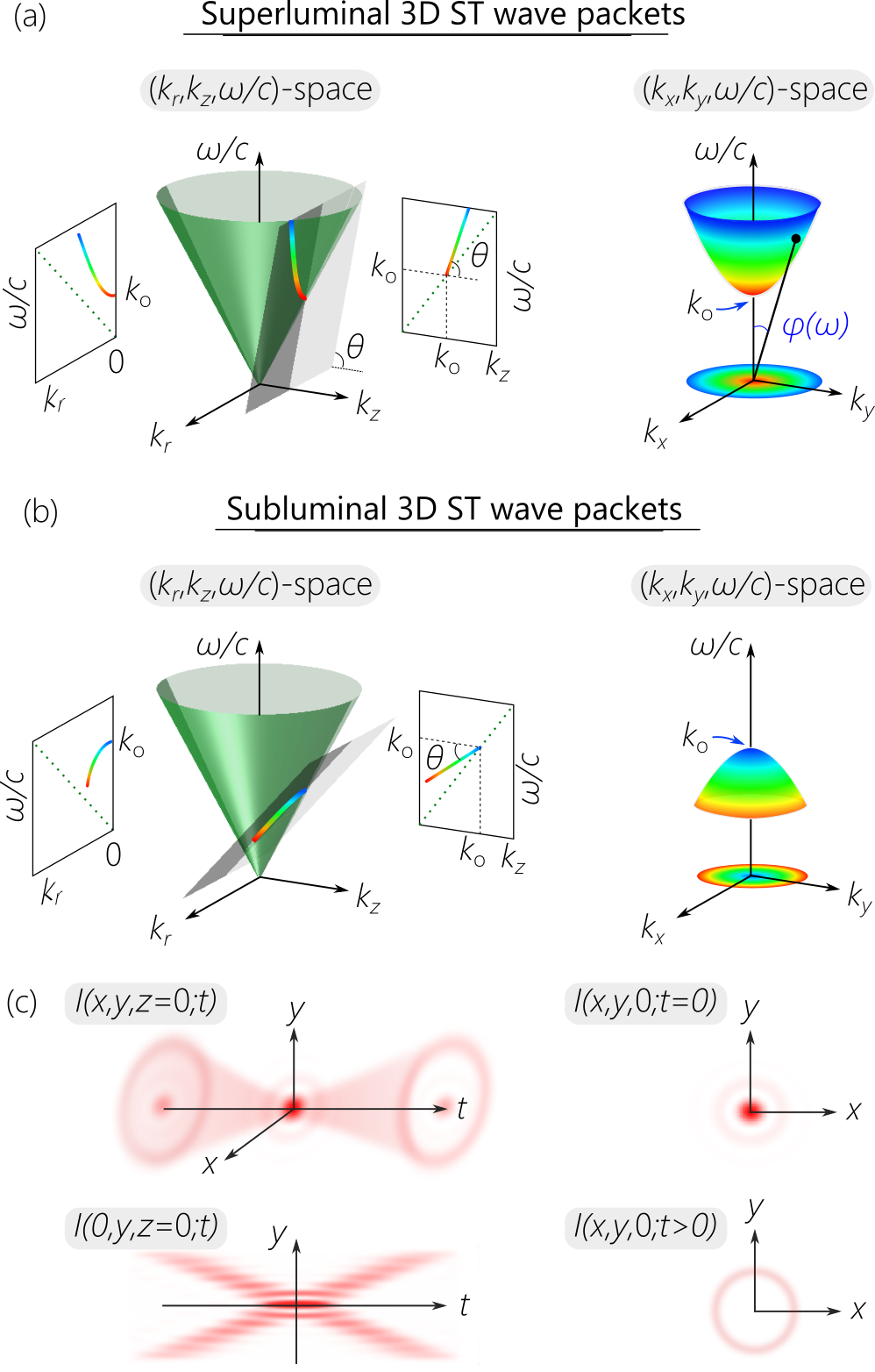}
\caption{\textbf{Visualization of the spectral support domain for 3D ST wave packets on the surface of the free-space light-cone.} \textbf{a} The spectral support domain for a superluminal 3D ST wave packet at the intersection of the light-cone $k_{r}^{2}+k_{z}^{2}\!=\!(\tfrac{\omega}{c})^{2}$ with a spectral plane that is parallel to the $k_{r}$-axis and makes an angle $\theta\!>\!45^{\circ}$ with the $k_{z}$-axis. The conic section at the intersection is a hyperbola. In $(k_{x},k_{y},\tfrac{\omega}{c})$-space the spectrum is one half of a two-sheet hyperboloid (an elliptic hyperboloid). \textbf{b} Same as \textbf{a} for a subluminal ST wave packet with $\theta\!<\!45^{\circ}$, where the spectral support domain on the light-cone in $(k_{r},k_{z},\tfrac{\omega}{c})$-space is an ellipse. In $(k_{x},k_{y},\tfrac{\omega}{c})$-space, the spectrum is an ellipsoid of revolution (a spheroid, which may be prolate or oblate according to the value of $\theta$). \textbf{c} Plot of the spatio-temporal intensity profile $I(x,y,z=0;t)$ at a fixed axial plane $z\!=\!0$, the intensity profile in a meridional plane $I(0,y,z=0;t)$, and the transverse profiles at the wave-packet center $I(x,y,0;0)$ and off-center $I(x,y,0;t>0)$.}
\label{Fig:Concept}
\end{figure*}

A useful conceptual tool for understanding the characteristics of ST wave packets and the requirements for their synthesis is to visualize their spectral support domain on the surface of the light-cone. The light-cone is the geometric representation of the free-space dispersion relationship $k_{x}^{2}+k_{y}^{2}+k_{z}^{2}\!=\!(\tfrac{\omega}{c})^{2}$, where $\omega$ is the temporal frequency, $c$ is the speed of light in vacuum, $(k_{x},k_{y},k_{z})$ are the components of the wave vector in the Cartesian coordinate system $(x,y,z)$, $x$ and $y$ are the transverse coordinate, and $z$ is the axial coordinate. Although this relationship corresponds to the surface of a four-dimensional hypercone, a useful representation follows from initially restricting our attention to azimuthally symmetric fields in which $k_{x}$ and $k_{y}$ are combined into a radial wave number $k_{r}\!=\!\sqrt{k_{x}^{2}+k_{y}^{2}}$, so that the light-cone can be then visualized in $(k_{r},k_{z},\tfrac{\omega}{c})$-space (Fig.~\ref{Fig:Concept}). The spectral support domain for 3D ST wave packets is restricted to the conic section at the intersection of the light-cone with a spectral plane that is parallel to the $k_{r}$-axis and makes an angle $\theta$ (the spectral tilt angle) with the $k_{z}$-axis, which is given by the equation $\Omega\!=\!(k_{z}-k_{\mathrm{o}})c\tan{\theta}$; here $\Omega\!=\!\omega-\omega_{\mathrm{o}}$, $\omega_{\mathrm{o}}$ is a carrier frequency, and $k_{\mathrm{o}}\!=\!\omega_{\mathrm{o}}/c$. It can be readily shown that such a construction in the narrowband paraxial regime results in a propagation-invariant 3D ST wave packet $E(r,z;t)\!=\!e^{i(k_{\mathrm{o}}z-\omega_{\mathrm{o}}t)}\psi(r,z;t)$, where the slowly varying envelope $\psi(r,z;t)$ travels rigidly at a group velocity $\widetilde{v}\!=\!c\tan{\theta}$, $\psi(r,z;t)\!=\!\psi(r,0;t-z/\widetilde{v})$, where $\psi(r,0;t)\!=\!\int\!dk_{r}\,\,k_{r}\widetilde{\psi}(k_{r})J_{0}(k_{r}r)e^{-i\Omega t}$, and $\widetilde{\psi}(k_{r})$ is the spectrum. Here $k_{r}$ and $\Omega$ are no longer independent variables, but are instead related via the particular spectral trajectory on the light-cone [Supplementary Note~1]. Although this spectral trajectory is a conic section whose kind is determined by the spectral tilt angle $\theta$, it can nevertheless be approximated in the narrowband paraxial regime by a parabola in the vicinity of $k_{r}\!=\!0$:
\begin{equation}\label{Eq:Parabola}
\frac{\Omega}{\omega_{\mathrm{o}}}=\frac{k_{r}^{2}}{2k_{\mathrm{o}}^{2}(1-\widetilde{n})},
\end{equation}
where $\widetilde{n}\!=\!\cot{\theta}$ is the wave-packet group index in free space. By setting $k_{r}\!=\!k\sin{\varphi(\omega)}$, where $\varphi(\omega)$ is the propagation angle for $\omega$ as shown in Fig.~\ref{Fig:Concept}a, we have $\varphi(\omega)\!\approx\!\eta\sqrt{\tfrac{\Omega}{\omega_{\mathrm{o}}}}$, which is not differentiable at $\Omega\!=\!0$ \cite{Hall21OE1,Hall21OE2}; here $\widetilde{n}\!=\!1-\tfrac{\sigma}{2}\eta^{2}$, $\sigma\!=\!1$ in the superluminal regime, and $\sigma\!=\!-1$ in the subluminal regime. In other words, non-differentiable AD is required to produce a propagation-invariant ST wave packet. This result is similar to that for ST light-sheets \cite{Kondakci17NP} except that the transverse coordinate $x$ is now replaced with the radial coordinate $r$.

The representation in Fig.~\ref{Fig:Concept} is particularly useful in identifying a path towards synthesizing 3D ST wave packets. When $45^{\circ}\!<\!\theta\!<\!90^{\circ}$, the ST wave packet is superluminal $\widetilde{v}\!>\!c$, $\Omega$ is positive, and $\omega_{\mathrm{o}}$ is the minimum allowable frequency in the spectrum. When viewed in $(k_{x},k_{y},\tfrac{\omega}{c})$-space, the wavelengths are arranged in concentric circles, with long wavelengths (low frequencies) at the center, and shorter wavelengths (higher frequencies) extending outward. On the other hand, when $0^{\circ}\!<\!\theta\!<\!45^{\circ}$, the ST wave packet is subluminal $\widetilde{v}\!<\!c$, $\Omega$ is negative, and $\omega_{\mathrm{o}}$ is the maximum allowable frequency in the spectrum. The wavelengths are again arranged in concentric circles in $(k_{x},k_{y},\tfrac{\omega}{c})$-space -- but in the opposite order: short wavelengths are close to the center and longer wavelengths extend outward. For both subluminal and superluminal 3D ST wave packets, each $\omega$ is associated with a single radial spatial frequency $k_{r}(\omega)$, and is related to it via the relationship in Eq.~\ref{Eq:Parabola}. This representation indicates the need for arranging the wavelengths in concentric circles with square-root radial chirp, and then converting the spatial spectrum into physical space via a spherical lens. Moreover, adding a spectral phase factor $e^{i\ell\chi}$, where $\ell$ is an integer and $\chi$ is the azimuthal angle in spectral space, produces OAM in physical space [Supplementary Note~1B].

Closed-form expressions can be obtained for 3D ST wave packets by applying Lorentz boosts to an appropriate initial field \cite{Belanger86JOSAA,Saari04PRE,Longhi04OE,Kondakci18PRL}. For example, starting with a monochromatic beam $E_{\mathrm{o}}(r,z;t)$, a subluminal 3D ST wave packet at a group velocity $\widetilde{v}$ is obtained by the Lorentz boost $E(r,z;t)\!=\!E_{\mathrm{o}}(r,\tfrac{z-\widetilde{v}t}{\sqrt{1-\beta^{2}}};\tfrac{t-\widetilde{v}z/c^{2}}{\sqrt{1-\beta^{2}}})$, where $\beta\!=\!\tfrac{\widetilde{v}}{c}$ is the Lorentz factor. On the other hand, closed-form expressions for superluminal 3D ST wave packets can be obtained by applying a Lorentz boost to the `needle beam' in \cite{Parker16OE}. The time-averaged intensity is $I(r,\varphi,z)\!=\!2\pi k_{\mathrm{o}}^2(1-\widetilde{n})\int dk_rk_r^2|\widetilde{\psi}(k_r)|^2J_\ell^2(k_rr)$, which is independent of $\varphi$ even if the field is endowed with OAM. In the case of 2D ST light-sheets, the time-averaged intensity separates into a sum of a constant background pedestal and a spatially localized feature at the center \cite{Kondakci17NP}. A similar decomposition is not possible for 3D ST wave packets. However, using the asymptotic form for Bessel functions that is valid far from $r\!=\!0$, we have:
\begin{equation}
I(r)\approx\frac{\!2\pi k_{\mathrm{0}}^2(1-\widetilde{n})}{\pi r}\int\!dk_{r} \sqrt{k_{\mathrm{o}}^{2}+k_{r}^{2}}|\widetilde{\psi}(k_{r})|^{2}+\frac{\!2\pi k_{\mathrm{o}}^2(1-\widetilde{n})(-1)^{\ell}}{\pi r}\int\!dk_{r}\sqrt{k_{\mathrm{o}}^{2}+k_{r}^{2}}|\widetilde{\psi}(k_{r})|^{2}\sin{(2k_{r}r)},
\end{equation}
where the first term is a pedestal decaying at a rate of $\tfrac{1}{r}$, and the second term tends to be localized closer to the beam center. In the vicinity of $r\!=\!0$, the two terms merge and cannot be separated. The spatio-temporal intensity profile of such a 3D ST wave packet is depicted in Fig.~\ref{Fig:Concept}c: two conic field structures emanate from the wave-packet center, such that the profile is X-shaped in any meridional plane containing the optical axis, and the intensity profile is circularly symmetric in any transverse plane.


\subsection*{Synthesizing ST wave packets localized in all dimensions}

\begin{figure*}[t!]
\centering
\includegraphics[width=8.5cm]{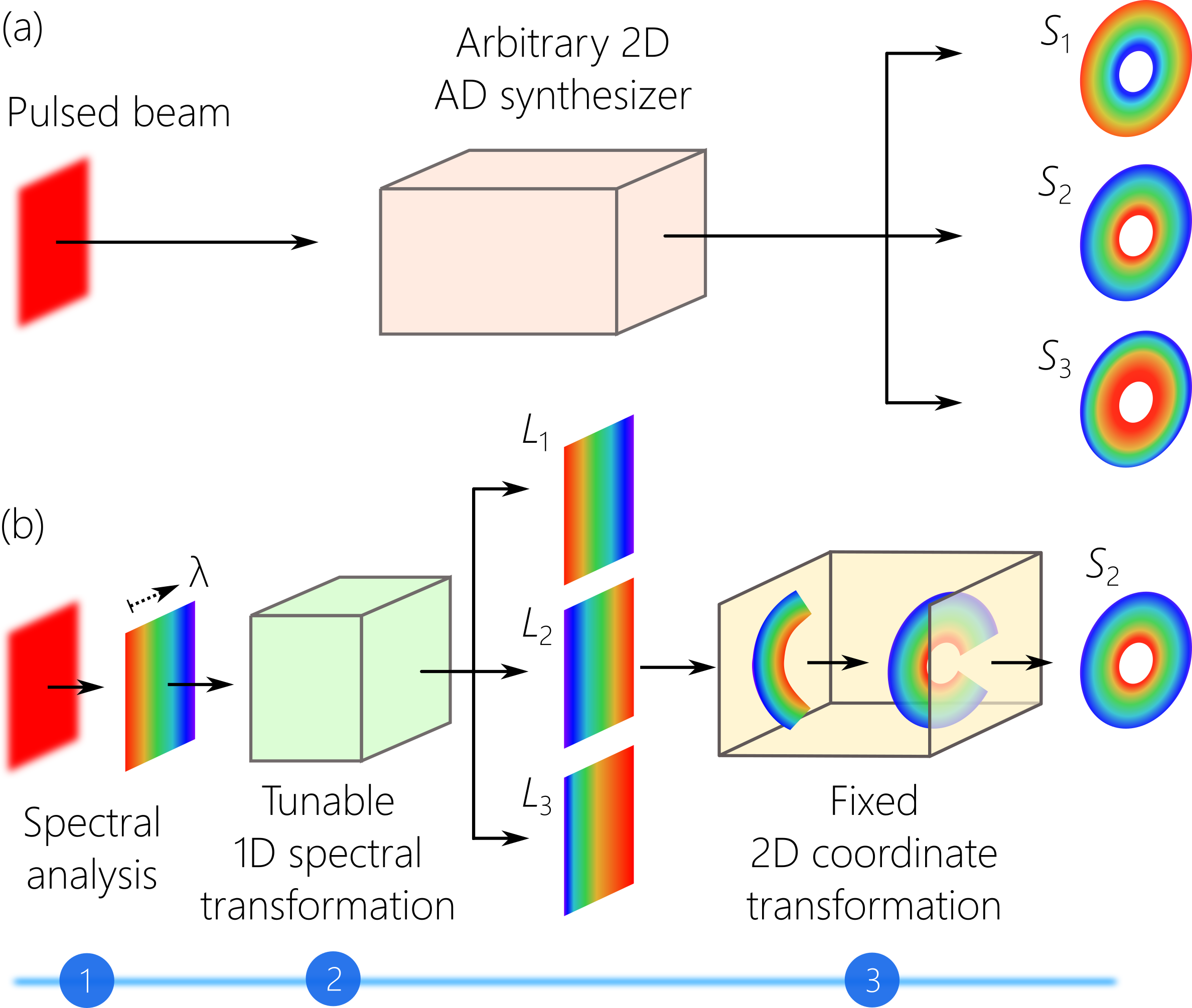}
\caption{\textbf{Synthesis strategy for 3D ST wave packets.} \textbf{a} Starting with a generic plane-wave pulse, we aim at constructing an angular-dispersion synthesizer in two dimensions that arranges the wavelengths in circles in a prescribed order. $S_{1}$ corresponds to a subluminal wave packet, whereas $S_{2}$ and $S_{3}$ correspond to superluminal wave packets of different group velocities. \textbf{b} The proposed strategy comprises spectral analysis followed by a tunable 1D spectral transformation that rearranges the initial wavelength sequence in the spectrally resolved wave front. The 1D spectra $L_{1}$, $L_{2}$, and $L_{3}$ are rectilinear counterparts of $S_{1}$, $S_{2}$, and $S_{3}$ in \textbf{a}. In the third stage, a fixed 2D conformal coordinate transformation converts vertical lines into circles, thereby realizing the targeted spatio-temporal spectra $S_{1}$, $S_{2}$, and $S_{3}$.}
\label{Fig:SynthesisMethod}
\end{figure*}

Central to converting a generic pulsed beam into a ST wave packet localized in all dimensions is the construction of an optical scheme that can associate each wavelength $\lambda$ with a particular azimuthally symmetric spatial frequency $k_{r}(\lambda)$ and arrange the wavelengths in concentric circles with the order prescribed in Eq.~\ref{Eq:Parabola} [Fig.~\ref{Fig:SynthesisMethod}a]. This system realizes two functionalities, producing a particular wavelength sequence, and changing the coordinate system, which are implemented in succession via the three-stage strategy outlined in Fig.~\ref{Fig:SynthesisMethod}b. In the first stage, the spectrum of a plane-wave pulse is resolved along one spatial dimension. At this point, the field is endowed with linear spatial chirp and the wavelengths are arranged in a fixed sequence. The second stage rearranges the wavelengths in a new prescribed sequence. This spectral transformation is tunable; that is, a wide range of spectral structures can be obtained from a fixed input. In the third stage, a 2D conformal transformation converts the coordinate system to map the rectilinear chirp into a radial chirp; i.e., lines corresponding to different wavelengths at the input are converted into circles at the output \cite{Bryngdahl74JOSA,Hossack87JOMO}. Because the spectral transformation in the second stage is tunable, the 2D coordinate transformation can be held fixed. In this way, we obtain arbitrary (including non-differentiable) AD in two dimensions.

The layout of the experimental setup is depicted in Fig.~\ref{Fig:Setup}. We start off in the first stage with pulses from a Ti:sapphire laser (pulse width $\approx\!100$~fs and bandwidth $\approx\!10$~nm at a central wavelength of $\approx\!800$~nm). Because a flat-phase front is critical for successfully implementing the subsequent transformations, the use of conventional surface gratings is precluded, and we utilize instead a double-pass configuration through a volume chirped Bragg grating (CBG). The CBG resolves the spectrum horizontally along the $x$-axis and introduces linear spatial chirp so that $x_{1}(\lambda)\!=\!\alpha(\lambda-\lambda_{\mathrm{o}})$; where $\alpha$ is the linear spatial chirp rate \cite{Glebov14SPIE}, $\lambda_{\mathrm{o}}$ is a fixed wavelength, and the bandwidth utilized is $\Delta\lambda\!\approx\!0.3$~nm. It is crucial that this task be achieved with high spectral resolution. Previous studies have shown that the critical parameter determining the propagation distance of ST wave packets is the 'spectral uncertainty' $\delta\lambda$, which is the finite spectral uncertainty in the association between spatial and temporal frequencies  \cite{Yessenov19OE}. Our measurements indicate that the optimal spectral uncertainty after the CBG arrangement is $\delta\lambda\!\sim\!35$~pm, which is achieved for a 2-mm input beam width [Supplementary Fig.~11].

The second stage of the synthesis strategy is a 1D spatial transformation along the $x$-axis to rearrange the wavelength sequence, thereby implementing a spectral transformation. Specifically, each wavelength $\lambda$ is transposed from $x_{1}(\lambda)$ at the input via a logarithmic mapping to $x_{2}(\lambda)\!=\!A\ln{(\tfrac{x_{1}(\lambda)}{B})}$ at the output. This transformation is realized via two phase patterns implemented by a pair of spatial light modulators (SLMs) to enable tuning the transformation parameters $A$ and $B$. This particular `reshuffling' of the wavelength sequence pre-compensates the exponentiation included in the subsequent coordinate transformation. By tuning the value of $B$, we can vary the group velocity $\widetilde{v}$ over the subluminal and superluminal regimes [Supplementary Table~1].

In the third stage we perform a log-polar-to-Cartesian coordinate transformation: $(x_{2},y_{2})\rightarrow(r,\varphi)$ via the 2D mapping: $r(\lambda)\!=\!C\exp{(-\tfrac{x_{2}(\lambda)}{D})}$ and $\varphi=\tfrac{y_{2}}{D}$ \cite{Bryngdahl74JOSA,Hossack87JOMO}. The exponentiation here is pre-compensated by the logarithmic mapping in the 1D spectral transformation, and the wavelength at position $x_{2}(\lambda)$ at the input is converted into a circle of radius $r(\lambda)\!\propto\!(\lambda-\lambda_{\mathrm{o}})^{A/D}$ at the output. This 2D coordinate transformation was developed decades ago \cite{Bryngdahl74JOSA,Hossack87JOMO}, and was recently revived as a methodology for sorting OAM modes \cite{Berkhout10PRL,Lavery12OE}. We operate the system in reverse (lines-to-circles, rather than the more typical circles-to-lines \cite{Berkhout10PRL}), and we make use of a polychromatic field (rather than monochromatic field). The exponent of the chirp rate depends only on the ratio $\tfrac{A}{D}$, so that setting $D\!=\!2A$ yields $r(\lambda)\!\propto\!\sqrt{\lambda-\lambda_{\mathrm{o}}}$ in accordance with Eq.~\ref{Eq:Parabola}. The wavelengths are arranged with square-root radial chirp, thereby realizing the required non-differentiable AD. Finally, a spherical converging lens of focal length $f$ generates the 3D ST wave packets in physical space, equivalently mapping $r\rightarrow\!k_{r}\!=\!k\tfrac{r}{f}$.

The 2D coordinate transformation is performed with two different embodiments: using a pair of diamond-machined refractive phase plates \cite{Lavery12OE}, and using a pair of diffractive phase plates \cite{Li19OE}, which yielded similar performance. Because both of these realizations are stationary, the values of $C$ and $D$ are fixed. The data reported in Fig.~\ref{Fig:SpectralMeasurements} through Fig.~\ref{Fig:2DFieldMeasurements} made use of the refractive phase plates with $C\!=\!4.77$~mm and $D\!=\!1$~mm. Moreover, fixing the value of $D$ entails in turn fixing the value of $A$ to maintain $A\!=\!D/2$. The group velocity $\widetilde{v}\!=\!c/\widetilde{n}$ is tuned over the subluminal and superluminal regimes by varying $B$, whereby $\widetilde{n}\!\approx\!1-\tfrac{2.24}{B}$, with $B$ in units of mm [Supplementary Note~2].

This experimental strategy provides two pathways for introducing OAM into the 3D ST wave packet. One may utilize a conventional spiral phase plate to imprint an OAM order $\ell$ after the 2D coordinate transformation and before the final Fourier-transforming lens. Another approach, which we implemented here, is to add at the output of the 1D spectral transformation a linear phase distribution along $y$ extending from 0 to $2\pi\ell$, which is subsequently wrapped around the azimuthal direction after traversing the 2D coordinate transformation, thereby realizing OAM of order $\ell$ \cite{Li19OE}.

For the sake of benchmarking, we also synthesized pulsed Bessel beams with separable spatio-temporal spectrum by circumventing the spectral analysis and 1D spectral transformation, and sending the input laser pulses directly to the 2D coordinate transformation. To match the temporal bandwidth of the pulsed Bessel beams to that of the 3D ST wave packets, we spectrally filter $\Delta\lambda\!=\!0.3$~nm from the input spectrum via a planar Fabry-P{\'e}rot cavity.


\subsection*{Characterizing 3D ST wave packets}

To verify the structure of the synthesized 3D ST wave packet, we characterize the field in four distinct domains: (1) the spatio-temporal spectrum to verify the square-root radial chirp [Fig.~\ref{Fig:SpectralMeasurements}]; (2) the time-averaged intensity to confirm diffraction-free propagation along $z$ [Fig.~\ref{Fig:Time-averagedIntensity}]; (3) time-resolved intensity measurements to reconstruct the wave-packet spatio-temporal profile and estimate the group velocity [Fig.~\ref{Fig:Time-ResolvedIntensity}]; and (4) complex-field measurements to resolve the spiral phase of the ST-OAM wave packets [Fig.~\ref{Fig:2DFieldMeasurements}]. 

\noindent\textbf{Spectral-domain characterization.} We measure the spatio-temporal spectrum by scanning a single-mode fiber connected to an optical spectrum analyzer across the spectrally resolved field profile. We scan the fiber along $x_{1}$ after the spectral analysis stage and verify the linear spatial chirp [Supplementary Fig.~10], and then scan the fiber along $x_{2}$ after the 1D spectral transformation to confirm the implemented change in spatial chirp. The measurement is repeated for superluminal ($B=10$~mm, $\widetilde{v}\!\approx\!1.37c$) and subluminal ($B=-10$~mm, $\widetilde{v}\!\approx\!0.83c$) wave packets, both with temporal bandwidth $\Delta\lambda\approx0.3$~nm, pulse width of $\sim6$~ps, and $\lambda_{\mathrm{o}}\!=\!796.1$~nm. After the 2D coordinate transformation, the spectrum is arranged radially along an annulus rather than a rectilinear domain, as shown in Fig.~\ref{Fig:SpectralMeasurements}a. By calibrating the conversion $x_{2}\!\rightarrow\!r$ engendered by the 2D coordinate transformation, and combining with the measured spatial chirp $x_{2}(\lambda)$ at its input, we obtain the radial chirp $k_{r}(\lambda)$ as shown in Fig.~\ref{Fig:SpectralMeasurements}b [Supplementary Fig.~15]. We find at each radial position a narrow spectrum ($\delta\lambda\!\approx\!50$~pm) whose central wavelength $\lambda_{\mathrm{c}}$ shifts quadratically with $r$, but with differently signed curvature for the superluminal and subluminal cases [Fig.~\ref{Fig:SpectralMeasurements}c].

\begin{figure}[t!]
\centering
\includegraphics[width=17.6cm]{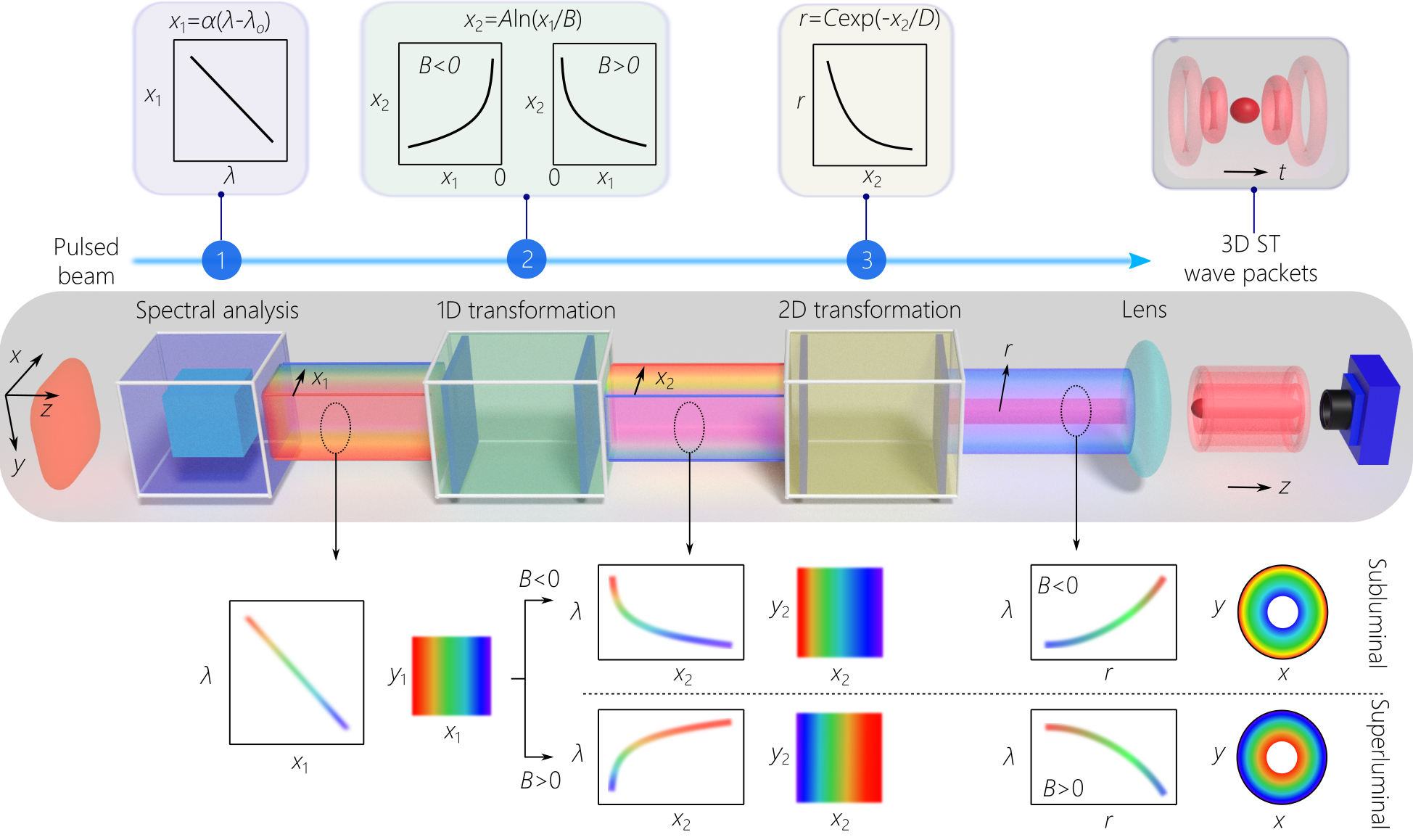}
\caption{\textbf{Schematic of the setup for synthesizing 3D ST wave packets.} Starting with a plane-wave pulse on the left, spectral analysis resolves the spectrum in space and produces linear spatial chirp, $x_{1}(\lambda)\!=\!\alpha(\lambda-\lambda_{\mathrm{o}})$. The spectrally resolved field enters a tunable 1D spectral transformation formed of two spatial light modulators `reshuffles' the wavelengths, $x_{2}(\lambda)\!=\!A\ln(\tfrac{x_{1}(\lambda)}{B})$. Opposite signs of chirp along $x_{2}$ are required for subluminal and superluminal wave packets. Next, a fixed 2D coordinate transformation (implemented with two fixed phase plates) converts the vertical lines corresponding to different wavelengths into circles of radius $r(\lambda)\!=\!C\exp(-\tfrac{x_{2}(\lambda)}{D})$. Finally, a converging spherical lens produces the 3D ST wave packet. On the top, we plot the implemented spectral and spatial transformations; on the bottom, we illustrate the field structure at different points along the setup.}
\label{Fig:Setup}
\end{figure}

\begin{figure}[t!]
\centering
\includegraphics[width=8.5cm]{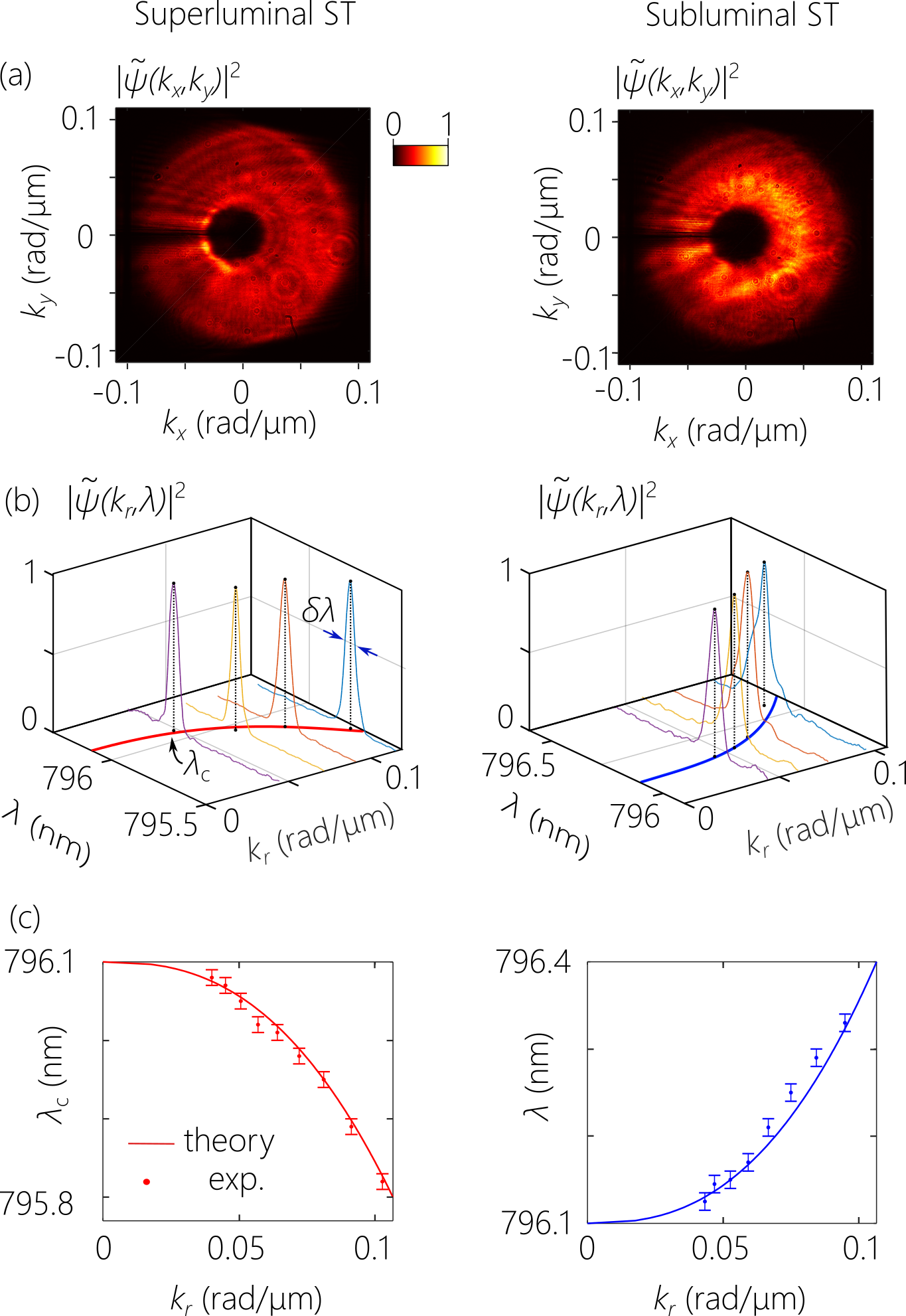}
\caption{\textbf{The spatio-temporal spectral structure of 3D ST wave packets.} Measurements for a superluminal wave packet ($\widetilde{v}\!\approx\!1.37c$) are plotted in the left column, and those for its subluminal counterpart ($\widetilde{v}\!\approx\!0.83c$) are plotted in the column on the right. \textbf{a} Measured spatial spectrum $|\widetilde{\psi}(k_{x},k_{y},\lambda)|^{2}$ by a wavelength-insensitive camera showing an annular structure. \textbf{b} Measured temporal spectra at selected radial positions revealing the radial chirp and the spectral uncertainty. \textbf{c} Measured radial chirp by plotting the central wavelength $\lambda_{\mathrm{c}}$ of the spectrum with radial spatial frequency $k_{r}$. Error bars in \textbf{c} represent the spectral resolution of the optical spectrum analyzer (OSA; Advantest AQ6317B) we made use to perform spectral measurements. }
\label{Fig:SpectralMeasurements}
\end{figure}

\begin{figure}[t!]
\centering
\includegraphics[width=16cm]{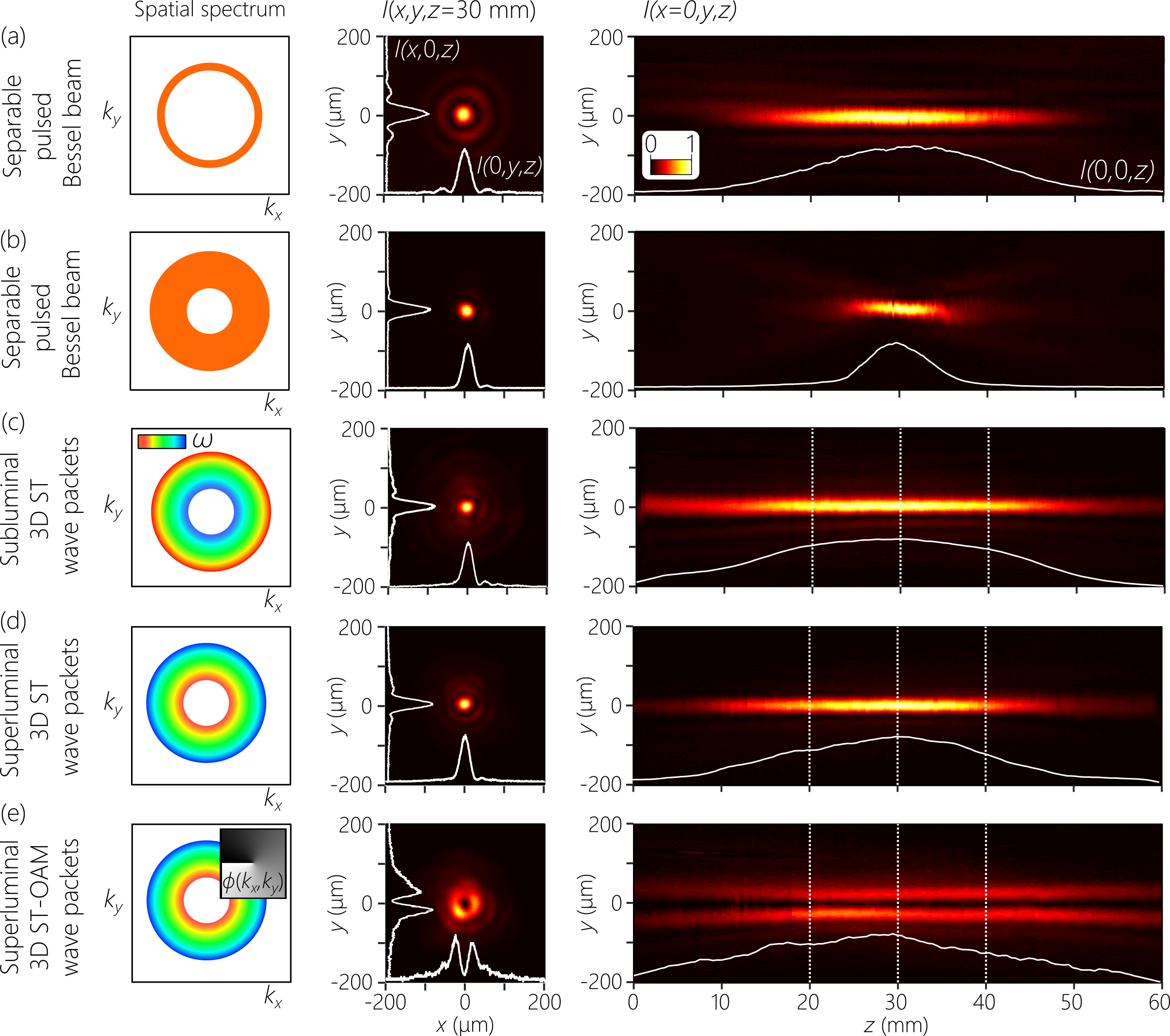}
\caption{\textbf{Measured transverse and axial time-averaged intensity for separable pulsed Bessel beams and 3D ST wave packets.} In the first column, we illustrate the spatio-temporal structure; in the second, we plot the measured transverse intensity $I(x,y,z)$ at $z\!=\!30$~mm, in addition to sections through $x\!=\!0$ and $y\!=\!0$ (white curves); and, in the third, we plot the measured intensity in a meridional plane $I(0,y,z)$. The white curve at the bottom of the panels in the last column is the on-axis intensity $I(0,0,z)$, except in \textbf{e} where we use $y\!=\!30$~$\mu$m. For all cases, $\Delta\lambda\!=\!0.3$~nm. \textbf{a} A separable pulsed Bessel beam with $\Delta k_{r}\!=\!0.02$~rad/$\mu$m. \textbf{b} A pulsed Bessel beam with $\Delta k_{r}\!=\!0.07$~rad/$\mu$m. \textbf{c-e} In all cases $\Delta k_{r}\!=\!0.07$~rad/$\mu$m as in \textbf{b}. \textbf{c} A subluminal ($\widetilde{v}\!=\!0.83c$) 3D ST wave packet; \textbf{d} a superluminal ($\widetilde{v}\!=\!1.37c$) 3D ST wave packet; and \textbf{e} a superluminal ($\widetilde{v}\!=\!1.16c$) 3D ST-OAM wave packet with $\ell\!=\!1$ (the inset in the first column is the associated transverse spectral phase distribution). The dotted vertical white lines in the third column in \textbf{c-e} identify the axial planes for the time-resolved measurements in Fig.~\ref{Fig:Time-ResolvedIntensity}.}
\label{Fig:Time-averagedIntensity}
\end{figure}

\begin{figure}[t!]
\centering
\includegraphics[width=8.6cm]{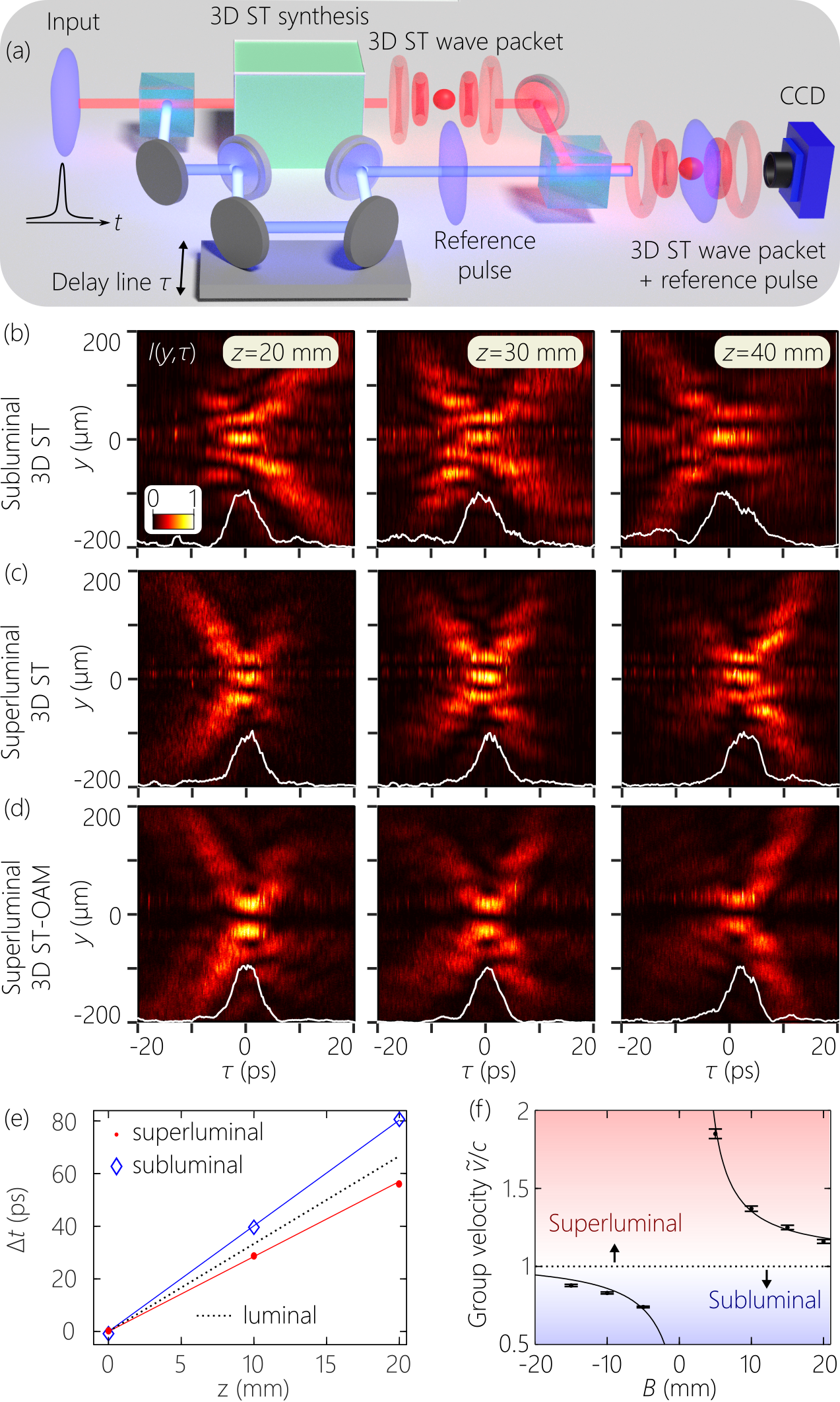}
\caption{\textbf{Reconstructing the spatio-temporal intensity profile and estimating the group velocity for 3D ST wave packets.} \textbf{a} Schematic of the interferometric configuration for reconstructing $I(x,y,z;t)$ and estimating $\widetilde{v}$. \textbf{b} Measured $I(0,y,z;\tau)$ at $z\!=\!20$, 30, and 40~mm for a subluminal ($\widetilde{v}\!=\!0.83c$) wave packet; \textbf{c} for a superluminal ($\widetilde{v}\!=\!1.37c$) wave packet; and \textbf{d} for a superluminal ($\widetilde{v}\!=\!1.16c$) wave packet endowed with the OAM mode $\ell\!=\!1$. We also plot the section $y\!=\!0$ through the intensity profile (white curve at the bottom of each panel), except in \textbf{d} where we use $y\!=\!30$~$\mu$m. \textbf{e} Measured group delay $\Delta t$ at different axial planes for subluminal and superluminal 3D ST wave packets. The straight lines are theoretical expectations and the symbols are data points. \textbf{f} Plot of the estimated group velocity $\widetilde{v}$ with the 1D spectral transformation parameter $B$. The curve is the theoretical expectation $\widetilde{v}\!=\!c/\widetilde{n}$, with $\widetilde{n}\!\approx\!1-\tfrac{2.24}{\mathrm{B}}$ ($B$ in mm). Error bars correspond to the uncertainty in the measurement of $\widetilde{v}$ due to the finite pulse width of 3D ST wave packets; see Supplementary Note~3C.}
\label{Fig:Time-ResolvedIntensity}
\end{figure}

\begin{figure}[t!]
\centering
\includegraphics[width=8.6cm]{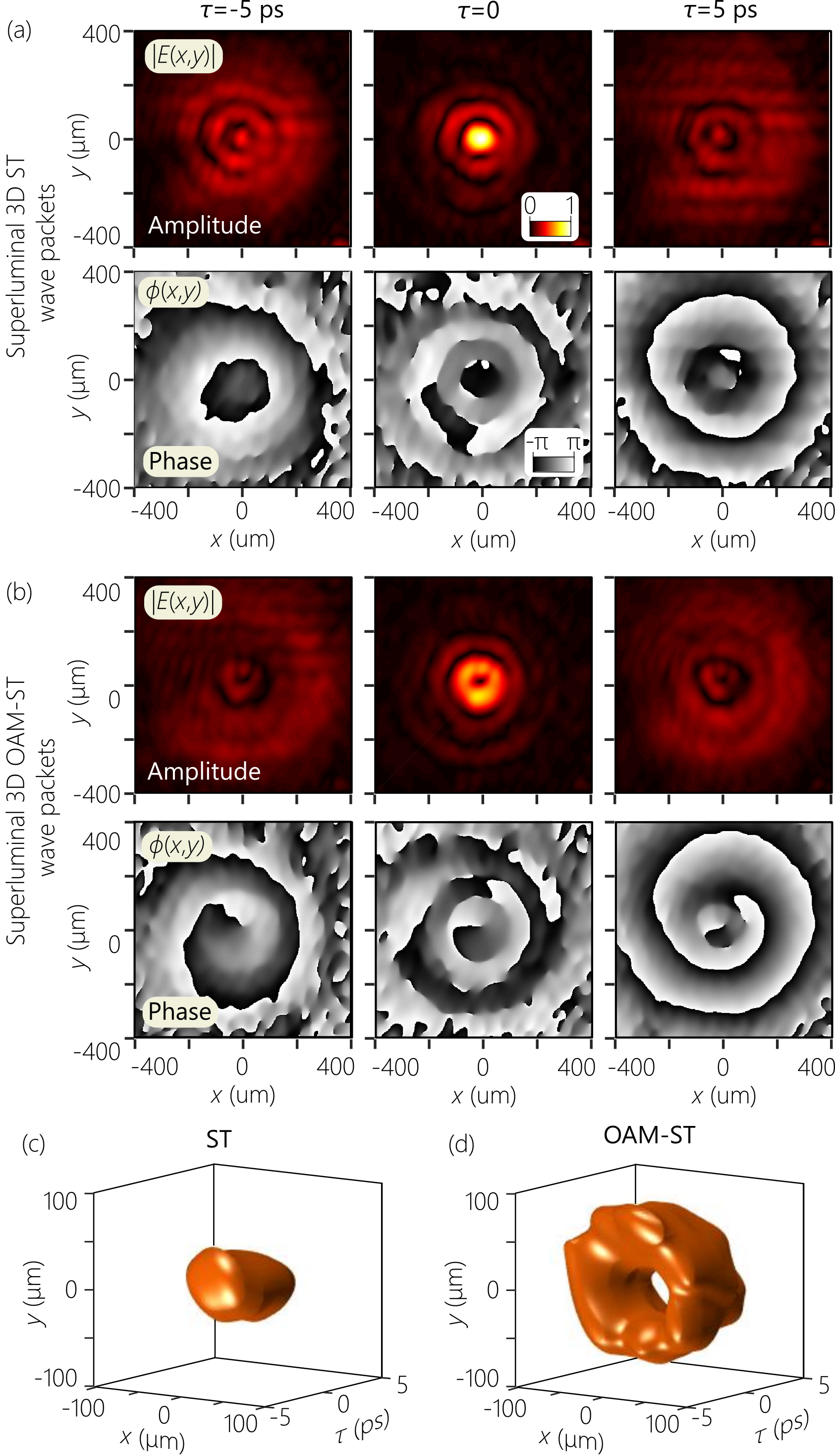}
\caption{\textbf{Measured complex-field amplitude and phase profiles for 3D ST wave packets with and without OAM.} \textbf{a} Measured amplitude $|\psi(x,y)|$ (first row) and phase $\phi(x,y)$ (second row) at a fixed axial plane $z\!=\!30$~mm (see Fig.~\ref{Fig:Time-averagedIntensity} and Fig.~\ref{Fig:Time-ResolvedIntensity}) at delays $\tau\!=\!-5$~ps, $\tau\!=\!0$ corresponding to the wave-packet center, and $\tau\!=\!5$~ps for a superluminal 3D ST wave packet with $\ell\!=\!0$ and \textbf{a} $\ell\!=\!1$. \textbf{c} Iso-amplitude contour $I=0.6I_{\mathrm{max}}$ for the 3D ST wave packet from \textbf{a} and ST-OAM from \textbf{b}. \textbf{d} Same as \textbf{c} but for the iso-amplitude contour $I=0.15I_{\mathrm{max}}$.}
\label{Fig:2DFieldMeasurements}
\end{figure}

\noindent\textbf{Propagation-invariance of the intensity distribution.} The time-averaged intensity profile $I(x,y,z)\propto\int\!dt|E(x,y,z;t)|^{2}$ is captured by scanning a CCD camera along the propagation axis $z$ after the Fourier transforming lens (Fig.~\ref{Fig:Setup}). For each wave packet, we plot in Fig.~\ref{Fig:Time-averagedIntensity} the intensity distribution (at a fixed axial plane $z\!=\!30$~mm) in transverse and meridional planes. As a point of reference, we start with a pulsed Bessel beam whose spatio-temporal spectrum is separable, where the spatial bandwidth is $\Delta k_{r}\!=\!0.02$~rad/$\mu$m and is centered at $k_{r}\!\approx\!0.06$~rad/$\mu$m [Fig.~\ref{Fig:Time-averagedIntensity}a]. Here, the full temporal bandwidth $\Delta\lambda$ is associated with each spatial frequency $k_{r}$. The finite spatial bandwidth $\Delta k_{r}$ renders the propagation distance finite \cite{Durnin87PRL}, and we observe a Bessel beam comprising a main lobe of width $\Delta r\approx30~\mu$m (FWHM) accompanied by several side lobes, which propagates for a distance $L_{\mathrm{max}}\approx50$~mm. For comparison, the Rayleigh range of a Gaussian beam with a similar size and central wavelength is $z_{\mathrm{R}}\!\approx\!1$~mm. By further increasing $\Delta k_{r}$ to 0.07~rad/$\mu$m while remaining centered at $k_{r}\!\approx\!0.06$~rad/$\mu$m as shown in Fig.~\ref{Fig:Time-averagedIntensity}b, the axial propagation distance is reduced proportionately to $L_{\mathrm{max}}\approx15$~mm, and the side lobes are diminished.

Now, rather than the separable spatio-temporal spectra for pulsed Bessel beams [Fig.~\ref{Fig:Time-averagedIntensity}(a,b)], we utilize the structured spatio-temporal spectra associated with 3D ST wave packets in which each $k_{r}$ is associated with a single $\lambda$ [Fig.~\ref{Fig:SpectralMeasurements}], whose spatial bandwidths are all $\Delta k_{r}\!=\!0.07$~rad/$\mu$m centered at $k_{r}\!\approx\!0.06$~rad/$\mu$m, similarly to the pulsed Bessel beam in Fig.~\ref{Fig:Time-averagedIntensity}b. Despite the large spatial bandwidth, the one-to-one correspondence between $k_{r}$ and $\lambda$ curtails the effect of diffraction, leading to an increase in the propagation distance [Fig.~\ref{Fig:Time-averagedIntensity}(c-e)]. The subluminal 3D ST wave packet ($\widetilde{v}\!=\!0.83c$) in Fig.~\ref{Fig:Time-averagedIntensity}c propagates for $L_{\mathrm{max}}\approx60$~mm, which is a $4\times$ improvement compared with the separable Bessel beam and a $60\times$ improvement compared with a Gaussian beam  of the same spatial bandwidth. We observe a similar behavior for a superluminal 3D ST wave packet ($\widetilde{v}\!=\!1.37c$) in Fig.~\ref{Fig:Time-averagedIntensity}d, and a superluminal ST-OAM wave packet ($\widetilde{v}\!=\!1.16c$) with $\ell=1$ in Fig.~\ref{Fig:Time-averagedIntensity}e. 

\noindent\textbf{Reconstructing the spatio-temporal profile and measuring the group velocity.} The spatio-temporal intensity profile $I(x,y,z;t)\!=\!|E(x,y,z;t)|^{2}$ of the 3D ST wave packet is reconstructed by placing the synthesizer (Fig.~\ref{Fig:Setup}) in one arm of a Mach-Zehnder interferometer, while the initial $100$-fs plane-wave pulses from the laser traverse an optical delay line $\tau$ in the reference arm [Fig.~\ref{Fig:Time-ResolvedIntensity}a]. By scanning $\tau$ we reconstruct the spatio-temporal intensity profile in a meridional plane from the visibility of spatially-resolved interference fringes recorded by a CCD camera when the 3D ST wave packet and the reference pulse overlap in space and time. The reconstructed time-resolved intensity profile $I(0,y,z;t)$ of the 3D ST wave packets corresponding to those in Fig.~\ref{Fig:Time-averagedIntensity}(c-e) are plotted in Fig.~\ref{Fig:Time-ResolvedIntensity}(b-d) at multiple axial planes, which reveal clearly the expected X-shaped profile that remains invariant over the propagation distance $L_{\mathrm{max}}$. In all cases, the on-axis pulse width, taken as the FWHM of $I(0,0,0;t)$, is $\Delta t\!\approx\!6$~ps. The spatio-temporal intensity profile of the superluminal ST-OAM wave packet with $\ell=1$ in Fig.~\ref{Fig:Time-ResolvedIntensity}d reveals a similar X-shaped profile, but with a central null instead of a peak, as expected from the helical phase structure associated with the OAM mode.

A subtle distinction emerges between the subluminal and superluminal wave packets regarding the axial evolution of their spatio-temporal profile. It can be shown that in presence of finite spectral uncertainty $\delta\lambda$, the realized ST wave packet can be separated into the product of an ideal ST wave packet traveling indefinitely at $\widetilde{v}$ and a long `pilot envelope' traveling at $c$. The finite propagation distance $L_{\mathrm{max}}$ is then a consequence of temporal walk-off between the ST wave packet and the pilot envelope \cite{Yessenov19OE}. For subluminal ST wave packets, this results initially in a `clipping' of the leading edge of the wave packet in [Fig.~\ref{Fig:Time-ResolvedIntensity}b at $z\!=\!20$~mm], and ultimately a clipping of the trailing edge of the ST wave packet as the faster pilot envelope catches up with it [Fig.~\ref{Fig:Time-ResolvedIntensity}b at $z\!=\!40$~mm]. The opposite behavior occurs for the superluminal ST wave packet in Fig.~\ref{Fig:Time-ResolvedIntensity}(c,d).

This experimental methodology also enables us to estimate the group velocity $\widetilde{v}$ \cite{Kondakci19NC,Bhaduri20NatPhot}. After displacing the CCD camera until the interference fringes are lost due to the mismatch between $\widetilde{v}\!=\!c\tan{\theta}$ for the ST wave packets and the reference pulses traveling at $\widetilde{v}\!=\!c$, we restore the interference by inserting a delay $\Delta t$ [Fig.~\ref{Fig:Time-ResolvedIntensity}e], which allows us to estimate $\widetilde{v}$ for the 3D ST wave packet. By tuning $B$, we record a broad span of group velocities in the range from $\widetilde{v}\approx0.7c$ to $\widetilde{v}\approx1.8c$ in free space [Fig.~\ref{Fig:Time-ResolvedIntensity}f]. The continuous tunability of the group velocity of 3D ST wave packets over the subluminal and superluminal ranges allows them to be exploited in applications previously proposed for ST light-sheets, such as for constructing in-line optical delay lines for all-optical communications \cite{Yessenov20NC}, whereby the localization of 3D ST wave packets in both transverse dimensions can provide a significant advantage with regards to efficiently coupling into optical fibers.

\noindent\textbf{Field amplitude and phase measurements.} Lastly, we modify the measurement system in Fig.~\ref{Fig:Time-ResolvedIntensity}a by adding a small relative angle between the propagation directions of the 3D ST wave packets and the reference pulses, and make use of off-axis digital holography \cite{Sanchez-Ortiga14AO} to reconstruct the amplitude $|\psi(x,y,z;\tau)|$ and phase $\phi(x,y,z;\tau)$ of their complex field envelope $\psi(x,y,z;t)\!=\!|\psi(x,y,z;t)|e^{i\phi(x,y,z;t)}$ [Supplementary Note~3D]. We reconstruct the complex field at a fixed axial plane $z\!=\!30$~mm for the time delays: $\tau=-5$, 0, and 5~ps [Fig.~\ref{Fig:2DFieldMeasurements}]. First, we plot the results for $|\psi(x,y,z;\tau)|$ and phase $\phi(x,y,z;\tau)$ for a superluminal 3D ST wave packet ($\widetilde{v}=1.1c$) with no OAM ($\ell\!=\!0$). At the pulse center $\tau=0$, the field is localized on the optical axis, whereas at $\tau=\pm5$~ps the field spreads away from the center [Fig.~\ref{Fig:2DFieldMeasurements}a]. For $\tau\!\neq\!0$ we find a spherical transverse phase distribution that is almost flat at $\tau\!=\!0$, similar to what one finds during the axial evolution of a Gaussian beam in space through its focal plane \cite{Porras17OL}.

After adding the OAM mode $\ell\!=\!1$ to the field structure, a similar overall behavior is observed for the superluminal ST-OAM wave packet except for two significant features. First, a dip is observed on-axis in Fig.~\ref{Fig:2DFieldMeasurements}b, in lieu of the central peak in Fig.~\ref{Fig:2DFieldMeasurements}a, as a result of the phase singularity associated with the OAM mode. Second, the phase at the wave-packet center $\phi(x,y,z;0)$ at $z\!=\!30$~mm is almost flat, while a helical phase front corresponding to OAM of order $\ell\!=\!1$ emerges as we move away from $\tau\!=\!0$. Finally, we plot in Fig.~\ref{Fig:2DFieldMeasurements}(c,d) iso-amplitude surface contours ($0.6\times$ and $0.15\times$ the maximum amplitude $I_{\mathrm{max}}$) for the two 3D ST wave packets in Fig.~\ref{Fig:2DFieldMeasurements}(a,b). We find a closed surface in Fig.~\ref{Fig:2DFieldMeasurements}c when $\ell\!=\!0$, and a doughnut structure in Fig.~\ref{Fig:2DFieldMeasurements}d when $\ell\!=\!1$ for the first contour $I\!=\!0.6I_{\mathrm{max}}$ that captures the structure of the wave-packet center. The second contour for $I\!=\!0.15I_{\mathrm{max}}$ captures the conic structure emanating from the wave-packet center that is responsible for the characteristic X-shaped spatio-temporal profile of all propagation-invariant wave packets in the paraxial regime. 


\section*{Discussion}

We have demonstrated a general procedure for spatio-temporal spectral modulation of pulsed optical fields that is capable of synthesizing 3D ST wave packets localized in all dimensions. At the heart of our experimental methodology lies the ability to sculpt the angular dispersion of a generic optical pulse in two transverse dimensions. Crucially, this approach produces the non-differentiable angular-dispersion necessary for propagation invariance \cite{Yessenov21ACSPhot}. Because such a capability has proven elusive to date, AD-free X-waves have been the sole class of 3D propagation-invariant wave packets conclusively produced in free space. Unfortunately, X-waves can exhibit only minuscule changes in the group velocity with respect to $c$ (typically $\Delta\widetilde{v}\!\sim\!0.001c$) in the paraxial regime, and only superluminal group velocities are supported. Furthermore, ultrashort pulses of width $<\!20$~fs are required to observe a clear X-shaped profile \cite{Grunwald2003PRA}, and OAM-carrying X-waves have not been realized to date. Even more stringent requirements are necessary for producing focus-wave modes, and consequently they have not been synthesized in three dimensions to date. By realizing instead propagation-invariant 3D ST wave packets, an unprecedented tunable span of group velocities has been realized, clear X-shaped profiles are observed with pulse widths in the picosecond regime, and they outperformed spectrally separable pulsed Bessel beams of the same spatial bandwidth with respect to their propagation distance and transverse side-lobe structure. In addition, we demonstrated propagation-invariant ST-OAM wave packets with tunable group velocity in free space.

Further optimization of the experimental layout is possible. We made use of four phase patterns to produce the target spatio-temporal spectral structure. It is conceivable that this spectral modulation scheme can be performed with only three phase patterns, or perhaps even fewer. Excitingly, a new theoretical proposal suggests that a single non-local nanophotonic structure can produce 3D ST wave packets through a process of spatio-temporal spectral filtering \cite{Guo21Light}. This theoretical proposal indicates the role nanophotonics is poised to play in reducing the complexity of the synthesis system, potentially without recourse to filtering strategies.

Finally, efforts in the near future will be directed to reducing the spectral uncertainty $\delta\lambda$ and concomitantly approaching $\theta\!\rightarrow\!45^{\circ}$ to increase the propagation length to the kilometer range \cite{Bhaduri19OL}. The experimental procedure presented here can in principle be extended to the synthesis of other exotic variants of ST wave packet, such as abruptly focusing needle pulses \cite{Wong17ACSP1} among other possibilities \cite{Wong20AdvSc,Wong21OE}.
With access to 3D ST wave packets, previous work on guided ST modes in planar wave-guides \cite{Shiri20NC} can be extended to conventional single-mode and multi-mode waveguides \cite{Guo21PRR}, and potentially to optical fibers\cite{Ruano2021JO,Bejot2021ACSP,Kibler2021PRL,Bejot2022Arxiv}. Moreover, the localization in both transverse dimensions provided by 3D ST wave packets opens new avenues for nonlinear optics by increasing the intensity with respect to 2D ST wave packets, for introducing topological features such as spin texture in momentum space \cite{Guo21Light}, and for the exploration of spatio-temporal vortices and polarization singularities \cite{Bliokh12PRA}. Our findings point therefore to profound new opportunities provided by the emerging field of space-time optics \cite{SaintMarie17Optica, Froula2018NP,Shaltout19Science,Shiri20NC,Guo21Light,Guo21PRR,Zdagkas2021arxiv}. 

\clearpage

\section*{Methods}

The 2D transformation used to construct the 3D STWP can be implemented by making use of \textit{diffractive} optics \cite{Berkhout10PRL,Lavery11IOP,Berkhout11OL,Li19OE} or \textit{refractive} optics \cite{Lavery12OE}. We exploited both types of phase plates in our experiments to imprint the desired phase profiles: diamond-edged refractive phase plates \cite{Lavery12OE} and analog diffractive phase plates \cite{Li19OE}.

\subsection*{Refractive phase plates}

The refractive optical elements used in our experiments are similar to those outlined by Lavery \textit{et al.} in \cite{Lavery12OE}, in which the transformation parameters are $C=4.77$~mm, $D\!=\!\tfrac{3.2}{\pi}\approx\!1$~mm, and $d_{2}=310$~mm. Each phase plate is made of the polymer PMMA (Poly methyl methacrylate) with accurately manufactured height profiles $Z_{1}(x_{3},y_{3})$ and $Z_{2}(x_{4},y_{4})$ to imprint the required phase profiles. The phase encountered by light at a wavelength $\lambda$ traversing a height $Z$ of a material of refractive index $n$ -- with respect to the phase encountered over the same distance in vacuum -- is given by $\Phi\!=\!2\pi(n-1) Z/\lambda$. Thus, the height profile of the first element is $Z_{1}(x_{3},y_{3})=\frac{\lambda}{2\pi(n-1)}\Phi_{3}(x_{3},y_{3})$ [Supplementary Fig.~14(a)] and that of the profile of the second element is $Z_{2}(x_{4},y_{4})=\frac{\lambda}{2\pi(n-1)}\Phi_{4}(x_{4},y_{4})$ [Supplementary Fig.~14(b)]. Note that each surface height is wavelength-independent, and dispersion effects in the material manifest themselves as a change in the focal length $d_{2}$ of the integrated lens for different wavelengths. Hence, in the experiment the system can be tuned to a specific wavelength by changing the distance between the two elements.

The elements were diamond-machined using a Natotech, 3-axis (X,Z,C) ultra precision lathe (UPL) in combination with a Nanotech NFTS6000 fast tool servo (FTS) system. The machined PMMA surfaces had a radius of 5.64~mm, angular spacing $1^{\circ}$, radial spacing of 5~$\mu$m, a spindle speed of 500~RPM, a roughing feed rate 5~mm/minute with a cut depth of 20~$\mu$m, and a finishing feed rate of 1~mm/minute with a cut depth of 10~$\mu$m \cite{Dow91PE}. The total sag height difference for each part was relatively small ($\approx\!115$~$\mu$m for surface 1 and $\approx\!144$~$\mu$m for surface 2). The transmission efficiency of the combination of the elements is $\approx85\%$.

\subsubsection*{Diffractive phase plates}

The diffractive phase plates were fabricated in fused silica using Clemson University facilities. The fabrication process is outlined in \cite{Sung06AO}, which involves writing a binary phase grating on a stepper mask with an electron-beam and subsequently transferring this analog mask into a fused silica substrate with projection lithography. The phase grating period is designed to be larger than the cutoff period of the projection stepper for higher diffraction orders, so only the zeroth-order diffracted light from the stepper can be transmitted. The transmission coefficient of the stepper light is then a function of the duty cycle of the electron-beam-patterned binary phase grating. The spatial intensity distribution of light in the wafer plane can be controlled with a spatial duty cycle function, which then exposes the I-line resist with a spatially varying analog intensity profile. This allows fabrication of analog diffractive optics with a single exposure from the stepper rather than binary $2^n$ diffractive optics, resulting in high-efficiency optics. The transmission efficiency of the combination of the two faces is $\approx\!92\%$.

The design parameters for the analog diffractive phase plates are chosen as follows: $D\!=\!\frac{7}{\pi}\!\approx\!2.2$~mm, $C\!=\!6$~mm, $\lambda_{\mathrm{o}}\!=\!798$~nm, and $d_{2}\!=\!225$~mm. These design parameters were optimized so the paraxial approximation remains valid over the desired transformation range of 5~mm.

\vspace{4mm}
\noindent
\textbf{Data availability}\\
The data that support the plots within this paper and other findings of this study are available from the corresponding author upon reasonable request.

\vspace{2mm}
\noindent
\textbf{Code availability}\\
The software code used for data acquisition and data analysis are available from the corresponding author upon reasonable request.



%

\vspace{2mm}
\noindent
\textbf{Acknowledgments}\\
We thank OptiGrate Company for making volume Bragg gratings, and Dr. Peter J. Delfyett and Dr. Ivan Divliansky for lending equipment. We thank L. A. Hall, A. Shiri, K. L. Schepler, L. Mach, M. G. Vazimali, I. Hatipoglu, and M. Eshaghi for useful discussions. M.Y. and A.F.A. were supported by the U.S. Office of Naval Research (ONR) under contracts N00014-17-1-2458, N00014-19-1-2192, and N00014-20-1-2789. J.F. and E.G.J. were supported by ONR contract N00014-20-1-2558. M.A.A. was funded by the Excellence Initiative of Aix Marseille University -- A*MIDEX, a French `Investissements d'Avenir' programme.

\vspace{2mm}
\noindent
\textbf{Author contributions}\\
\noindent
A.F.A. and M.Y developed the concept. M.Y. designed the experiments, carried out the measurements, and analyzed the data.  Z.C. and M.P.J.L designed and manufactured the diamond-machined refractive elements for the 2D coordinate transformation. J.F and E.G.J. designed and fabricated the analog diffractive phase plates for the 2D coordinate transformation. M.A.A. and A.F.A. developed the theoretical aspects. A.F.A supervised the research. All the authors contributed to writing the paper.

\noindent
Correspondence and requests for materials should be addressed to A.F.A.\\(email: raddy@creol.ucf.edu)

\vspace{2mm}
\noindent
\textbf{Competing interests:} The authors declare no competing interests.

\clearpage

\end{document}


\title{Supplementary Information:\\Space-time wave packets localized in all dimensions}

\author{Murat Yessenov$^{1,\dagger}$, Justin Free$^{2}$, Zhaozhong Chen$^{3}$, Eric G. Johnson$^{2}$, Martin P.~J. Lavery$^{3}$, Miguel A. Alonso$^{4,5}$, and Ayman F. Abouraddy$^{1,}$ }
\email{Corresponding authors: raddy@creol.ucf.edu}
\email{$^{\dagger}$yessenov@knights.ucf.edu}
\affiliation{$^{1}$ CREOL, The College of Optics \& Photonics, University of Central Florida, Orlando, Florida 32816, USA \\
$^{2}$ Micro-Photonics Laboratory, the Holcombe Department of Electrical and Computer Engineering, Clemson University, Clemson, South Carolina 29634, USA \\
$^{3}$ James Watt School of Engineering, University of Glasgow, UK\\
$^{4}$ CNRS, Centrale Marseille, Institut Fresnel, Aix Marseille Univ., Marseille, France\\
$^{5}$ The Institute of Optics, University of Rochester, Rochester, NY, USA }

\date{\today}

\renewcommand{\thesection}{Supplementary Note~\arabic{section}}   
\renewcommand{\thefigure}{Supplementary Fig.~\arabic{figure}}
\renewcommand{\theequation}{S\arabic{equation}}
\renewcommand{\thetable}{Supplementary Table~\arabic{table}}

\maketitle

\tableofcontents
\clearpage

\section{Representation of 3D space-time wave packets on the surface of the light-cone}

In our previous work on space-time (ST) wave packets in the form of light-sheets \cite{Kondakci16OE,Kondakci17NP,Kondakci19NC,Yessenov19PRA,Yessenov19OE,Yessenov19OPN}, we make heavy use of the representation of the wave-packet spectral support domain on the surface of the light-cone. This is a useful visualization tool that provides physical intuition with regards to the structure and behavior of ST wave packets. In this Section, we briefly review this representation in the reduced-dimension case of ST light sheets \cite{Kondakci17NP}, where the field is localized along one transverse dimension and extended uniformly along the other. We refer to these field structures as 2D ST wave packets (one transverse dimension and one longitudinal dimension). We then proceed to show that such a representation can also be gainfully employed with minor changes for ST wave packets localized in all dimensions. We refer to such field structures as 3D ST wave packets (two transverse dimensions and one longitudinal dimensions). Therefore, the wealth of results that have amassed over the past few years based on this conceptual framework \cite{Yessenov19OPN} can be appropriated for the new 3D ST wave packets investigated here.

\subsection{Light-cone representation for 2D ST wave packets (light sheets)}

\begin{figure}[t!]
  \begin{center}
  \includegraphics[width=8.5cm]{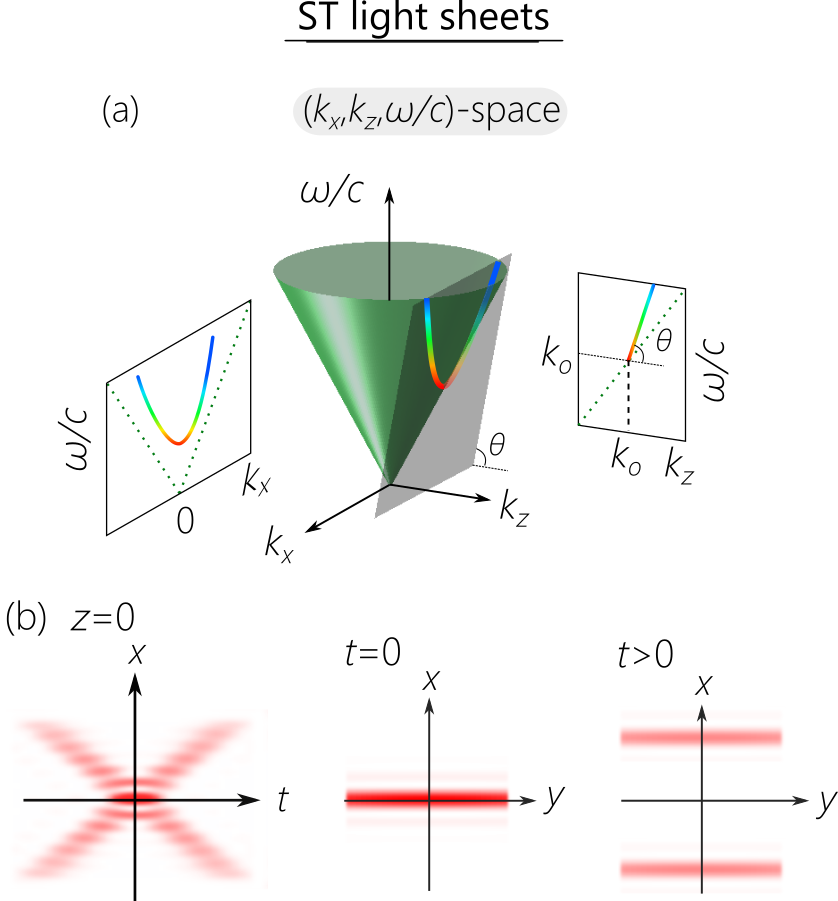}
  \end{center}\vspace{-5mm}
  \caption{\textbf{Supplementary Fig.~1.} Representation of the spectral support domain for 2D ST wave packets (light sheets) on the light-cone surface. (a) The spatio-temporal spectrum is restricted to the intersection of the light-cone $k_{x}^{2}+k_{z}^{2}\!=\!(\tfrac{\omega}{c})^{2}$ with a tilted spectral plane in the region $k_{z}\!>\!0$. The spectral projection onto the $(k_{z},\tfrac{\omega}{c})$-plane is a straight line, and that onto the $(k_{x},\tfrac{\omega}{c})$-plane is a segment of a conic section. (b) The spatio-temporal intensity profile in the plane $z\!=\!0$, $I(x,z=0;t)$, and the transverse intensity profile in the same $z\!=\!0$ plane at the wave-packet center $t\!=\!0$, and off-center $t\!>\!0$. In the $(x,t)$-plane, the spatio-temporal intensity profile is X-shaped. In the $(x,y)$-plane, the 2D ST wave packet at $t\!=\!0$ takes the form of a light sheet that is localized along the $x$-axis and extends uniformly along the $y$-axis.}
  \label{Fig:STLightSheet}
  \vspace{12mm}
\end{figure}

When the field is held uniform along one transverse dimension (say $y$), then the dispersion relationship in free space is $k_{x}^{2}+k_{z}^{2}\!=\!(\tfrac{\omega}{c})^{2}$, where $k_{x}$ and $k_{z}$ are the transverse and longitudinal components of the wave vector along $x$ and $z$, respectively, $\omega$ is the temporal frequency, and $c$ is the speed of light in vacuum. This relationship is represented geometrically by the surface of a cone that we refer to as the light-cone. A monochromatic plane wave $e^{i(k_{x}x+k_{z}z-\omega t)}$ is represented by a point on the surface of the light-cone (\ref{Fig:STLightSheet}). In general, a pulsed beam $E(x,z;t)$ is expressed as a product of a slowly varying envelope $\psi(x,z;t)$ and a carrier term $e^{i(k_{\mathrm{o}}z-\omega_{\mathrm{o}}t)}$, where $\omega_{\mathrm{o}}$ is a fixed temporal frequency, and $k_{\mathrm{o}}\!=\!\omega_{\mathrm{o}}/c$ is its associated wave number. The envelope is written in terms of an angular spectrum as follows:
\begin{equation}\label{Eq:2DGeneral}
\psi(x,z;t)=\iint\!dk_{x}d\Omega\widetilde{\psi}(k_{x},\Omega)e^{i\{k_{x}x+(k_{z}-k_{\mathrm{o}})z-\Omega t\}},
\end{equation}
where $\Omega\!=\!\omega-\omega_{\mathrm{o}}$, and the spatio-temporal spectrum $\widetilde{\psi}(k_{x},\Omega)$ is the 2D Fourier transform of $\psi(x,0;t)$. The spectral support domain for a pulsed beam or wave packet corresponds in general to a 2D area on the surface of the light-cone \cite{Kondakci17NP}; see \ref{Fig:GaussianBeam}.

For the special case of a propagation-invariant 2D ST wave packet, the spectral support domain is confined to the intersection of the light-cone with a spectral plane that is parallel to the $k_{x}$-axis and makes an angle $\theta$ with the $k_{z}$-axis, which is thus given by the equation:
\begin{equation}
\Omega=(k_{z}-k_{\mathrm{o}})c\tan{\theta},
\end{equation}
where we refer to $\theta$ as the spectral tilt angle. Consequently, there is a one-to-one relationship between $\Omega$ and $|k_{x}|$, which takes the form of a parabola in the narrowband ($\Delta\omega\!\ll\!\omega_{\mathrm{o}}$, where $\Delta\omega$ is the temporal bandwidth) and paraxial ($\Delta k_{x}\!\ll\!k_{\mathrm{o}}$, where $\Delta k_{x}$ is the spatial bandwidth) limits:
\begin{equation}\label{Eq:Parabola1D}
\frac{\Omega}{\omega_{\mathrm{o}}}=\frac{k_{x}^{2}}{2k_{\mathrm{o}}^{2}(1-\widetilde{n})},
\end{equation}
where $\widetilde{n}\!=\!\cot{\theta}$ is the wave packet group index in free space. Because of this correspondence between $\Omega$ and $|k_{x}|$, the spatio-temporal spectrum has a reduced dimensionality with respect to a conventional pulsed beam $\widetilde{\psi}(k_{x},\Omega)\!\rightarrow\!\widetilde{\psi}(k_{x})\delta(\Omega-\Omega(k_{x}))$, where $\Omega\!=\!\Omega(k_{x})$ is given by Eq.~\ref{Eq:Parabola1D}. The wave-packet envelope now takes the simpler integral form:
\begin{equation}\label{Eq:2DSTenvelope}
\psi(x,z;t)=\int\!dk_{x}\,\,\widetilde{\psi}(k_{x})\,\,e^{ik_{x}x}\,\,e^{-i(t-z/\widetilde{v})\Omega(k_{x})}=\psi(x,0;t-z/\widetilde{v}),
\end{equation}
which is a 2D ST wave packet that travels rigidly in free space at a group velocity $\widetilde{v}\!=\!c\tan{\theta}$.

The spectral projection of the 2D ST wave packet onto the $(k_{z},\tfrac{\omega}{c})$-plane is a straight line that makes an angle $\theta$ with the $k_{z}$-axis and intersects with the light-line $k_{z}\!=\!\tfrac{\omega}{c}$ at the point $(k_{z},\tfrac{\omega}{c})\!=\!(k_{\mathrm{o}},k_{\mathrm{o}})$. The corresponding spectral projection onto the $(k_{x},\tfrac{\omega}{c})$-plane is a segment of a conic section: an ellipse when $0^{\circ}\!<\!\theta\!<\!45^{\circ}$ or $135^{\circ}\!<\!\theta\!<\!180^{\circ}$, whereupon $|\widetilde{v}|\!<\!c$; a hyperbola when $45^{\circ}\!<\!\theta\!<\!135^{\circ}$, whereupon $|\widetilde{v}|\!>\!c$; a straight tangent line when $\theta\!=\!45^{\circ}$ and $\widetilde{v}\!=\!c$, corresponding to a plane-wave pulse; and a parabola when $\theta\!=\!135^{\circ}$ and $\widetilde{v}\!=\!-c$. We refer to ST wave packets associated with the range $0^{\circ}\!<\!\theta\!<\!45^{\circ}$ as subluminal, with $45^{\circ}\!<\!\theta\!<\!90^{\circ}$ as superluminal, and with $\theta\!>\!90^{\circ}$ (whereupon $\widetilde{v}\!<\!0$) as negative-$\widetilde{v}$ ST wave packets.

Note that causal emission and propagation require that only the values $k_{z}\!>\!0$ be considered, so that the light-cone half corresponding to the acausal backward-propagating components $k_{z}\!<\!0$ is eliminated from consideration \cite{Shaarawi00JPA,Yessenov19PRA}.

The one-to-one correspondence between $|k_{x}|$ and $\omega$ has a crucial impact on the form of the axial evolution of the time-averaged intensity $I(x,z)\!=\!\int\!dt\,|E(x,z;t)|^{2}\!=\!\int\!dt|\,\psi(x,z;t)|^{2}$. Substituting for $\psi(x,z;t)$ from Eq.~\ref{Eq:2DSTenvelope} we reach:
\begin{equation}
I(x,z)=2\pi k_{\mathrm{o}}^2 (1-\tilde{n})\left[ \int\!dk_{x}|\widetilde{\psi}(k_{x})|^{2}+\int\!dk_{x}\widetilde{\psi}(k_{x})\widetilde{\psi}^{*}(-k_{x})e^{i2k_{x}x}\right]=I_{\mathrm{o}}+I(2x),
\end{equation}
where $I_{\mathrm{o}}\!=\!2\pi k_{\mathrm{o}}^2 (1-\tilde{n})\int\!dk_{x}|\widetilde{\psi}(k_{x})|^{2}$ and $I(x)$ is the Fourier transform of $\widetilde{\psi}(k_{x})\widetilde{\psi}^{*}(-k_{x})$. In other words, $I(x,z)$ is altogether independent of the axial coordinate $z$, and is formed of the sum of a constant background pedestal term $I_{\mathrm{o}}$ and a localized spatial feature at $x\!=\!0$. Moreover, the height of the localized spatial feature cannot exceed the height of the pedestal. This structure of the intensity profile for 2D ST wave packets has been borne out in previous measurements \cite{Kondakci17NP,Yessenov19OE}. It is important to note that this structure is unique to 2D ST wave packets. The time-averaged intensity of 3D ST wave packets cannot be separated into a sum of a pedestal and a localized central feature (see main text).

\subsection{Spectral representation for 3D fields on the light-cone surface}

\subsubsection{Conventional optical fields}

When both transverse coordinates $x$ and $y$ are retained, we have the general dispersion relationship $k_{x}^{2}+k_{y}^{2}+k_{z}^{2}\!=\!(\tfrac{\omega}{c})^{2}$ in free space. This relationship is represented mathematically by a hypercone in 4D, which cannot be visualized in 3D space. However, because we are mostly interested in cylindrically symmetric fields, we can write the dispersion relationship as $k_{r}^{2}+k_{z}^{2}\!=\!(\tfrac{\omega}{c})^{2}$, where $k_{r}\!=\!\sqrt{k_{x}^{2}+k_{y}^{2}}$ is the radial wave number. This dispersion relationship can indeed be represented in 3D space. Because $k_{r}$ is positive-valued only, in contrast to $k_{x}$ in the case of the 2D ST wave packets that can take on either positive or negative values, only the quarter of the light-cone corresponding to $k_{r}\!>\!0$ \textit{and} $k_{z}\!>\!0$ need be retained here.

A wave packet is once again given in terms of a carrier term and a slowly varying envelope $E(x,y,z;t)\!=Re\left[\!e^{i(k_{\mathrm{o}}z-\omega_{\mathrm{o}}t)}\psi(x,y,z;t)\right]$, where:
\begin{equation}
\psi(x,y,z;t)=\iiint\!dk_{x}dk_{y}d\Omega\,\,\widetilde{\psi}(k_{x},k_{y},\Omega)\,\,e^{i\{k_{x}x+k_{y}y+(k_{z}-k_{\mathrm{o}})z-\Omega t\}},
\end{equation}
and $\widetilde{\psi}(k_{x},k_{y},\Omega)$ is the 3D Fourier transform of $\psi(x,y,0;t)$. We switch to transverse polar coordinates $(r,\varphi)$ in physical space and to the corresponding polar coordinates $(k_{r},\chi)$ in Fourier space. In physical space we have the relationships:
\begin{equation}
r=\sqrt{x^{2}+y^{2}},\:\:\:\varphi=\arctan{\left(\frac{y}{x}\right)},\:\:\: x=r\sin{\varphi},\:\:\:y=r\cos{\varphi};   
\end{equation}
and in Fourier space we have 
\begin{equation}
k_{r}=\sqrt{k_{x}^{2}+k_{y}^{2}},\:\:\:\chi=\arctan{\left(\frac{k_{y}}{k_{x}}\right)},\:\:\:k_{x}=k_{r}\sin{\chi},\:\:\:k_{y}=k_{r}\cos{\chi}.
\end{equation}
We can thus rewrite the angular spectrum of the envelope as follows:
\begin{equation}
\psi(r,\varphi,z;t)=\iiint\!dk_{r}d\chi d\Omega\,\,\, k_{r}\widetilde{\psi}(k_{r},\chi,\Omega)\,\,e^{ik_{r}r\cos{(\varphi-\chi)}}\,\,e^{i(k_{z}-k_{\mathrm{o}})z}\,\,e^{-i\Omega t};
\end{equation}
where the integral over $\chi$ extends from 0 to $2\pi$, that over $k_{r}$ extends from 0 to $\infty$, and that for $\Omega$ from $-\Delta\omega/2$ to $\Delta\omega/2$.

We can separate the spatio-temporal spectrum $\widetilde{\psi}(k_{r},\chi,\Omega)$ with respect to the radial and azimuthal coordinates $k_{r}$ and $\xi$, respectively, as follows:
\begin{equation}
\widetilde{\psi}(k_{r},\chi,\Omega)=\frac{1}{2\pi}\sum_{\ell=-\infty}^{\infty}\widetilde{\psi}_{\ell}(k_{r},\Omega)e^{i\ell\chi},
\end{equation}
whereupon the wave packet envelope can be expressed as:
\begin{equation}
\psi(r,\varphi,z;t)=\sum_{\ell=-\infty}^{\infty}e^{i\ell\varphi}\iint\!dk_{r}d\Omega\;\;k_{r}\widetilde{\psi}(k_{r},\Omega)J_{\ell}(k_{r}r)e^{i(k_{z}-k_{\mathrm{o}})z}e^{-i\Omega t};
\end{equation}
where we have made use of the identity $2\pi J_{\ell}(x)=\int_{-\pi}^{\pi}dy\,e^{i(x\sin{y}-\ell y)}$, and $J_{\ell}(\cdot)$ is the $\ell^{\mathrm{th}}$-order Bessel function of the first kind.

 \begin{figure}[t!]
  \begin{center}
  \includegraphics[width=8.5cm]{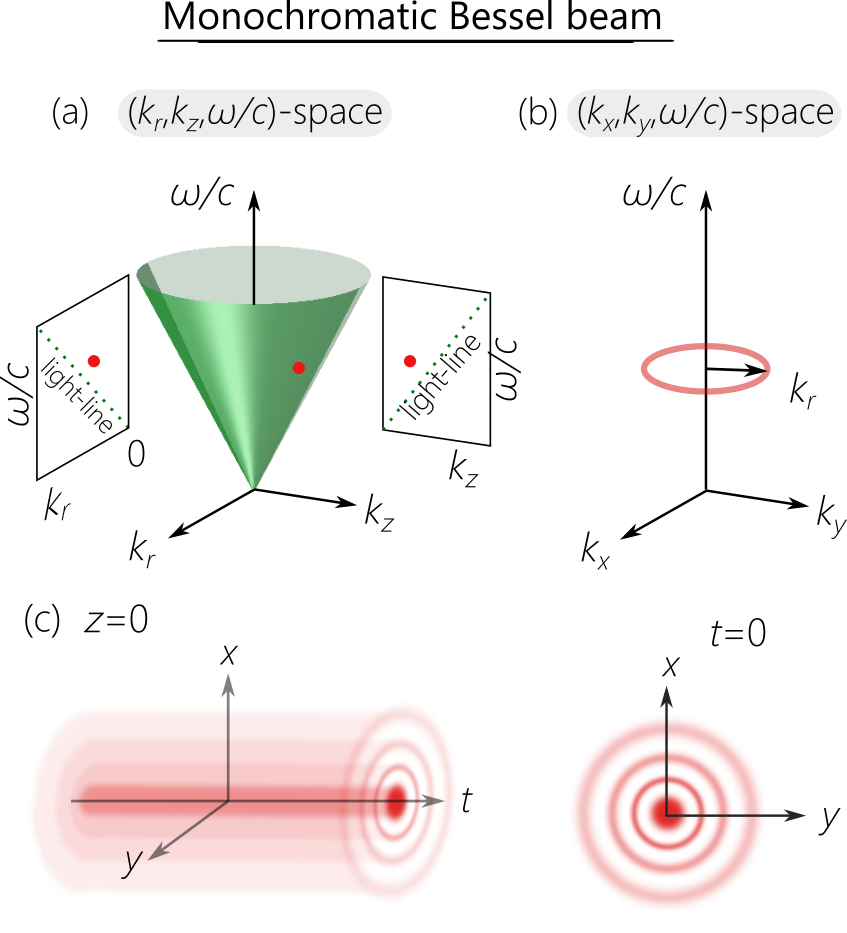}
  \end{center}\vspace{-5mm}
  \caption{\textbf{Supplementary Fig.~2.} Representation of the spectral support domain of a monochromatic Bessel beam in 3D space on the surface of the light-cone. (a) A point on the surface of the light-cone $k_{r}^{2}+k_{z}^{2}\!=\!(\tfrac{\omega}{c})^{2}$ at coordinates $(k_{r},k_{z},\tfrac{\omega}{c})$ corresponds to a monochromatic Bessel beam of the form $E(r,z;t)\!=\!J_{0}(k_{r}r)e^{i(k_{z}z-\omega t)}$. (b) In $(k_{x},k_{y},\tfrac{\omega}{c})$-space, the point on the light-cone in (a) takes the form of a horizontal iso-frequency circle of radius $k_{r}$ at a height $\tfrac{\omega}{c}$. (c) In physical space at an axial plane $z\!=\!0$, the intensity $I(x,y,0;t)$ is uniform along $t$ and takes the form of a Bessel function $J_{0}^{2}(k_{r}r)$ in the $(x,y)$-plane at any $t$.}
  \label{Fig:BesselBeam}
  \vspace{12mm}
\end{figure}

 \begin{figure}[]
  \begin{center}
  \includegraphics[width=8.5cm]{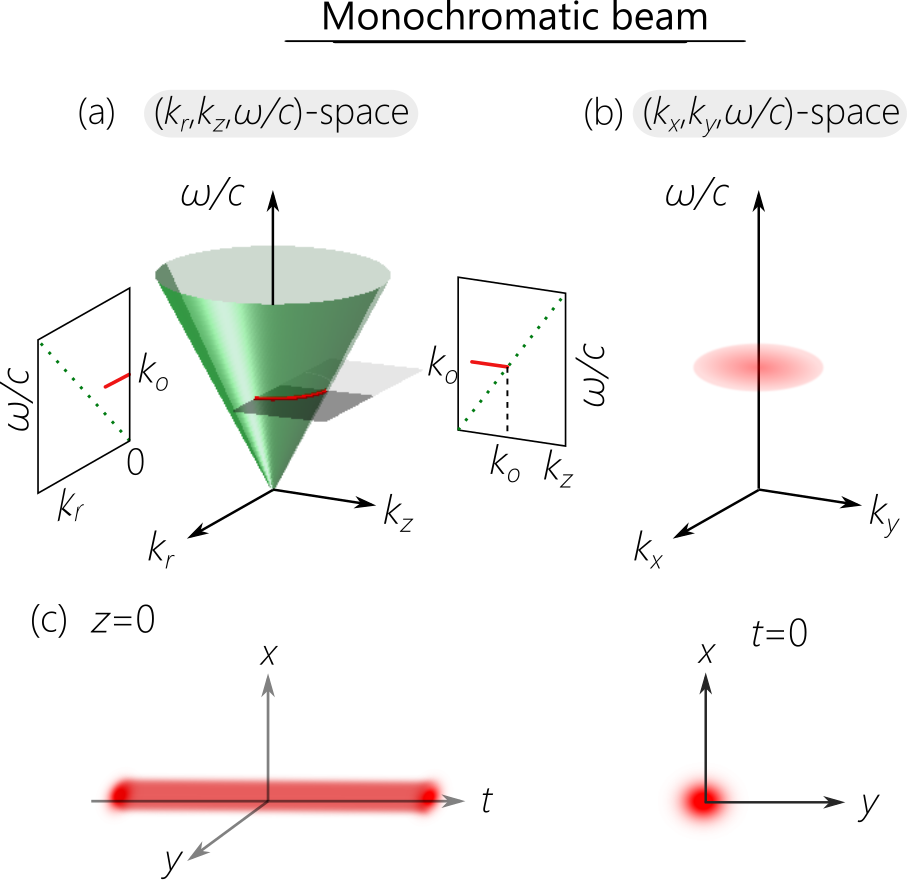}
  \end{center}\vspace{-5mm}
  \caption{\textbf{Supplementary Fig.~3.} Representation of the spectral support domain for a monochromatic beam in 3D space on the surface of the light-cone. (a) The spectral support domain is restricted to the circle on the light-cone surface at its intersection with a horizontal iso-frequency plane $\omega\!=\!\omega_{\mathrm{o}}$. (b) In $(k_{x},k_{y},\tfrac{\omega}{c})$-space, the spectral support domain is a horizontal iso-frequency disc at $\omega\!=\!\omega_{\mathrm{o}}$. The disc radius corresponds to the maximum spatial frequency in (a). (c) In physical space at $z\!=\!0$, the intensity $I(x,y,0;t)$ is uniform along $t$ with the transverse beam profile given by the plot in the $(x,y)$-plane at any $t$.}
  \label{Fig:GaussianBeam}
  \vspace{12mm}
\end{figure}

 \begin{figure}[]
  \begin{center}
  \includegraphics[width=8.5cm]{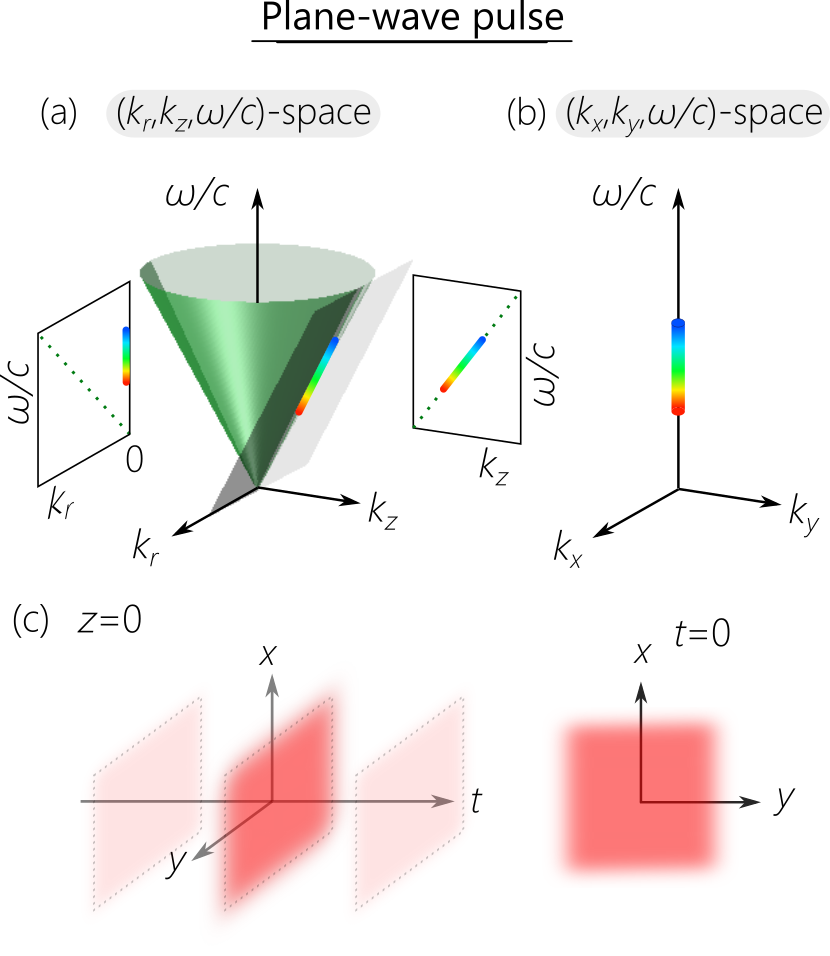}
  \end{center}\vspace{-5mm}
  \caption{\textbf{Supplementary Fig.~4.} Representation of the spectral support domain for a plane-wave pulse in 3D space on the surface of the light-cone. (a) The spectral support domain is restricted to the straight-line tangent at $k_{r}\!=\!0$ on the light-cone surface. (b) In $(k_{x},k_{y},\tfrac{\omega}{c})$-space, the spectral support domain is a vertical line along the $\tfrac{\omega}{c}$-axis. (c) In physical space at an axial plane $z\!=\!0$, the intensity $I(x,y,0;t)$ is uniform everywhere at fixed $t$. The overall intensity drops away from $t\!=\!0$ according to the pulse linewidth.}
  \label{Fig:PulsedPlaneWave}
  \vspace{12mm}
\end{figure}

 \begin{figure}[]
  \begin{center}
  \includegraphics[width=8.5cm]{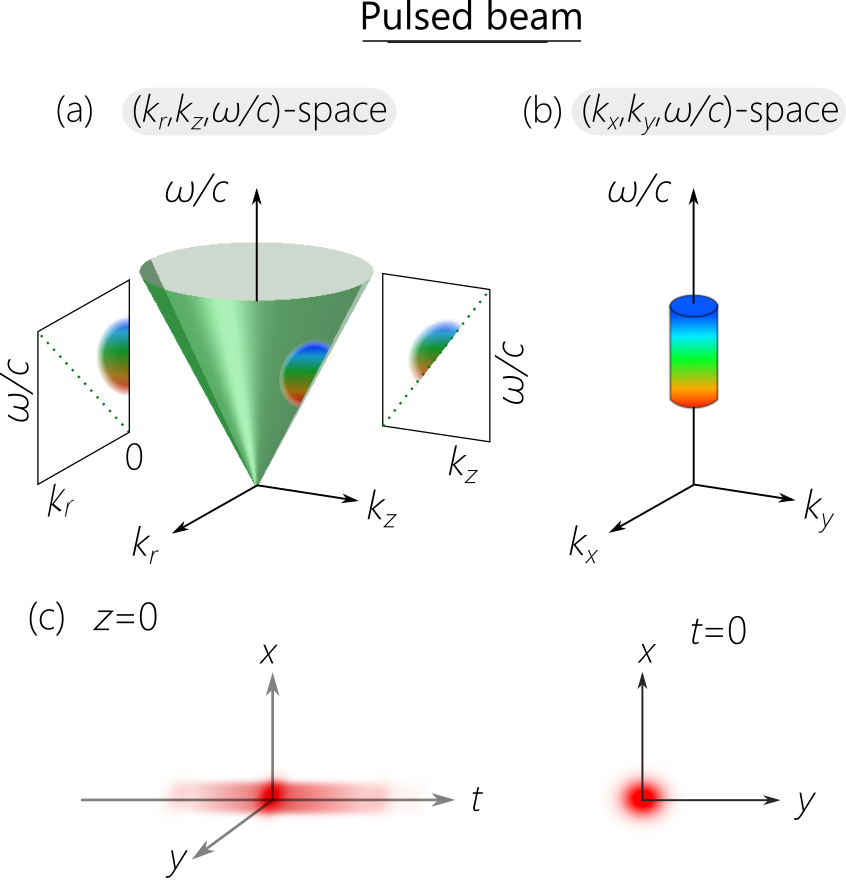}
  \end{center}\vspace{-5mm}
  \caption{\textbf{Supplementary Fig.~5.} Representation of the spectral support domain for a conventional pulsed beam in 3D space on the surface of the light-cone. (a) The spectral support domain is a 2D area on the light-cone surface. (b) In $(k_{x},k_{y},\tfrac{\omega}{c})$-space, the spectral support domain is a cylindrical volume whose radius and height correspond to the spatial and temporal bandwidths of the wave packet. (c) In physical space at $z\!=\!0$, the intensity $I(x,y,0;t)$ varies in intensity along $t$ according to the pulse linewidth, with the transverse beam profile as prescribed in the plot in the $(x,y)$-plane at $t\!=\!0$.}
  \label{Fig:PulsedBeam}
  \vspace{12mm}
\end{figure}

One simplification is obtained by assuming that the spatio-temporal spectrum is azimuthally symmetric, which corresponds to setting $\ell\!=\!0$, and the summation over $\ell$ is removed. This assumption is equivalent to setting $\widetilde{\psi}(k_{r},\chi,\Omega)\!\rightarrow\!\widetilde{\psi}(k_{r},\Omega)$, which is thus independent of $\chi$, whereupon:
\begin{equation}\label{Eq:3DGeneralAzimuthallySymmetric}
\psi(r,\varphi,z;t)=\iint\!dk_{r}d\Omega \,\,\,k_{r}\widetilde{\psi}(k_{r},\Omega)J_{0}(k_{r}r)\,\,e^{i(k_{z}-k_{\mathrm{o}})z}\,\,e^{-i\Omega t}=\psi(r,z;t),
\end{equation}
where $J_{0}(\cdot)$ is the zeroth-order Bessel function of the first kind. As a result of the azimuthal symmetry of the spectrum, the wave-packet envelope is also azimuthally symmetric in physical space. We will address shortly the more general case of fields with azimuthal variation.

Comparing Eq.~\ref{Eq:2DGeneral} for $\psi(x,z;t)$ to Eq.~\ref{Eq:3DGeneralAzimuthallySymmetric} for $\psi(r,z;t)$, we find that both have 2D spatio-temporal spectra: $\widetilde{\psi}(k_{x},\Omega)$ for the former and $\widetilde{\psi}(k_{r},\Omega)$ for the latter. The light-cone in $(k_{r},k_{z},\tfrac{\omega}{c})$-space can thus be used to represent the spectral support domain for $\psi(r,z;t)$. Here, each point on the light-cone at coordinates $(k_{r},k_{z},\tfrac{\omega}{c})$ corresponds to a monochromatic Bessel beam $E(r,z;t)\!=\!J_{0}(k_{r}r)e^{i(k_{z}z-\omega t)}$ rather than a monochromatic plane wave; see \ref{Fig:BesselBeam}.

In the case of light sheets in which the field is uniform along $y$, the light-cone in \ref{Fig:STLightSheet}(a) suffices to capture the complete description of the spectral support domain. On the other hand, in the case of fields in 3D space, the light-cone in \ref{Fig:BesselBeam}(a) does \textit{not} convey the whole picture because we collapsed all the plane waves having spatial-frequency pairs $(k_{x},k_{y})$ into their radial counterpart $k_{r}$. The picture can be completed by adding a second spectral representation in $(k_{x},k_{y},\tfrac{\omega}{c})$-space. Each point with coordinates $(k_{x},k_{z},\tfrac{\omega}{c})$ corresponds to a monochromatic plane wave $e^{i(k_{x}x+k_{y}y+k_{z}z-\omega t)}$, where $k_{z}\!=\!\sqrt{(\tfrac{\omega}{c})^{2}-k_{x}^{2}-k_{y}^{2}}$. Although $k_{z}$ is not represented in this space explicitly, it can nevertheless be found by referring to the light-cone in \ref{Fig:BesselBeam}(a). The single point representing a monochromatic Bessel beam on the light-cone in \ref{Fig:BesselBeam}(a) corresponds to a horizontal circle of radius $k_{r}$ in $(k_{x},k_{y},\tfrac{\omega}{c})$-space in \ref{Fig:BesselBeam}(b).

The representation in $(k_{x},k_{y},\tfrac{\omega}{c})$-space is particularly useful for wave packets in 3D space because it provides the spatio-temporal spectral structure required to synthesize the field in question. For a monochromatic Bessel beam, \ref{Fig:BesselBeam}(b) points to the well-known approach for producing such a beam by inserting a spatial filter in the Fourier domain in the form of a thin annulus, followed by a spherical converging lens \cite{Durnin87PRL}.

More generally, the spectral support domain for a monochromatic beam at frequency $\omega_{\mathrm{o}}$ is restricted to the circle at the intersection of the light-cone with a horizontal iso-frequency plane $\omega\!=\!\omega_{\mathrm{o}}$; see \ref{Fig:GaussianBeam}(a). In $(k_{x},k_{y},\tfrac{\omega}{c})$-space, the spectral support domain lies in the horizontal plane $\omega\!=\!\omega_{\mathrm{o}}$, and takes the form of a disc. The central point at $k_{x}\!=\!k_{y}\!=\!0$ corresponds to the point on the light-line in \ref{Fig:GaussianBeam}(a). The radius of the disc in \ref{Fig:GaussianBeam}(b) corresponds to the maximum radial spatial frequency in \ref{Fig:GaussianBeam}(a).

As another example, consider a pulsed plane wave that lacks transverse spatial features. The spectral support domain of this field in $(k_{r},k_{z},\tfrac{\omega}{c})$-space lies along the light-line $k_{z}\!=\!\tfrac{\omega}{c}$ [\ref{Fig:PulsedPlaneWave}(a)], whereupon $\widetilde{\psi}(k_{r},\Omega)\!\rightarrow\!\widetilde{\psi}(\Omega)\frac{\delta (k_{r})}{k_{r}}$, so that the pulse is a plane wave with no transverse features, and whose longitudinal and temporal shape is determined exclusively by $\widetilde{\psi}(\Omega)$, therefore:
\begin{equation}
\psi(z;t)=\int\!d\Omega\widetilde{\psi}(\Omega)e^{i\Omega(t-z/c)}=\psi(0;t-z/c).
\end{equation}
The spectral support domain in $(k_{x},k_{y},\tfrac{\omega}{c})$-space lies along the $\tfrac{\omega}{c}$-axis, with $k_{x}\!=\!k_{y}\!=\!0$ [\ref{Fig:PulsedPlaneWave}(b)].

Finally, the most general conventional azimuthally symmetric pulsed beam or wave packet has a 2D spectral support domain on the surface of the light-cone in $(k_{r},k_{z},\tfrac{\omega}{c})$-space as shown in \ref{Fig:PulsedBeam}(a). Typically, $\widetilde{\psi}(k_{r},\Omega)$ is separable with respect to $k_{r}$ and $\Omega$, $\widetilde{\psi}(k_{r},\Omega)\!\rightarrow\!\widetilde{\psi}_{r}(k_{r})\widetilde{\psi}_{t}(\Omega)$. The spectral support domain in $(k_{x},k_{y},\tfrac{\omega}{c})$-space as shown in \ref{Fig:PulsedBeam}(b) takes the form approximately of a cylindrical volume centered on the $\tfrac{\omega}{c}$-axis, whose radius is the maximum width of spatial spectrum $\widetilde{\psi}_{r}(k_{r})$, and whose heights is the extent of the temporal spectrum $\widetilde{\psi}_{t}(\Omega)$. 

\subsubsection{3D ST wave packets}

 \begin{figure}[]
  \begin{center}
  \includegraphics[width=8.5cm]{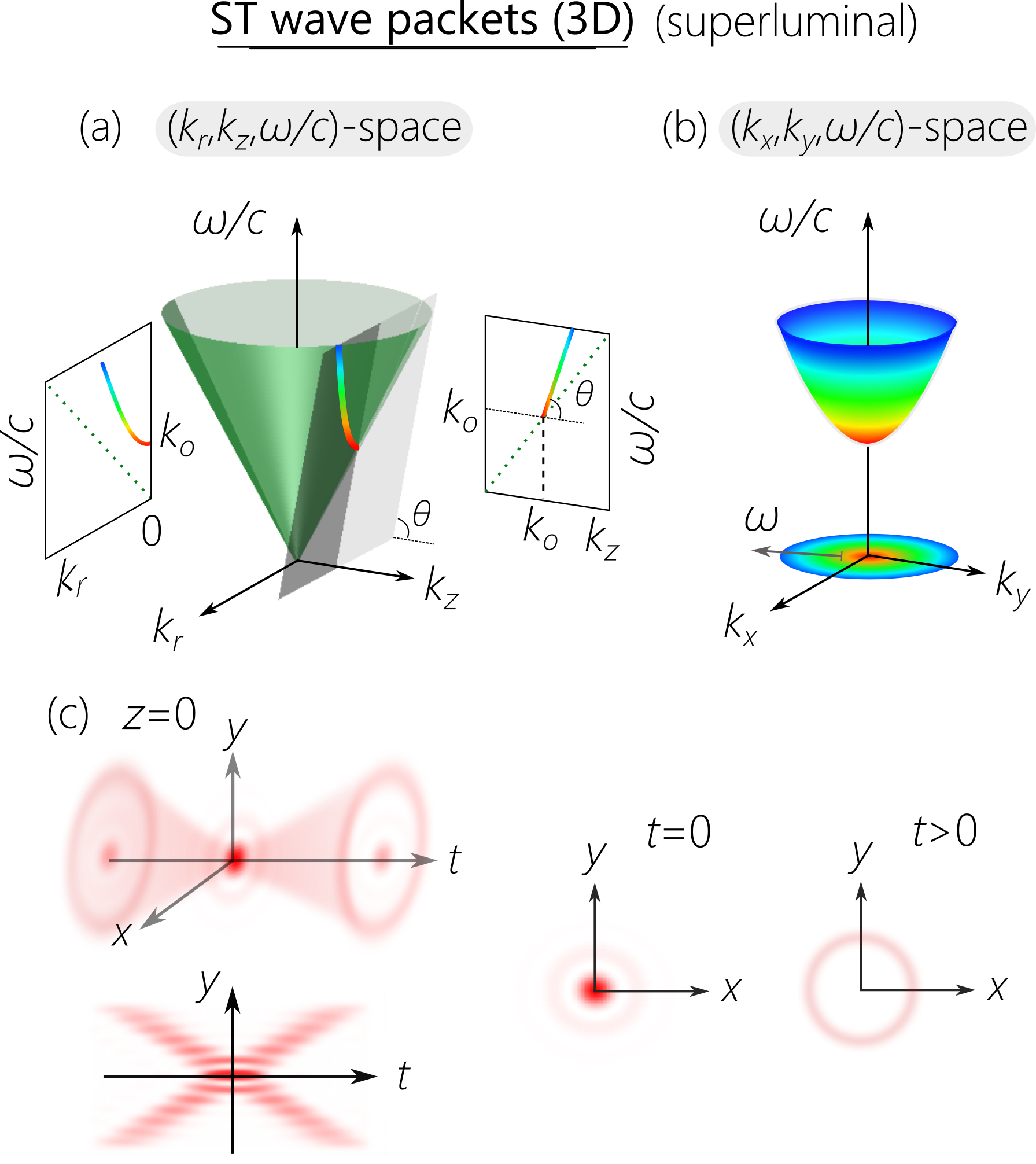}
  \end{center}\vspace{-5mm}
  \caption{\textbf{Supplementary Fig.~6.} Representation of the spectral support domain for a superluminal 3D ST wave packet on the surface of the light-cone. (a) The spectral support domain is restricted to the hyperbola on the light-cone surface at its intersection with a spectral plane that is parallel to the $k_{r}$-axis and makes an angle $\theta$ with the $k_{z}$-axis ($45^{\circ}\!<\!\theta\!<\!90^{\circ}$). (b) In $(k_{x},k_{y},\tfrac{\omega}{c})$-space, the spectral support domain is a 2D surface in the form of one half of a two-sheet hyperboloid (an elliptic hyperboloid). (c) In physical space at $z\!=\!0$, the intensity $I(x,y,0;t)$ takes the form of two cones emanating from the central feature at $t\!=\!0$. In any meridional plane such as $x\!=\!0$, the spatio-temporal intensity profile is X-shaped. In the $(x,y)$-plane at $t\!=\!0$, the intensity profile is centered at $x\!=\!y\!=\!0$. At $t\!\neq\!0$, the profile in the $(x,y)$-plane is an annulus whose radius increases with $t$.}
  \label{Fig:SuperlumST}
  \vspace{12mm}
\end{figure}

\begin{figure}[]
  \begin{center}
  \includegraphics[width=8.5cm]{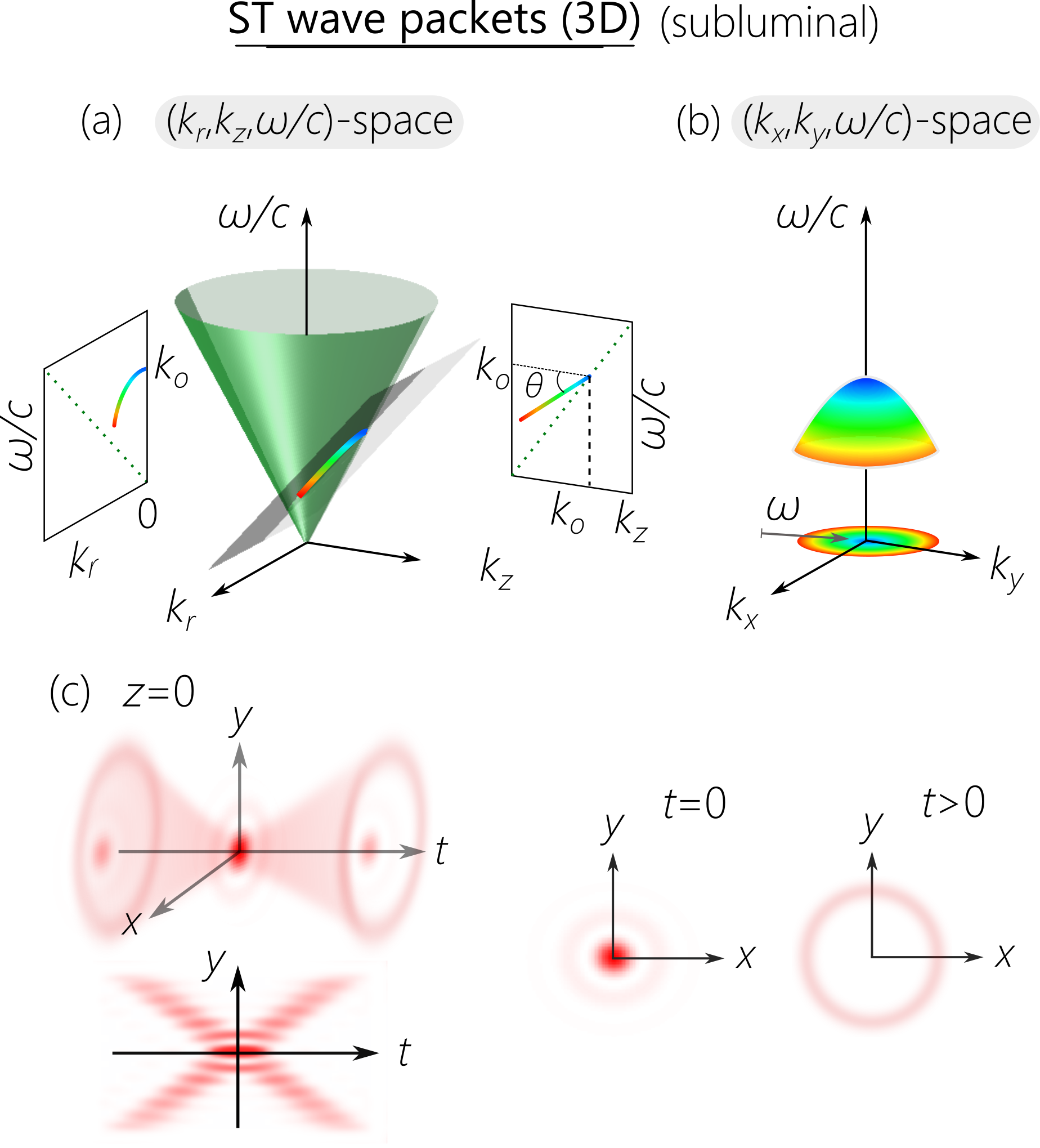}
  \end{center}\vspace{-5mm}
  \caption{\textbf{Supplementary Fig.~7.} Representation of the spectral support domain for a subluminal 3D ST wave packet on the surface of the light-cone. (a) The spectral support domain is restricted to the ellipse on the light-cone surface at its intersection with a spectral plane that is parallel to the $k_{r}$-axis and makes an angle $\theta$ with the $k_{z}$-axis ($0^{\circ}\!<\!\theta\!<\!45^{\circ}$). (b) In $(k_{x},k_{y},\tfrac{\omega}{c})$-space, the spectral support domain is a 2D surface in the form of a ellipsoid of revolution (a spheroid). (c) In physical space at $z\!=\!0$, the intensity $I(x,y,0;t)$ takes the form of two cones emanating from the central feature at $t\!=\!0$. The structure of the spatio-temporal intensity profile is similar to that in \ref{Fig:SuperlumST}(c).}
  \label{Fig:SublumST}
  \vspace{12mm}
\end{figure}

When considering 3D ST wave packets that are localized along all dimensions, such that the time-averaged intensity takes the form of an axial needle \cite{Turunen10PO,FigueroaBook14,Parker16OE}, the spectral support domain is once again restricted to the intersection of the light-cone $k_{r}^{2}+k_{z}^{2}\!=\!(\tfrac{\omega}{c})^{2}$ with the spectral plane $\Omega\!=\!(k_{z}-k_{\mathrm{o}})c\tan{\theta}$, as in the case of the 2D ST wave packets, where the plane is parallel to the $k_{r}$-axis and makes an angle $\theta$ with the $k_{z}$-axis. The group velocity of the propagation-invariant 3D ST wave packet is $\widetilde{v}\!=\!c\tan{\theta}$:
\begin{equation}
\psi(r,\varphi,z;t)=\sum_{\ell=-\infty}^{\infty}e^{i\ell\varphi}\int\!dk_{r}\,\,k_{r}\widetilde{\psi}(k_{r})J_{\ell}(k_{r}r)e^{-i\Omega(t-z/\widetilde{v})}=\psi(r,\varphi,0;t-z/\widetilde{v}).
\end{equation}
The impact of $\theta$ on the shape of the conic section resulting from this intersection is the same as that for the 2D ST wave packet or light sheets. We depict the superluminal scenario in \ref{Fig:SuperlumST}(a) where $45^{\circ}\!<\!\theta\!<\!90^{\circ}$, $\widetilde{v}\!>\!c$, and the spectral support domain is a hyperbola. Alternatively, a subluminal wave packet where $0^{\circ}\!<\!\theta\!<\!45^{\circ}$ and $\widetilde{v}\!<\!c$ is shown in \ref{Fig:SublumST}(a), which has an ellipse for its spectral support domain. In the narrowband paraxial regime, the conic section can be approximated in the vicinity of $k_{r}\!=\!0$ by a parabola:
\begin{equation}\label{Eq:Parabola3D}
\frac{\Omega}{\omega_{\mathrm{o}}}=\frac{k_{r}^{2}}{2k_{\mathrm{o}}^{2}(1-\widetilde{n})}.
\end{equation}
Introducing this quadratic relationship between $\Omega$ and $k_{r}$ into a pulsed beam is the goal of our synthesis methodology. Note that for small bandwidths, this also entails a quadratic relationship between $\lambda-\lambda_{\mathrm{o}}$ and $k_{r}$, where $\lambda_{\mathrm{o}}\!=\!\tfrac{2\pi}{k_{\mathrm{o}}}$.

In the case of the 3D ST wave packets, the structure of spectral representation in $(k_{x},k_{y},\tfrac{\omega}{c})$-space is particularly instructive, as shown in \ref{Fig:SuperlumST}(b) for the superluminal case and in \ref{Fig:SublumST}(b) for its subluminal counterpart. Consider first the superluminal 3D ST wave packet in \ref{Fig:SuperlumST}(b). Each point $(k_{r},k_{z},\tfrac{\omega}{c})$ along the hyperbola on the surface of the light-cone in \ref{Fig:SuperlumST}(a) corresponds to a circle of radius $k_{r}$ at a height $\tfrac{\omega}{c}$ in \ref{Fig:SuperlumST}(b). Because each $k_{r}$ is associated with a different $\omega$, the circles in \ref{Fig:SuperlumST}(b) are all located at different heights, thus forming a 2D surface rather than the 3D volume in \ref{Fig:PulsedBeam}(b) for a conventional pulsed beam. Because the spectral trajectory on the light-cone surface in \ref{Fig:SuperlumST}(a) is a hyperbola, the surface in \ref{Fig:SuperlumST}(b) is one half of a two-sheet hyperboloid (an elliptic hyperboloid) centered on the $\tfrac{\omega}{c}$-axis.

In physical space, as shown in \ref{Fig:SuperlumST}(c), the spatio-temporal intensity profile at a fixed axial profile $z\!=\!0$ takes the form of a central peak at $x\!=\!y\!=\!0$ and $t\!=\!0$, with two cones centered on the $t$-axis emanating from this peak. Consequently, in any meridional plane, for example $x\!=\!0$, the intensity profile is X-shaped, similarly to its 2D ST wave packet counterpart [\ref{Fig:STLightSheet}(c)]. The shape of the cross section of the intensity profile is time-dependent: at $t\!=\!0$ it is localized at $x\!=\!y\!=\!0$, at $t\!\neq\!0$ it takes the form of an annulus whose radius increases with $t$.

The corresponding graphs for a subluminal 3D ST wave packet are plotted in \ref{Fig:SublumST}(b,c). Because the intersection of the spectral plane with the light-cone is an ellipse, the spectral support domain in $(k_{x},k_{y},\tfrac{\omega}{c})$-space is an ellipsoid of revolution, or spheroid, as shown in \ref{Fig:SublumST}(b). Depending on the value of $\theta$, this spheroid may be oblate of prolate. The threshold value of $\tan{\theta}\!=\!\tfrac{1}{\sqrt{2}}$ separates these two regimes. Indeed, at $\tan{\theta}\!=\!\tfrac{1}{\sqrt{2}}$, the projection of the spectral support domain on the light-cone surface onto the $(k_{r},\tfrac{\omega}{c})$-plane is a circle, and the spectral support domain in $(k_{x},k_{y},\tfrac{\omega}{c})$-space is a sphere. Of course, in all cases only the plane-wave components corresponding to $k_{z}\!>\!0$ are physically meaningful. 

Lastly, the spatio-temporal intensity profile of a subluminal 3D ST wave packet [\ref{Fig:SublumST}(c)] in general resembles that of its superluminal counterpart [\ref{Fig:SuperlumST}(c)].

The dynamics of 3D ST wave packets upon free space propagation are calculated by applying Fresnel propagation to the wave packets after introducing the space-time coupling given in Eq.~\ref{Eq:Parabola3D} into the optical field.

\clearpage

\section{Synthesis of 3D space-time wave packets}

The experimental methodology to synthesize 3D ST wave packets shown in Fig.~3 in the main text is expanded in more technical detail here in \ref{Fig:ST_setup}. Our strategy consists of three stages as outlined in \ref{Fig:ST_setup}(a):
\begin{enumerate}
    \item Spectral analysis
    \item A tunable 1D spectral transformation
    \item A fixed 2D coordinate transformation
\end{enumerate}

We start off with femtosecond plane-wave pulses from a mode-locked Ti:sapphire laser (Tsunami; Spectra Physics) of width $\approx\!100$~fs and bandwidth $\Delta\lambda\!\approx\!10$~nm centered at a wavelength of $\approx\!800$~nm. The pulses are directed to the first stage of the synthesis system: spectral analysis using a volume chirped Bragg grating (CBG), which spatially resolves the spectrum \cite{Kaim13SPIE,Glebov14SPIE}. A double-pass through the CBG produces a linear spatial chirp but with a flat phase front. The spectrally resolved wave front from the CBG arrangement is then fed to the second stage: a 1D conformal mapping implemented by a pair of spatial light modulators (SLM; Meadowlark 1920$\times$1080 series) that produces a logarithmic coordinate transformation to yield a logarithmic spatial chirp. The transformed wave front is then directed to a fixed log-polar-to-Cartesian coordinate transformation. This 2D coordinate transformation maps a line at its input into a circle at its output \cite{Bryngdahl74JOSA,Hossack87JOMO,Berkhout10PRL}, and is implemented by means of two refractive \cite{Lavery12OE} or diffractive \cite{Sung06AO} phase plates. The combination of the 1D spectral transformation and the 2D coordinate transformation produces a field endowed with a quadratic radial chirp. Finally, a spherical converging lens performs an optical Fourier transform along both transverse dimensions to yield 3D ST wave packets. We proceed to provide a detailed description of each stage of this novel spatio-temporal synthesis setup.

\subsection{Spectral analysis: Volume chirped Bragg grating (CBG)}

\subsubsection{Introducing spatial chirp}

The goal of the first stage in the setup is to spatially resolve the spectrum but retain a flat phase-front, which is necessary for the successful operation of the subsequent coordinate transformations. A conventional surface grating is therefore \textit{not} suitable for our purposes because the resolved spectrum does \textit{not} have a flat phase. Instead, we make use of an arrangement based on a volume CBG to achieve this goal.

The CBG is a reflective Bragg grating \cite{SalehBook07} with a multilayered structure having a linearly varying periodicity $\Lambda(z)$ along the longitudinal axis $z$. Consequently, different wavelengths are reflected from different depths $z$ within the grating volume \cite{Glebov14SPIE}. As a result, the CBG introduces a \textit{spectral chirp} into normally incident pulses, thus stretching the plane-wave pulse in time [\ref{Fig:CBG_configuration}(a)]. For this reason, volume CBGs are widely used in high-power chirped pulse amplification (CPA) systems, where they are well-known for their high damage threshold and their ability to introduce extremely large spectral chirp \cite{Liao07OE,Sun16OE}.

\begin{figure}[t!]
  \begin{center}
  \includegraphics[width=17cm]{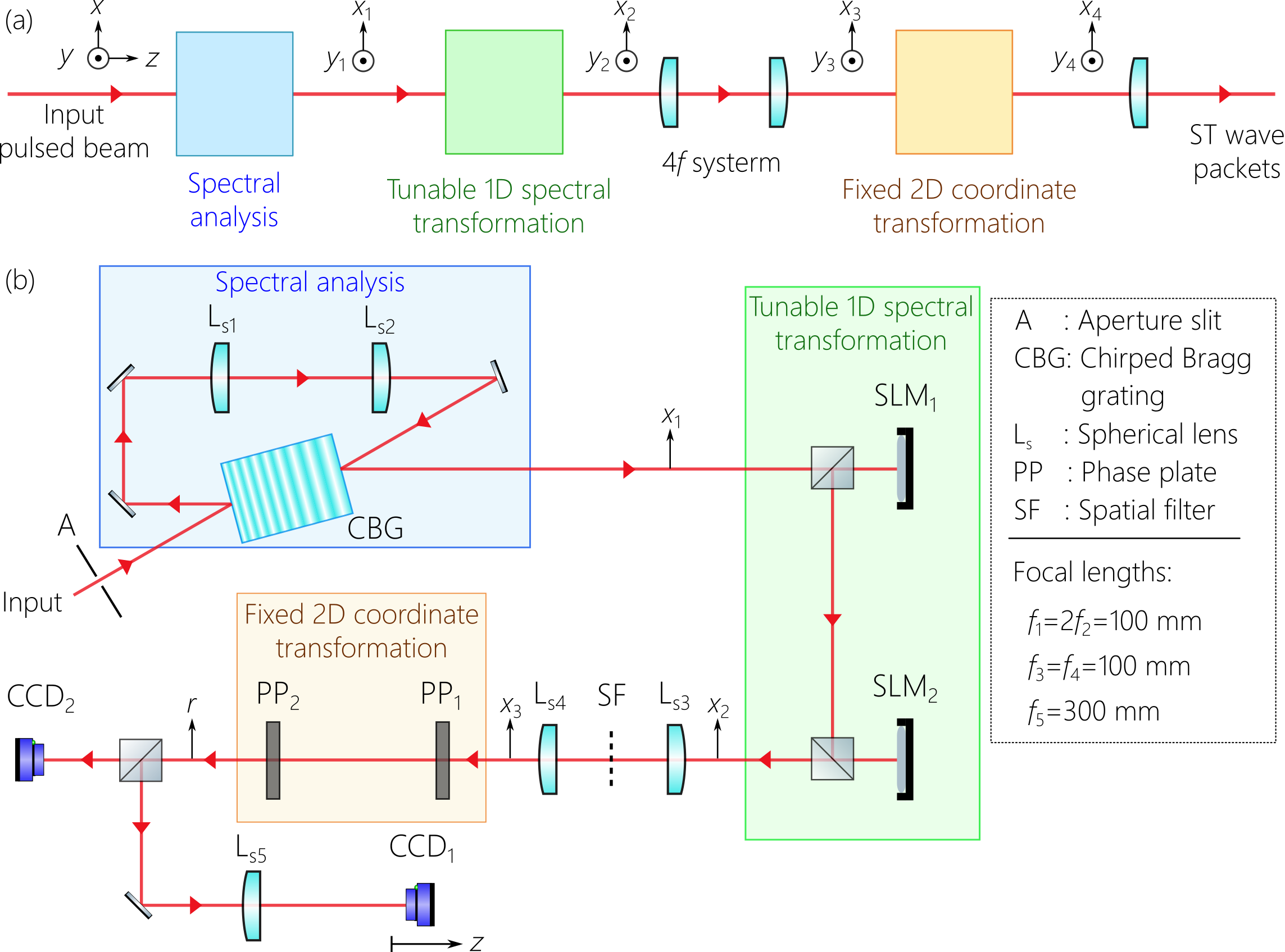}
  \end{center}
  \caption{\textbf{Supplementary Fig.~8.} Setup for synthesizing 3D ST wave packets localized in all dimensions. (a) A conceptual layout of the setup identifying its three main stages: spectral analysis, a tunable 1D spectral transformation, and a fixed 2D coordinate transformation. (b) Detailed setup with a chirped Bragg grating (CBG) in a double-pass configuration, followed by a tunable 1D spectral transformation implemented via SLM$_1$ and SLM$_2$, and a fixed 2D coordinate transformation implemented via phase plates PP$_1$ and PP$_2$. The camera CCD$_{1}$ characterizes the 3D ST wave packets in physical space $(x,y,z)$, and CCD$_{2}$ in the Fourier domain $(k_{x},k_{y},\lambda)$.}
  \label{Fig:ST_setup}
  \vspace{12mm}
\end{figure}

\begin{figure}[t!]
  \begin{center}
  \includegraphics[width=11cm]{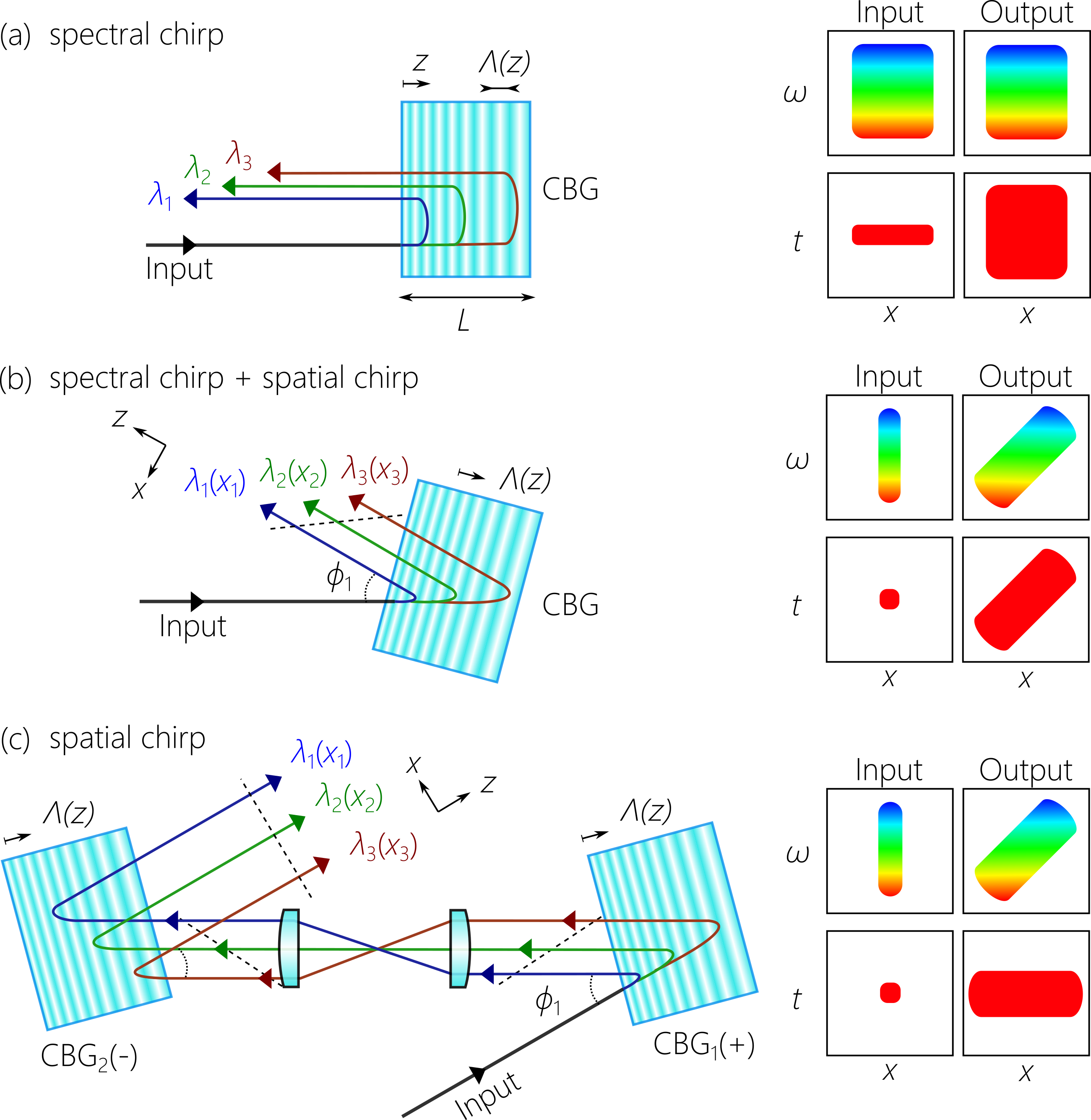}
  \end{center}\vspace{-5mm}
  \caption{\textbf{Supplementary Fig.~9.} Spectral analysis via volume chirped Bragg gratings (CBGs). (a) When a plane-wave pulse is normally incident on a CBG, different wavelengths are reflected from different depths within the CBG. Spectral chirp is thus introduced into the pulse, but with no spatial chirp. (b) When a pulse is obliquely incident on the CBG, both spectral and spatial chirps are introduced. (c) By traversing a pair of identical CBGs with reversed chirp rates, the spectral chirp acquired by the plane-wave pulse from the first CBG is undone by the second, while doubling the spatial chirp. On the right-hand side in each panel, we sketch the change in the spatial distribution of the spectrum and the pulse profile before and after the CBG configuration.}
  \label{Fig:CBG_configuration}
  \vspace{12mm}
\end{figure}

Our goal is to produce \textit{spatial} chirp rather than \textit{spectral} chirp. At oblique incidence, however, both spectral \textit{and} spatial chirps are introduced; i.e., the pulse is stretched longitudinally in time, and the spectrum is resolved transversely in space [\ref{Fig:CBG_configuration}(b)]. The spatial chirp can be determined easily from two parameters: (1) the chirp rate of the CBG structure $\beta\!=\!\frac{d\lambda}{dz}=\frac{1}{2n}\frac{d\Lambda}{dz}$, where $n$ is the average refractive index of the CBG; and (2) the incident angle $\phi_{1}$ with respect to the normal to the grating structure. After reflecting from the CBG at oblique incidence, the spectrum is spatially resolved along the $x$-axis such that each wavelength $\lambda$ is located at a position $x(\lambda)$ given by:
\begin{equation}
x(\lambda)=\frac{1}{\beta}(\lambda-\lambda_{o})\zeta(\phi_{1}),
\end{equation}
where $\zeta(\phi_{1})\!=\!\frac{n\sin{2\phi_1}}{n^{2}-\sin^{2}{\phi_1}}$, $\lambda_{\mathrm{o}}\!=\!2n\Lambda_{\mathrm{o}}$ is the central wavelength, $\Lambda_{\mathrm{o}}\!=\!\Lambda(L/2)$ is the central periodicity of the CBG, and $L$ is its length along the direction of the chirp. Here $x(\lambda)$ is the transverse spatial displacement each wavelength experiences with respect to $\lambda_{\mathrm{o}}$. 

For our purposes here, we aim at retaining the spatial chirp while eliminating the accompanying spectral chirp \cite{Kaim13SPIE}, which we achieve by directing the field to an identical CBG placed in a reversed geometry with respect to the first one \cite{Glebov14SPIE}; see \ref{Fig:CBG_configuration}(c). The output from CBG$_{1}$ first passes through a $4f$ imaging system that flips the field along $x$ and thus reverses the sign of the spatial chirp. Consequently, after passing through CBG$_{2}$ with an opposite sign of chirp $\beta_{2}\!=\!-\beta_{1}$, CBG$_{2}$ doubles the the spatial chirp, while cancelling out the spectral chirp introduced by CBG$_{1}$. As a result, we obtain a spatially resolved spectrum with a flat phase front.

In our setup, we used a folded configuration in which the beam is first incident obliquely on one port of the CBG, and the reflected and flipped field is then directed to the second port of the same device at the same incident angle $\phi_{1}$ [\ref{Fig:ST_setup}(b); Spectral analysis]. Because the spatial spread of the spectrum is doubled after the CBG, we design the $4f$ imaging system between the two passes to de-magnify the field, thus matching the beam size to the CBG aperture size upon incidence on the second port. 
This double-pass configuration produces a spatially resolved spectrum with the following distribution: 
\begin{equation}\label{Eq:CBG_spatial_chirp}
x_{1}(\lambda)=\frac{1}{\beta}(\lambda-\lambda_{o})\zeta(\phi_1)=\alpha(\lambda - \lambda_{o}),
\end{equation}
where $\alpha\!=\!\frac{1}{\beta}\zeta(\phi_{1})$ is the linear spatial chirp rate.

\begin{figure}[t!]
  \begin{center}
  \includegraphics[width=7cm]{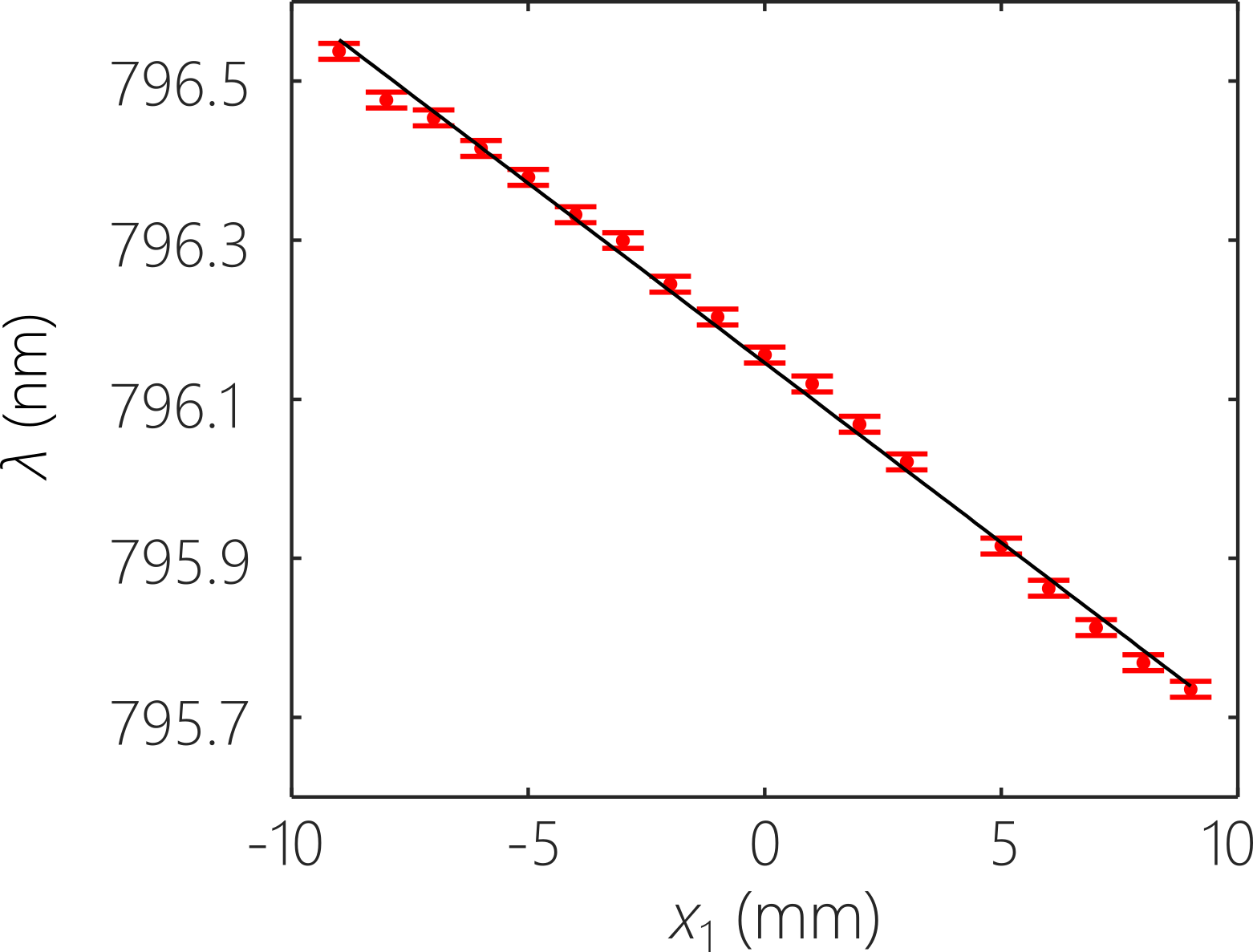}
  \end{center}\vspace{-5mm}
  \caption{\textbf{Supplementary Fig.~10.} The measured spatial spread of the spectrum after the spectral analysis stage in \ref{Fig:ST_setup}(b). The measurements confirm a linear spatial chirp over a temporal bandwidth of $\Delta\lambda\approx0.8$~nm spread over $\approx18$~mm in space horizontally along $x_{1}$. The straight line is a theoretical fit, and symbols are data points. Error bars correspond to the spectral resolution of the Optical Spectrum Analyzer (OSA) used in the measurements, $\delta\lambda=10$~pm. }
  \label{Fig:Spectrum_afterCBG}
  \vspace{12mm}
\end{figure}

In our experiments, we made use of a CBG (OptiGrate L1-021) with a central periodicity of $\Lambda_{\mathrm{o}}\!=\!270$~nm, a chirp rate of $\beta\!=\!-30$~pm/mm, an average refractive index of $n=1.5$, and a length of $L\!=\!34$~mm. The input beam is incident at an angle $\phi_{1}\!\approx\!16^{\circ}$ with respect to the normal to the CBG entrance surface. We measure the spatially resolved spectrum by scanning a single-mode fiber (Thorlabs 780HP) connected to an optical spectrum analyzer (OSA; Advantest AQ6317B). The measured spectrum after the double-pass configuration is plotted in \ref{Fig:Spectrum_afterCBG}, which verifies that the CBG produces a linear spatial chirp of rate $\alpha\!=\!-22.2$~mm/nm (from Eq.~\ref{Eq:CBG_spatial_chirp}) centered at $\lambda_{\mathrm{o}}\!=\!796.1$~nm. Therefore, at the output of the spectral-analysis stage we have a collimated, spectrally resolved optical field.

\subsubsection{Spectral uncertainty induced by CBGs}

A critical parameter characterizing ST wave packets in general is the so-called `spectral uncertainty' $\delta\lambda$: the unavoidable `fuzziness' in the association between spatial frequencies ($k_{x}$ for 2D ST wave packets and $k_{r}$ for 3D ST wave packets) and the wavelength $\lambda$. In the case of a CBG, its spectral resolving power (the smallest spectral shift that can be resolved) determines $\delta\lambda$ \cite{Loewen18Book}. In the case of surface gratings, the spectral uncertainty at a central wavelength $\lambda_{\mathrm{o}}$ is given by $\delta\lambda\!=\!\lambda_{o}/N$, where $N$ is the number of grating rulings covered by the incident beam. Therefore, $\delta\lambda$ here depends on the beam size and the grating period. To minimize $\delta\lambda$ of a surface grating with a fixed ruling density, it is therefore desirable to use the largest possible incident beam size. In contrast, the spectral resolving power of a CBG at normal incidence $\delta\lambda_{\mathrm{BG}}$ -- and thus the associated spectral uncertainty -- is determined by its refractive-index modulation contrast and the grating length \cite{Glebov14SPIE}. Furthermore, an additional contribution to the spectral uncertainty $\delta\lambda_{\mathrm{GE}}$ arises at oblique incidence due to a geometric effect stemming from the finite spatial width of the input beam. When a beam of finite transverse width $w$ is obliquely incident on the CBG, the shifted spectral intervals reflected from different points within the CBG will overlap, thereby leading to an additional contribution to the spectral uncertainty, which is proportional to the input beam width, $\delta\lambda_{\mathrm{GE}}\propto w$. The total spectral uncertainty of the CBG at oblique incidence is estimated to be $\delta\lambda_{\mathrm{CBG}}\!=\!\sqrt{(\delta\lambda_{\mathrm{BG}})^{2}+(\delta\lambda_{\mathrm{GE}})^{2}}$. When the beam passes through the CBG twice, a factor-of-2 \textit{drop} is expected in the spectral uncertainty due to the doubling of the total CBG length traversed by the beam. 

\begin{figure}[t!]
  \begin{center}
  \includegraphics[width=15cm]{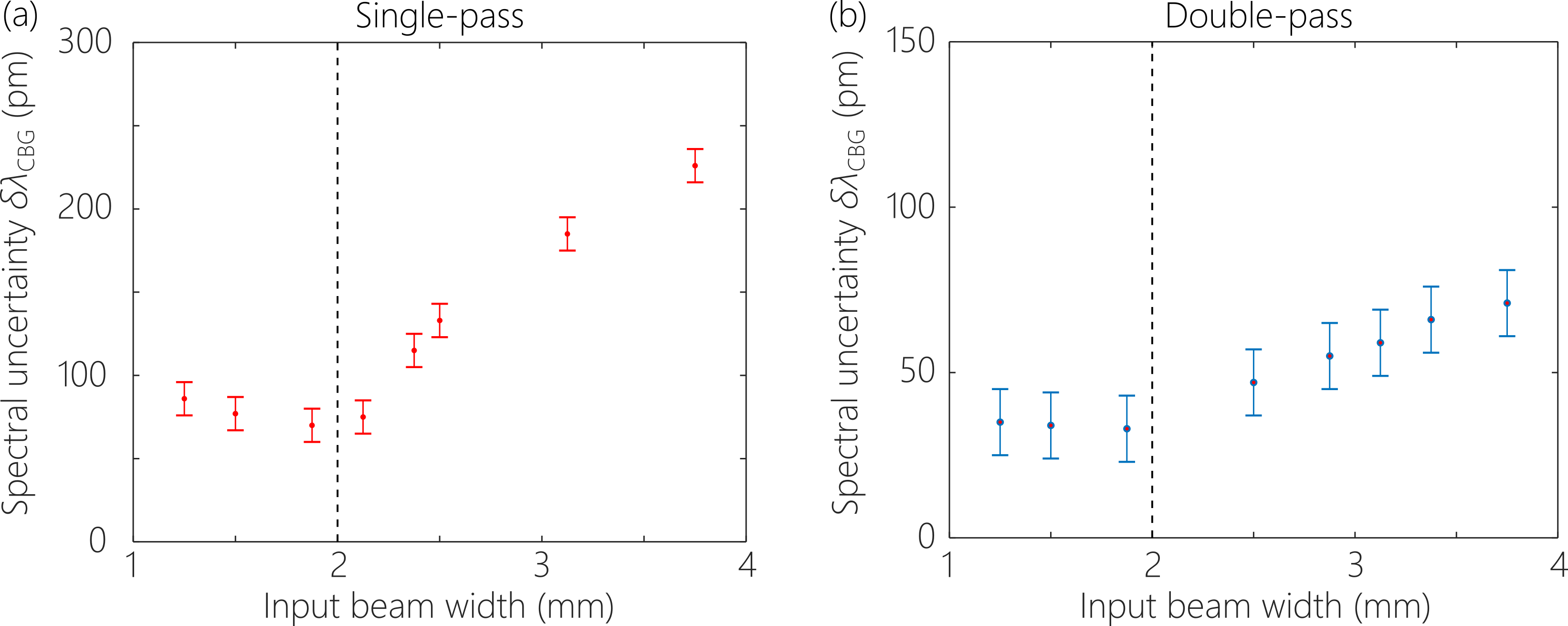}
  \end{center}\vspace{-5mm}
  \caption{\textbf{Supplementary Fig.~11.} Spectral uncertainty for the CBG. (a) The measured spectral uncertainty $\delta_{\mathrm{CBG}}$ from the CBG while varying the the input beam width in a single-pass configuration, and (b) in a double-pass configuration. The vertical dashed lines in (a) and (b) identify the operating point for our experiments. Error bars correspond to the spectral resolution of the OSA, $\delta\lambda=10$~pm.}
  \label{Fig:UncertaintyOfCBG}
  \vspace{12mm}
\end{figure}

In our experiments, we measured $\delta\lambda_{\mathrm{CBG}}$ as a function of the input beam width $w$ after traversing the CBG once [\ref{Fig:UncertaintyOfCBG}(a)] and twice [\ref{Fig:UncertaintyOfCBG}(b)]. The beam width $w$ is tuned using a variable-width aperture preceding the CBG [\ref{Fig:ST_setup}(b)]. The measurements confirm that $\delta\lambda_{\mathrm{CBG}}$ increases with $w$ as expected. However, this trend ends when $w\!\sim\!2$~mm. Below this beam size, the spectral uncertainty begins to increase rather than to decrease further, thereby leading to a valley in the plots in \ref{Fig:UncertaintyOfCBG}(a,b) in the vicinity of $w\!\sim\!2$~mm. This unexpected increase in $\delta_{\mathrm{CBG}}$ at small $w$ is most likely caused by diffraction inside the CBG resulting from the small input beam size. The optimum beam width is thus $w\!\approx\!2$~mm, which yields a minimum spectral uncertainty of $\delta\lambda_{\mathrm{CBG}}\approx35$~pm in the double-pass configuration. The synthesis experiments we performed made use of $w\!\approx\!2$~mm, and thus we take $\delta\lambda\geq35$~pm for the 3D ST wave packets produced.

\subsection{Tunable 1D spectral transformation}

The second stage of the 3D ST wave-packet synthesis setup is the tunable 1D spectral transformation. This is a coordinate transformation performed along the horizontal $x$-axis. Because the wavelengths are arranged linearly along $x$ after the CBG and the field is uniform along $y$, this 1D coordinate transformation results in a reshuffling of the arrangement of the wavelengths along $x$, thereby producing a spectral transformation. This spectral transformation aims at achieving two goals: (1) pre-compensating for the exponentiation in the subsequent 2D coordinate transformation; and (2) tuning the group velocity $\widetilde{v}$ of the synthesized 3D ST wave packet.

The $x$-axis at the input plane is labelled $x_{1}$ and that at the output plane $x_{2}$. The targeted transformation then takes the form:
\begin{subequations}\label{Eq:1DTransform}
\begin{align}
x_{2}=&A\,\,\ln\left(\frac{x_{1}}{B}\right),\\
y_{2}=&y_{1},
\end{align}
\end{subequations}
where $A$ and $B$ are the transformation parameters, the physical significance of which will be discussed shortly. The uniform field distribution along $y$ remains intact. This conformal mapping is implemented by means of two phase patterns placed at the input and output planes and separated by a distance $d_{1}$. The first phase distribution $\Phi_{1}(x_{1},y_{1})$ at the input plane performs the particular transformation given in Eq.~\ref{Eq:1DTransform}. In other words, the wavelength $\lambda$ located at position $x_{1}$ is now located at position $x_{2}$. However, such a transformation does not produce a collimated field. The second phase distribution $\Phi_{2}(x_{2},y_{2})$ placed at the output plane collimates the transformed wave front to yield an afocal transformation. Usually, a lens is placed between the two phase plates in a $2f$-configuration. In our experiments, we do not make use of a lens and instead distribute the quadratic phase associated with such a lens between the two phase plates. The required phase profiles $\Phi_{1}(x_1,y_{1})$ and $\Phi_{2}(x_{2},y_{2})$ -- both of which are independent of $y$ -- can be derived using the methodology outlined in \cite{Hossack87JOMO}, and they take the form:
\begin{subequations}\label{Eq:1DTPhase}
\begin{align}
\Phi_{1}(x_{1},y_{1})=&\frac{kA}{d_{1}}\left[ x_{1}\ln\left(\frac{x_{1}}{B}\right)-x_{1}\right] - \frac{k x_{1}^{2}}{2d_{1}},\\
\Phi_{2}(x_{2},y_{2})=&\frac{kAB}{d_{1}}\exp{\left(\frac{x_{2}}{A}\right)}-\frac{k x_{2}^{2}}{2d_{1}},
\end{align}
\end{subequations}
where $k=\frac{2\pi}{\lambda}$ is the wave number. In the setup, $\Phi_{1}(x_1,y_{1})$ and $\Phi_{2}(x_{2},y_{2})$ are displayed on two phase-only SLMs with $d_{1}\!=\!400$~mm [\ref{Fig:ST_setup}(b)]. Because the phase profiles according to Eq.\ref{Eq:1DTPhase} depend only on $x$ and not $y$, 1D SLMs can be used here in principle. However, as we show later, utilizing 2D SLMs provides the possibility of also modulating the field along $y$, which translates into an azimuthal modulation after the subsequent 2D coordinate transformation. 

\begin{figure}[t!]
  \begin{center}
  \includegraphics[width=13cm]{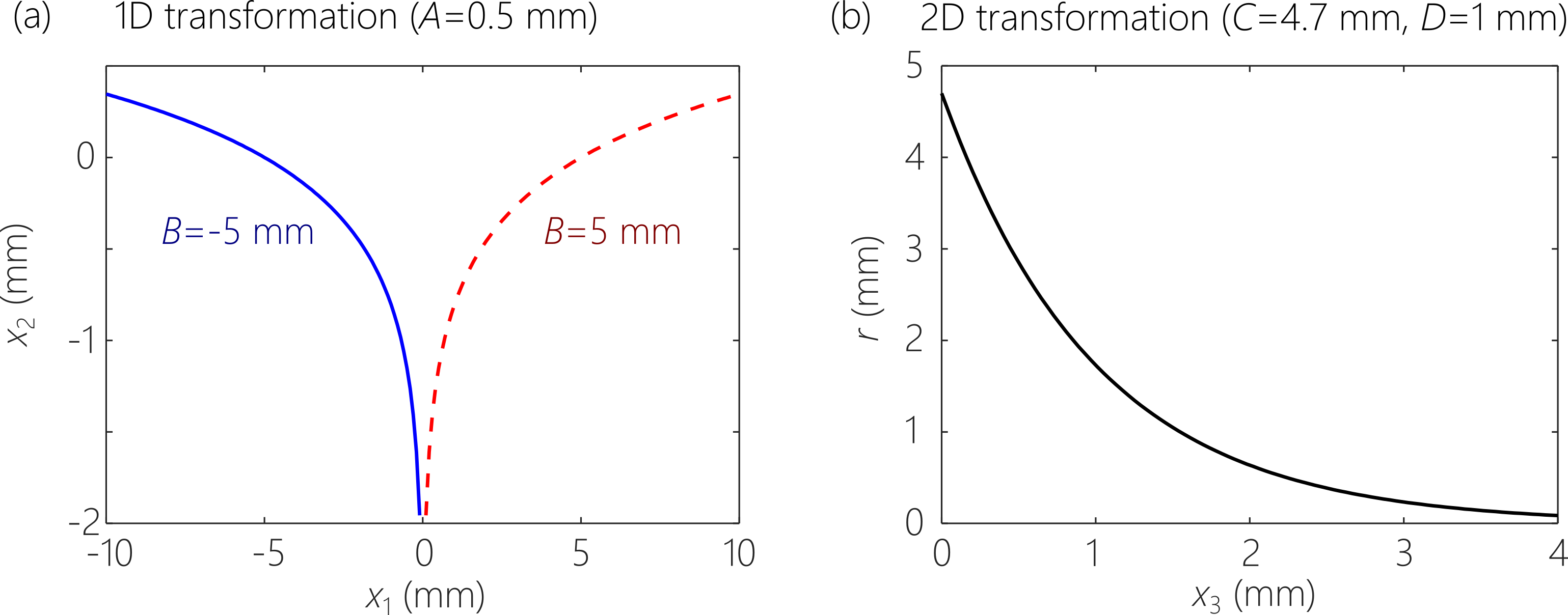}
  \end{center}\vspace{-5mm}
  \caption{\textbf{Supplementary Fig.~12.} Spatial transformations associated with the 1D spectral transformation and the 2D coordinate transformation. (a) Plot of the transformation from the input plane $x_{1}$ to the output plane $x_{2}$ of the tunable 1D spectral transformation. We make use of the parameter values $A\!=\!0.5$~mm, and $B\!=\!-5$~mm (blue line) or $B\!=\!5$~mm (dashed red line). (b) Plot of the relationship between $x_{3}$ at the input plane to the radius $r$ at the output plane for the fixed 2D coordinate transformation. We make use of the parameter values $D\!=\!2A\!=\!1$~mm and $C\!=\!4.77$~mm.}
  \label{Fig:TranformationPlot}
  \vspace{12mm}
\end{figure}

In our experiments, we fix $A\!=\!0.5$~mm and tune $B$ over the range $B\!=\![-15,20]$~mm to control the group velocity of the 3D ST wave packets, as explained in detail later. See \ref{Fig:TranformationPlot}(a) for a plot of the relationship between $x_{1}$ and $x_{2}$ when making use of these values.

The field after this 1D spectral transformation is imaged by a one-to-one $4f$ system comprising two spherical lenses $L_{\mathrm{s}3}$ and $L_{\mathrm{s}4}$, as shown in \ref{Fig:ST_setup}(b), from the output plane of the 1D spectral transformation $(x_{2},y_{2})$ to the input plane ($x_{3},y_{3}$) of the 2D coordinate transformation. Thus, the beam is flipped along the $x$ and $y$ axes after this $4f$ system: $x_{3}=-x_{2}$ and $y_{3}=-y_{2}$. This is important to keep in mind for determining the required transformation parameters $A$ and $B$ based on the desired group velocity $\widetilde{v}$. In addition, a spatial filter is placed in the Fourier plane of the $4f$ system to eliminate the undesired zeroth-order field component resulting from the limited efficiency of SLM$_{1}$ and SLM$_{2}$.

\subsection{Fixed 2D coordinate transformation}

 \begin{figure}[t!]
  \begin{center}
  \includegraphics[width=9cm]{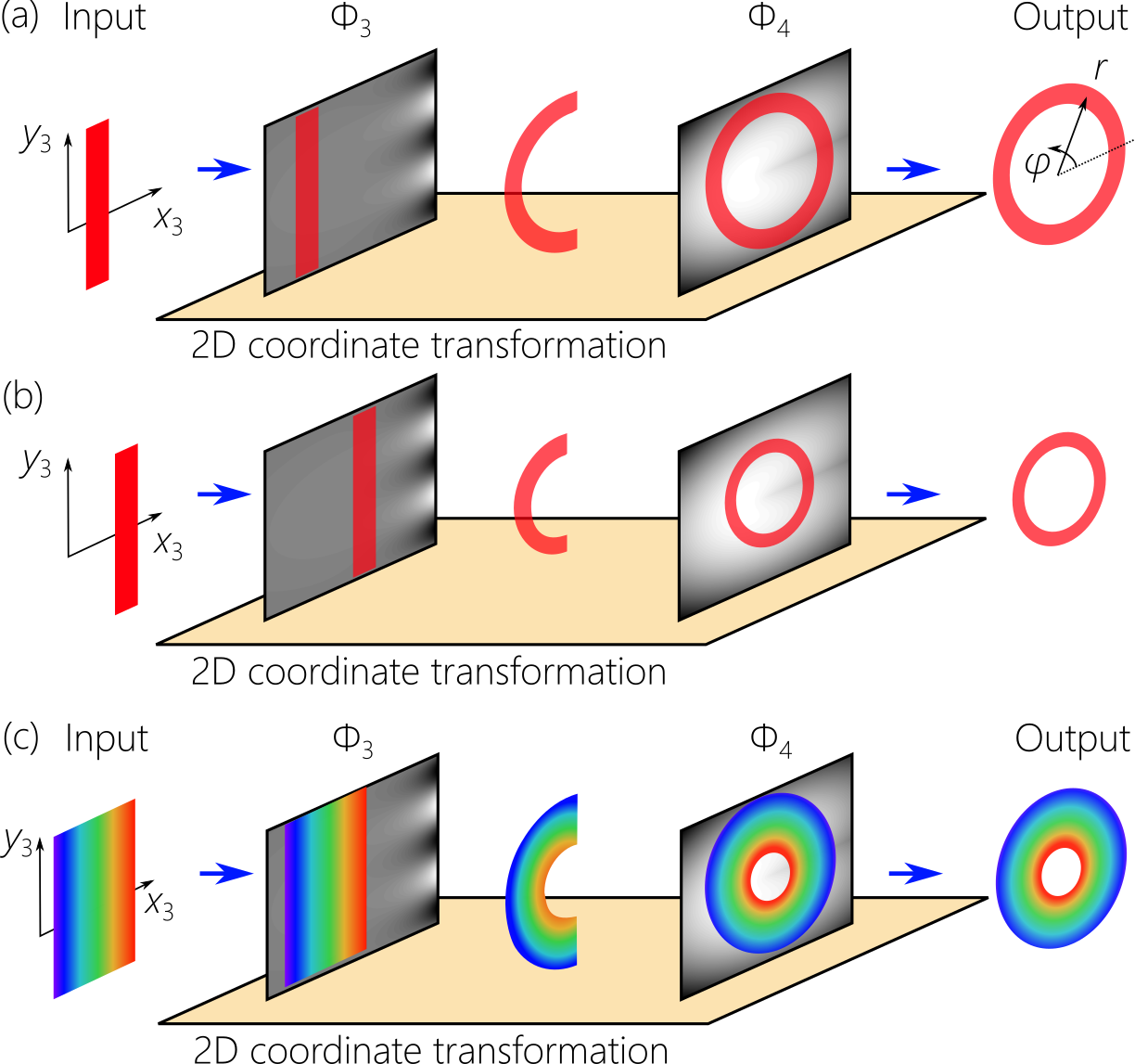}
  \end{center}\vspace{-5mm}
  \caption{\textbf{Supplementary Fig.~13.} Principle of operation for the fixed 2D coordinate transformation. (a) For a monochromatic input field in the form of a vertical rectangular strip, the 2D transformation converts it into an annulus. (b) When the vertical rectangular strip is shifted horizontally, an annulus of different radius is formed at the output, which is nevertheless concentric with the annulus in (a). (c) In our experiments, the input field is a spatially extended spectrally resolved wave front. The transformation converts the input into an annulus with the wavelengths arranged radially in circles.}
  \label{Fig:2DTransform}
  \vspace{12mm}
\end{figure}

The third stage of the spatio-temporal synthesis strategy for 3D ST wave packets is a fixed 2D coordinate transformation; see \ref{Fig:ST_setup}(b) and \ref{Fig:2DTransform}. This transformation performs a conformal mapping from log-polar to Cartesian coordinate systems. The input plane is spanned by Cartesian coordinates $(x_{3},y_{3})$ and the output by $(x_{4},y_{4})$. The transformation maps the input Cartesian coordinate system $(x_{3},y_{3})$ to a polar coordinate system at the output $(x_{3},y_{3})\rightarrow(r,\varphi)$, where $r\!=\!\sqrt{x_{4}^{2}+y_{4}^{2}}$ and $\varphi\!=\!\arctan{(\tfrac{y_{4}}{x_{4}})}$; see \ref{Fig:TranformationPlot}(b). The transformation is given explicitly in the following form \cite{Bryngdahl74JOSA,Hossack87JOMO,Saito83OC}:
\begin{subequations}\label{Eq:2DTransform}
\begin{align}
r=&C\exp{\left(-\frac{x_{3}}{D}\right)}, \\
\varphi=&\frac{y_{3}}{D}.
\end{align}
\end{subequations}
The transformation parameter $D$ is chosen to map the vertical size of the input field $y_{3}\!=\![-y_{3}^{\mathrm{max}},y_{3}^{\mathrm{max}}]$ to the range $\varphi\!=\![-\pi,\pi]$, and thus we impose $D=\frac{y_{3}^{\mathrm{max}}}{\pi}$; the value of C is selected based on the aperture size of the optics used. The 2D coordinate transformation therefore maps a vertical line located at $x_{3}$ at the input plane into a circle of radius $r$ at the output plane. Shifting the location of the line horizontally along $x_{3}$ at the input thus results in a radial expansion or shrinkage of the circle radius at the output according to the direction of the shift at the input [\ref{Fig:2DTransform}(a,b)]. If the input beam is a rectangle of width $\Delta x_{3}$, the transformed beam is an annulus of radial thickness $\Delta r$, where the inner and outer radii of the annulus correspond to the two vertical boundaries of the input rectangle.

This 2D coordinate transformation is implemented by two 2D phase patterns (both depending on the $x$ and $y$ coordinates) separated by a distance $d_{2}$: $\Phi_{3}$ at the input plane $(x_{3},y_{3})$ and $\Phi_{4}$ at the output plane $(x_{4},y_{4})$. For the mapping given in Eq.~\ref{Eq:2DTransform}, the phase patterns take the following form \cite{Hossack87JOMO,Berkhout10PRL,Lavery12OE}:
\begin{subequations}\label{Eq:2DTPhase}
\begin{align}
\Phi_{3}(x_{3},y_{3})=& - \frac{kCD}{d_{2}}\exp{\left(-\frac{x_{3}}{D}\right)}\cos{\left(\frac{y_{3}}{D}\right)} - \underbrace{\frac{k (x_{3}^{2}+y_{3}^{2})}{2d_{2}}}_{\textrm{lens term}},\\
\Phi_{4}(x_{4},y_{4})=& \frac{kD}{d_{2}}\left[\mathrm{atan2}(y_{4},x_{4})-x_{4}\ln{\left(\frac{\sqrt{(x_{4}^{2}+y_{4}^{2})}}{C}\right)} +x_{4}\right] - \underbrace{\frac{k (x_{4}^{2}+y_{4}^{2})}{2d_{2}}}_{\textrm{lens term}},\
\end{align}
\end{subequations}
where $\mathrm{atan2}(y_{4},x_{4})$ is 2-argument $\arctan$ function. Usually, a spherical lens is placed midway between the two phase patterns in a $2f$ configuration. Instead, the appropriate quadratic phases are added in $\Phi_{3}$ and $\Phi_{4}$, which are the last terms in Eq.~\ref{Eq:2DTPhase}(a) and Eq.~\ref{Eq:2DTPhase}(b) \cite{Lavery12OE}.

 \begin{figure}[t!]
  \begin{center}
  \includegraphics[width=12cm]{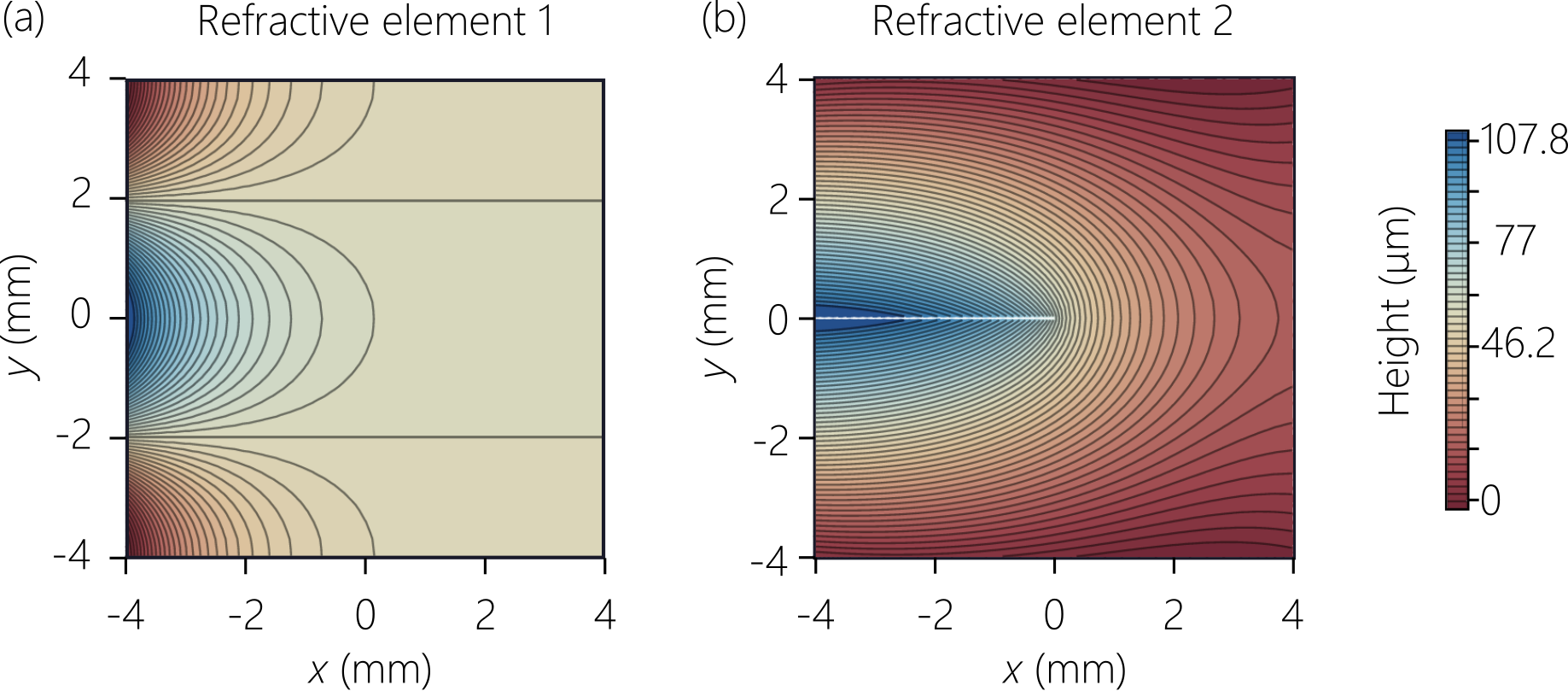}
  \end{center}\vspace{-5mm}
  \caption{\textbf{Supplementary Fig.~14.} Refractive phase plates for implementing the 2D coordinate transformation. (a) Height profile of the refractive element 1, and (b) of the refractive element 2. The phase plates are designed such that the aperture size is $2y_{3}^{\mathrm{max}}=8$~mm, the distance separating them in the setup is $d_{2}=310$~mm, and the transformation parameters are $C=4.77$~mm, $D=1$~mm.}
  \label{Fig:RefractivePP}
  \vspace{12mm}
\end{figure}

As discussed in the Methods Section in the main text, the 2D transformation was implemented using \textit{diffractive} optics \cite{Lavery11IOP,Berkhout11OL,Berkhout10PRL,Li19OE} and separately \textit{refractive} optics \cite{Lavery12OE} to imprint the phase profiles in Eq.~\ref{Eq:2DTPhase} via diamond-edged refractive phase plates \cite{Lavery12OE} and analog diffractive phase plates \cite{Li19OE}. The refractive optical elements used in our experiments are similar to those outlined by Lavery \textit{et al.} in \cite{Lavery12OE}, in which the transformation parameters are $C=4.77$~mm, $D\!=\!\tfrac{3.2}{\pi}\approx\!1$~mm, and $d_{2}=310$~mm. Each phase plate is made of the polymer PMMA (Poly methyl methacrylate) with accurately manufactured height profiles $Z_{1}(x_{3},y_{3})$ and $Z_{2}(x_{4},y_{4})$ to imprint the required phase profiles given in Eq.~\ref{Eq:2DTPhase}. The phase encountered by light at a wavelength $\lambda$ traversing a height $Z$ of a material of refractive index $n$ -- with respect to the phase encountered over the same distance in vacuum -- is given by $\Phi\!=\!2\pi(n-1) Z/\lambda$. Thus, the height profile of the first element is $Z_{1}(x_{3},y_{3})=\frac{\lambda}{2\pi(n-1)}\Phi_{3}(x_{3},y_{3})$ [\ref{Fig:RefractivePP}(a)] and that of the profile of the second element is $Z_{2}(x_{4},y_{4})=\frac{\lambda}{2\pi(n-1)}\Phi_{4}(x_{4},y_{4})$ [\ref{Fig:RefractivePP}(b)]. See Methods in the main text for details of the fabrication and characteristics of the fabricated elements
\cite{Dow91PE}. The diffractive phase plates were fabricated in fused silica using Clemson University facilities using the process outlined in \cite{Sung06AO}. See Methods in the main text for the design parameters of the phase plates.

\subsection{Spatial Fourier transform}

After the 2D coordinate transformation, the spatially resolved wavelengths are arranged radially $r(\lambda)$. A spherical converging lens L$_{\mathrm{s}5}$ [\ref{Fig:ST_setup}(b)] performs a 2D Fourier transform to produce the 3D ST wave packet in physical space. In effect, each spatial position is now mapped into a spatial frequency $k_{r}(\lambda)\!=\!k\tfrac{r}{f}$, where $f$ is the focal length of the lens. because of the small bandwidth $\Delta\lambda$ utilized in our experiments, we can use the approximation $k\!\approx\!k_{\mathrm{o}}$ in the scaling of the radial spatial frequency.

\subsection{Full system analysis}

We can now combine all the stages in our synthesis system to identify the role played by the various parameters involved: the chirp rate $\alpha$ from the CBG; the parameters $A$ and $B$ from the tunable 1D spectral transformation; the parameters $C$ and $D$ from the fixed 2D coordinate transformation; and the focal length $f$ of the Fourier-transforming lens. After the CBG we have $x_{1}(\lambda)\!=\!\alpha(\lambda-\lambda_{\mathrm{o}})$; after the tunable 1D spectral transformation we have $x_{2}(\lambda)\!=\!A\ln{\{\tfrac{x_{1}(\lambda)}{B}\}}$; after the $4f$ imaging system we have $x_{3}\!=\!-x_{2}$; after the fixed 2D coordinate transformation we have $r(\lambda)\!=\!C\exp{\{-\tfrac{x_{3}(\lambda)}{D}\}}$; and, finally, after the Fourier-transforming lens we have $k_{r}\!=\!k_{\mathrm{o}}\tfrac{r(\lambda)}{f}$. Combining all these steps, we have:
\begin{equation}
k_{r}(\lambda)=C\frac{k_{\mathrm{o}}}{f}\left(\frac{\alpha}{B}(\lambda-\lambda_{\mathrm{o}})\right)^{A/D}.
\end{equation}

\begin{table}[b!]
\begin{center} 
\begin{tabular}{|c|c|c|}
\hline
      & Group velocity $\widetilde{v}$ & $B$  \\ \hline\hline
    Positive-subluminal & $0<\widetilde{v}<c$ & $B<0$ \\ \hline
    Positive-luminal    & $\widetilde{v}=c$  & $B=\infty$ \\ \hline
    Positive-superluminal    & $c<\widetilde{v}<\infty$  & $\frac{|\alpha|\lambda_{o}C^2}{2f^{2}}<B<\infty$ \\ \hline
    Infinite-$\widetilde{v}$    & $\widetilde{v}=\infty$  & $B=\frac{|\alpha|\lambda_{o}C^2}{2f^{2}}$ \\ \hline
    Negative-superluminal    & $\widetilde{v}<-c$  & $\frac{|\alpha|\lambda_{o}C^2}{4f^{2}}<B<\frac{|\alpha|\lambda_{o}C^2}{2f^{2}}$\\ \hline
    Negative-luminal    & $\widetilde{v}=-c$  & $B=\frac{|\alpha|\lambda_{o}C^2}{4f^{2}}$ \\ \hline
    Negative-subluminal & $-c<\widetilde{v}<0$ & $0<B<\frac{|\alpha|\lambda_{o}C^2}{4f^{2}}$ \\ \hline
\end{tabular} 
\end{center}
    \caption{\textbf{Supplementary Table~1.} Conversion from the group velocity to the parameter $B$. This Table identifies the range taken by the values of of the parameter $B$ (third column) corresponding to the target range of the values of the STWP group velocity $\widetilde{v}$ (second column).}
    \label{Table:vgVsB}
\end{table}

To obtain the desired radial chirp from Eq.~\ref{Eq:Parabola3D}, we must impose the constraint $\tfrac{D}{A}\!=\!2$, whereupon:
\begin{equation}\label{Eq:SpectralCorrelation}
k_{r}(\lambda)=C\frac{k_{\mathrm{o}}}{f}\sqrt{\frac{\alpha}{B}(\lambda-\lambda_{\mathrm{o}})}.
\end{equation}
The parameter $C\!=\!4.77$~mm is determined by the size of the optics (the aperture size is $2y_{3}^{\mathrm{max}}\!=\!8$~mm), $f\!=\!300$~mm, $\alpha\!=\!-22.2$~mm/nm is determined by the CBG, and $\lambda_{\mathrm{o}}\!\approx\!798$~nm. This leaves $B$ as the sole free parameter determining the chirp rate, and thus also the group velocity $\widetilde{v}$. 



By comparing the result in Eq.~\ref{Eq:SpectralCorrelation} to the target spatio-temporal spectrum for 3D ST wave packets given in Eq.~\ref{Eq:Parabola3D}, we can obtain the group index:
\begin{equation}\label{Eq:ExpGroupIndex}
\widetilde{n}=1+\left(\frac{C^2\alpha\lambda_{o}}{2f^2}\right)\frac{1}{B}=1-\frac{2.24~\mathrm{mm}}{B}
\end{equation}
As a result, we can generate ST wave packets with group velocities from subluminal $(B<0)$ to superluminal $(B>C^{2}|\alpha|\lambda_{o}/2f^2)$, and in principle negative group velocities $(0<B<C^{2}|\alpha|\lambda_{o}/2f^2)$. See Table~\ref{Table:vgVsB} for the conversion between the desired group velocity ranges and the required value of the parameter $B$. 

\subsection{Synthesizing pulsed Bessel beams with separable spatio-temporal spectrum}

To compare the performance of 3D ST wave packets with that of the conventional pulsed Bessel beams shown in Fig.~5(a-b) in the main text, we bypass the spectral analysis and 1D spectral transformation stages and send the input laser pulses directly to the 2D transformation. Consequently, separable pulsed Bessel beams are produced. In addition, we spectrally filter $\Delta\lambda\approx0.3$~nm of the input pulses to obtain a spectral bandwidth comparable to that of the 3D ST wave packets.  

\clearpage

\section{Characterization of space-time wave packets}

We characterize the 3D ST wave packets in four domains:
\begin{enumerate}
    \item The spatio-temporal spectral plane $(k_{x},k_{y},\lambda)$ or $(k_{r},\lambda)$ to confirm the presence of desired spectral structure.
    \item In physical space we follow the evolution of the time-averaged intensity $I(x,y,z)$ along the $z$-axis to verify the diffraction-free propagation of the 3D ST wave packets.
    \item We reconstruct the spatio-temporal intensity profile $I(x,y,z;t)$ using time-resolved linear interferometry, which also enables us to estimate the group velocity $\widetilde{v}$.
    \item The amplitude and phase of the complex-field envelope $\psi(x,y,z;t)\!=\!|\psi(x,y,z;t)|e^{i\phi(x,y,z;t)}$ is reconstructed using off-axis digital holography. 
\end{enumerate}

\subsection{Spatio-temporal spectrum measurements}

\begin{figure}[t!]
  \begin{center}
  \includegraphics[width=17cm]{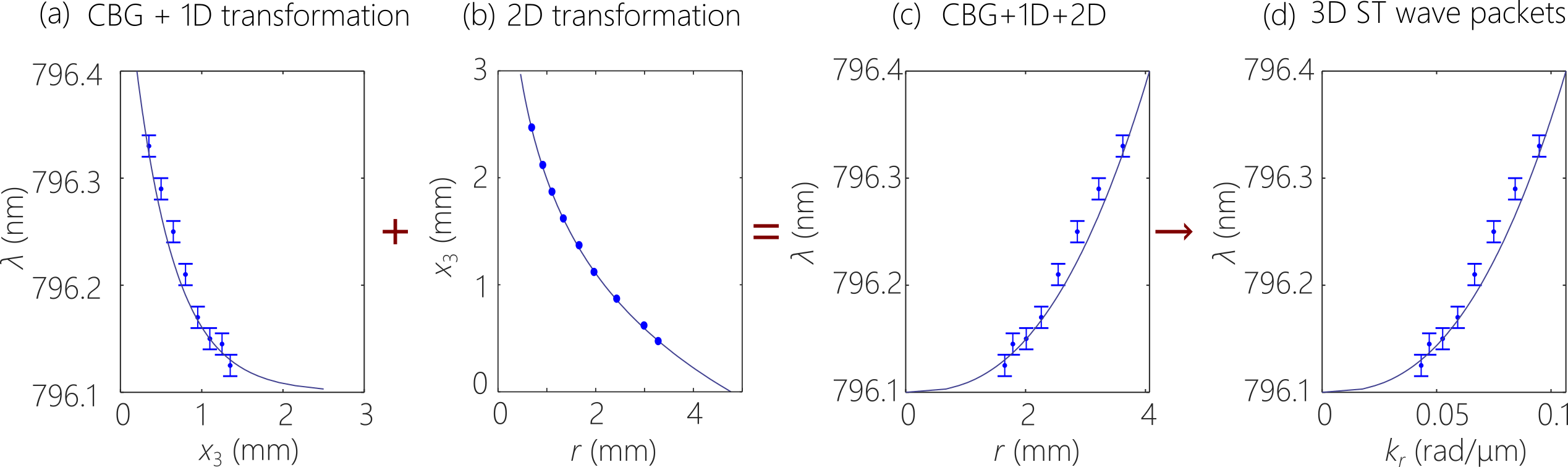}
  \end{center}\vspace{-5mm}
  \caption{\textbf{Supplementary Fig.~15.} Evolution of the spatio-temporal spectrum of the 3D ST wave packets through the synthesis setup. (a) Measured relationship between $\lambda$ and the transverse coordinate $x_{3}$ after the 1D spectral transformation. (b) Measured relationship between $x_{3}$ before the 2D coordinate transformation and $r$ after it. (c) By combining (a) and (b), we obtain the relationship between $\lambda$ and $r$ after the 2D spectral transformation. (d) Transforming the relationship in (c) to one between $\lambda$ and the radial spatial frequency $k_{r}$. Error bars in (a,c,d) correspond to the spectral resolution of the OSA, $\delta\lambda=10$~pm}
  \label{Fig:SpectrumCalibration}
  \vspace{12mm}
\end{figure}

The spatio-temporal spectrum $|\widetilde{\psi}(k_{x},k_{y};\lambda)|^{2}$ is captured at CCD$_2$ (The ImagingSource, DMK 33UX178) as shown in \ref{Fig:ST_setup}, which corresponds to the Fourier plane ($k_{x},k_{y})$ of the synthesized 3D ST wave packets [Fig.~4(a) in the main text]. Because the camera cannot distinguish between the various wavelengths, we resolve the temporal spectrum in two steps. We first scan the fiber tip connected to an OSA along the horizontal axis $x_{2}$ after the 1D spectral transformation and determine the spatial chirp $x_{2}(\lambda)$. The spatial chirp after the $4f$ imaging system is identical to $x_{2}(\lambda)$ except for a spatial flip [\ref{Fig:SpectrumCalibration}(a)]. In a second step, we verify experimentally the impact of the 2D coordinate transformation by scanning a vertical slit horizontally along $x_{3}$ and measure the radius of the annulus formed at the output [\ref{Fig:SpectrumCalibration}(b)]. By combining these two measurements we obtain the spatial chirp along the radial direction $r(\lambda)$ after the 2D coordinate transformation [\ref{Fig:SpectrumCalibration}(c)]. Finally, we obtain spatio-temporal spectrum $k_{r}(\lambda)$ [\ref{Fig:SpectrumCalibration}(d) and Fig.~4(iii) in the main text] by converting from the physical space to Fourier space $k_{r}\!=\!k\frac{r}{f}$, where $k\!=\!\frac{2\pi}{\lambda}$ is the wave number, and $f\!=\!300$~mm is the focal lens of the Fourier-transforming lens L$_{\mathrm{s}5}$. The solid curves correspond to theoretical predictions and the dots correspond to data points. 

\subsection{Time-averaged intensity measurements}

The axial evolution of the time-averaged intensity profile $I(x,y,z)\!=\!\int\!dt|\psi(x,y,z;t)|^{2}$ is obtained in the physical space by scanning CCD$_{1}$ (The ImagingSource, DMK 27BUP031) along the propagation axis $z$ [\ref{Fig:ST_setup}(b)]. This measurement confirms the diffraction-free propagation of the 3D ST wave packets, as shown in the Fig.~5 in the main text. 

\subsection{Time-resolved intensity measurements}

 \begin{figure}[t!]
  \begin{center}
  \includegraphics[width=14cm]{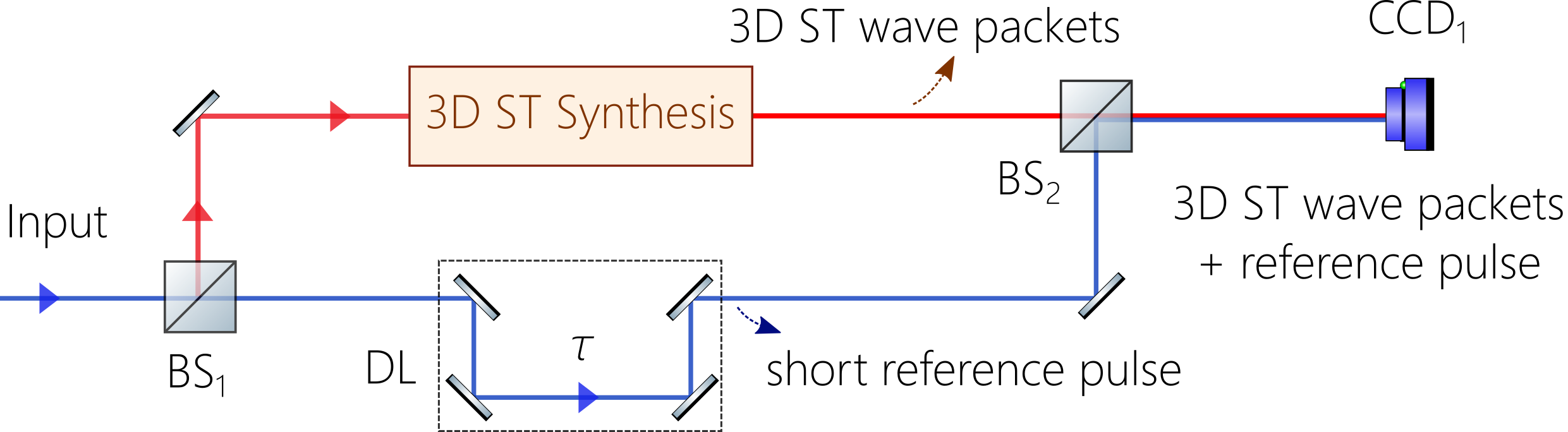}
  \end{center}\vspace{-5mm}
  \caption{\textbf{Supplementary Fig.~16.} Schematic illustration of the setup for reconstructing the spatio-temporal profile of 3D ST wave packets and estimating their group velocity. The spatio-temporal synthesis arrangement from \ref{Fig:ST_setup}(b) is placed in one arm of a Mach-Zehnder interferometer, and an optical delay $\tau$ is placed in the reference arm that is traversed by the short pulses from the initial laser source.}
  \label{Fig:ST_characterization}
  \vspace{12mm}
\end{figure}

The spatio-temporal intensity profile of the 3D ST wave packet is measured using a Mach-Zehnder interferometer in which the spatio-temporal synthesis arrangement is placed in one arm. The second (reference) arm is traversed by the original short plane-wave laser pulses from the source (pulse width $\sim\!100$~fs) that encounter an optical delay line $\tau$ [\ref{Fig:ST_characterization}]. The two wave packets (the 3D ST wave packet and the reference pulse) propagate co-linearly after they are merged at the beam splitter BS$_{2}$.  When the two wave packets overlap in space and time, spatially resolved interference fringes are recorded by CCD$_{1}$ at a fixed axial position $z$, from whose visibility we extract the spatio-temporal profile at that plane. By scanning the delay $\tau$, we can reconstruct the spatio-temporal intensity profile $I(x=0,y,z;\tau)$ at the fixed axial plane $z$. Using this approach, we obtained the wave packet profiles shown in Fig.~6(b-d) in the main text (see \cite{Kondakci19NC,Bhaduri20NatPhot} for further details).

To estimate the group velocity $\widetilde{v}$ of the 3D ST wave packet, we first arrange for the 3D ST wave packet and the reference pulse to overlap at $z\!=\!0$ as described above. We then axially displace CCD$_{1}$ to a different axial position $\Delta z$, which results in a loss of interference visibility due to the group-velocity mismatch between the 3D ST wave packet traveling at $\widetilde{v}\!=\!c\tan{\theta}$ (where $\theta$ is the spectral tilt angle) and the reference pulse traveling at $\widetilde{v}\!=\!c$. The interference is retrieved, however, by adjusting the delay by $\Delta t$ in the reference arm, from which we obtain the relative group delay between the two wave packets and thence the group velocity $\widetilde{v}\!=\!\tfrac{\Delta z}{\Delta t}$ for the 3D ST wave packet [Fig.~6(e) in the main text]. By repeating this procedure for ST wave packets with different $\theta$ (by varying the parameter $B$), we obtain the data plotted in Fig.~6(f) in the main text.

The uncertainty in the group-velocity measurement is estimated using the propagation-of-errors principle \cite{Bevington02BOOK}. In our case, the largest contribution to errors in estimating $\widetilde{v}$ stems from the uncertainty $\delta t$ in estimating the group delay $\Delta t$, which is limited by the pulse-width $\Delta T$, which we set at $\delta t\!=\!\Delta T/10$. The error in the estimated value of $\widetilde{v}$ is $\delta\widetilde{v}\!=\!|\tfrac{\partial\widetilde{v}}{\partial t}|\delta t\!=\!\tfrac{\widetilde{v}^2}{\Delta z}\delta t$. Using this relationship, we calculate the error bars in Fig.~6(f) in the main text. 

\subsection{Measurements of the field amplitude and phase for 3D ST wave packets}

\begin{figure}[t!]
  \begin{center}
  \includegraphics[width=17cm]{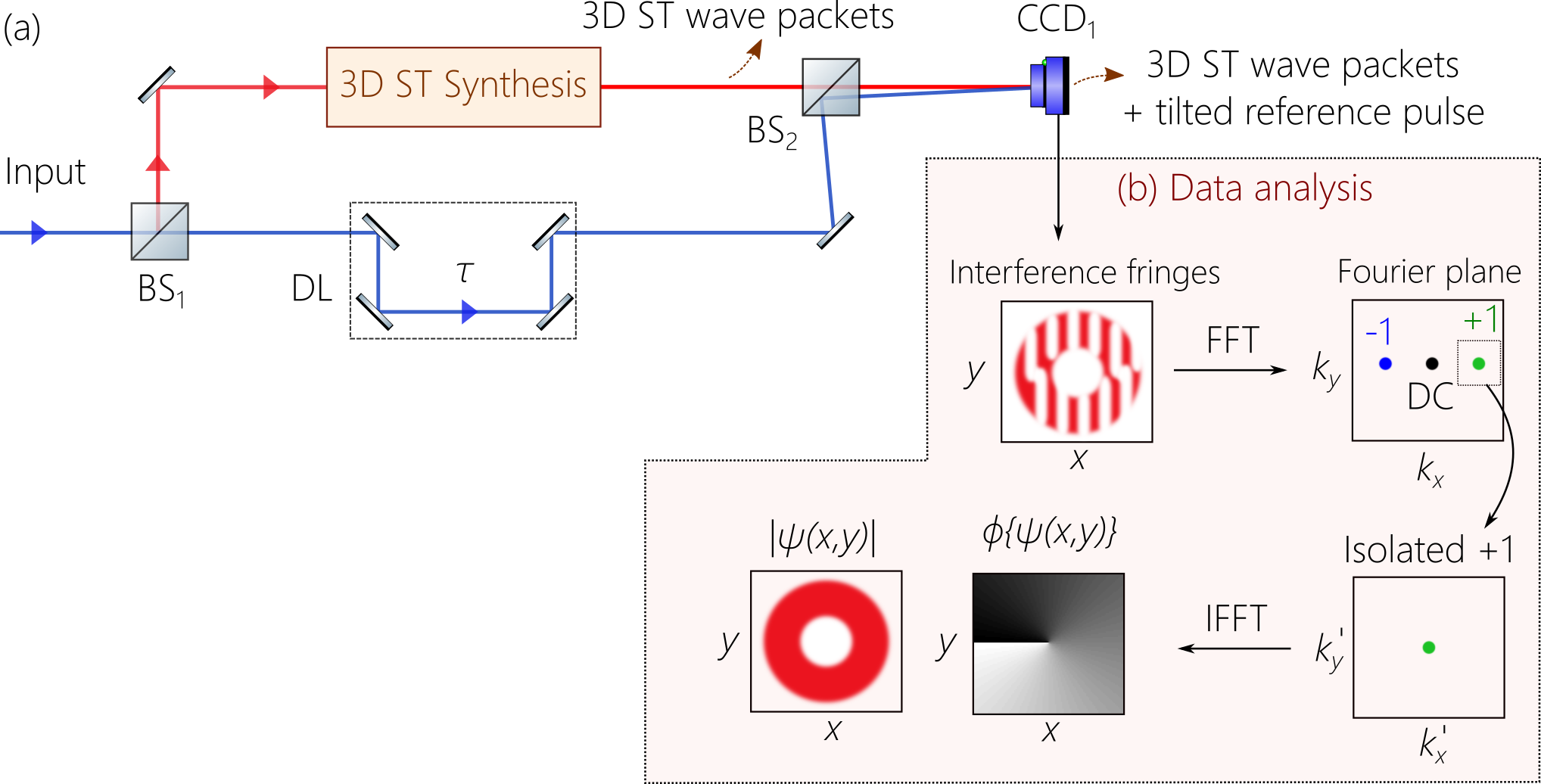}
  \end{center}\vspace{-5mm}
  \caption{\textbf{Supplementary Fig.~17.} Schematic depiction of the off-axis holography strategy for reconstructing the complex field for 3D ST wave packets. (a) The system is similar to that in \ref{Fig:ST_characterization} except that the reference pulse is spatially tilted with respect to the 3D ST wave packet. (b) Outline of the procedure for estimating the field amplitude and phase from the spatially resolved interference fringes.\\}
  \label{Fig:Phase_retrieval}
  \vspace{12mm}
\end{figure}

Finally, we make use of off-axis digital holography \cite{Cuche99OL,Cuche00AO,Sanchez-Ortiga14AO} to obtain the field amplitude and phase of 3D ST wave packets at fixed locations along $z$ and at fixed instances in time $\tau$. This is especially crucial to measure the phase of the OAM-carrying 3D ST wave packets shown in Fig.~7(b) in the main text. 

\begin{figure}[t!]
  \begin{center}
  \includegraphics[width=13cm]{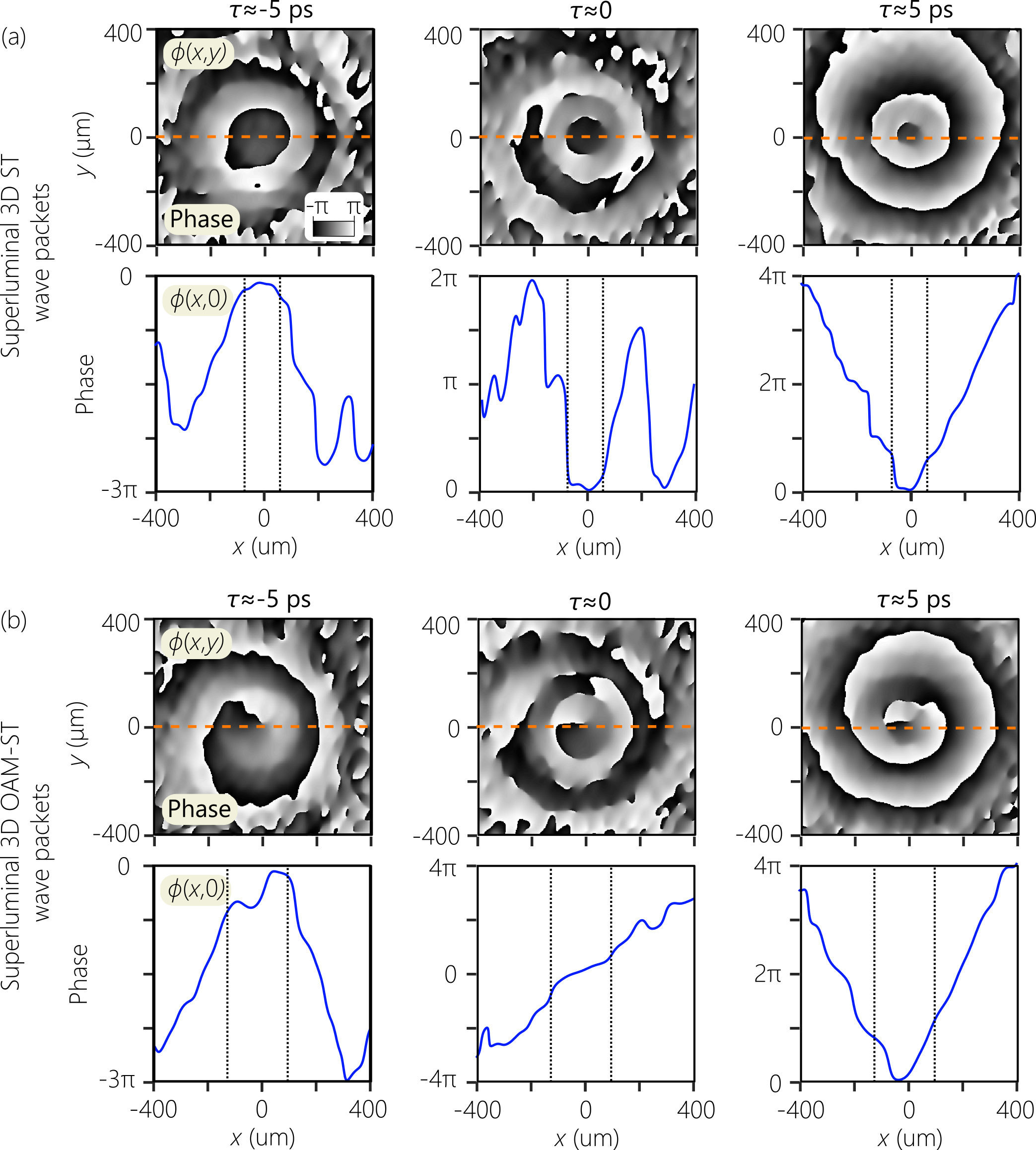}
  \end{center}\vspace{-5mm}
  \caption{\textbf{Supplementary Fig.~18.} Measured phase profiles for 3D ST wave packets with and without OAM. (a) Measured transverse phase profile $\phi(x,y)$ (first row) and on-axis phase profile $\phi(x,y=0)$ along the $x$-axis (second row) at delays $\tau\approx0,\pm5$~ps for a superluminal 3D ST wave packet. Here $z\!=\!30$~mm and $\ell=0$; i.e., there is no OAM structure. (b) Same as (a) except that $\ell=1$; i.e., OAM structure has been introduced into the field.}
  \label{Fig:Measured_phase}
  \vspace{12mm}
\end{figure}

We make use of the off-axis digital holography (ODH) methodology \cite{Cuche99OL,Cuche00AO,Sanchez-Ortiga14AO} to reconstruct the complex field of 3D ST wave packets $\psi(x,y,z;t)\!=\!|\psi(x,y,z;t)|e^{i\phi(x,y,z;t)}$. For this purpose, the same Mach-Zehnder configuration from the previous section is exploited but with a slight modification -- we add a small angle between the propagation directions of the short reference pulse and the 3D ST wave packets, which are arranged to overlap in space and time at a fixed axial position $z$ [\ref{Fig:Phase_retrieval}(a)]. The interference pattern captured on CCD$_{1}$ contains a constant background term and an interference term of interest [\ref{Fig:Phase_retrieval}(b)]. Following the ODH algorithm, we perform a digital fast Fourier transform (FFT)
to separate the constant background from the interference term. By digitally isolating the first diffraction order of the Fourier-transformed image, we access the interference term that contains the complex field. Finally, the inverse FFT of the centered first term gives the amplitude $|\psi(x,y,z;\tau)|$ and phase $\phi\{\psi(x,y,z;\tau)\}$ of the 3D ST wave packets at that axial location $z$ and time $\tau$ (see \cite{Cuche99OL,Cuche00AO,Sanchez-Ortiga14AO}  for more details). By repeating this procedure for several instances $\tau$ we obtain the data plotted in Fig.~7 of the Main text. 

The measured phase structure is plotted in \ref{Fig:Measured_phase}(a) for $\ell\!=\!0$ and in \ref{Fig:Measured_phase}(b) for $\ell\!=\!$. In the first case, in absence of OAM, the phase distribution takes a Gaussian form. The curvature increases as we move away from the center of the wave packet $\tau\!=\!0$. We highlight the phase along the center of the wave front $y\!=\!0$, and especially where the intensity of the field is appreciable (between the two vertical dotted lines). Here the phase is flat at $\tau\!=\!0$, and becomes Gaussian as $\tau$ increases. This is to be expected for a monochromatic Gaussian beam going through its focal point (where the phase front is flat). However, this is seen here in the time domain, which is a manifestation of so-called `time diffraction' \cite{Porras17OL,Kondakci18PRL,Yessenov20PRLveiled}.

In presence of non-zero OAM, superimposed on the above-described behavior of the phase front is a helical phase distribution. At $\tau\!=\!0$ the helical phase front flattens out, but appears as we move away from the wave packet center [\ref{Fig:Measured_phase}(b)]. Once again, this is the expected behavior in space for a diffracting OAM mode when examined from the beam waist outward, but observed here in the time domain.

\clearpage

%